\documentclass[%
reprint,
superscriptaddress,
 amsmath,
 amssymb,
 aps,
prab,
]{revtex4-2}

\usepackage{graphicx}
\usepackage{dcolumn}
\usepackage{bm}
\usepackage{algorithm,algcompatible}
\usepackage{textcomp}
\usepackage[capitalize]{cleveref}
\usepackage[acronym]{glossaries}
\usepackage[dvipsnames]{xcolor}

\newacronym[sort=a]{AI}{AI}{Artificial Intelligence}
\newacronym[sort=a]{ML}{ML}{Machine Learning}
\newacronym[sort=a]{NN}{NN}{Neural Network}
\newacronym[sort=a]{BO}{BO}{Bayesian Optimization}

\begin{document}

\preprint{APS/123-QED}

\title{Bayesian Optimization Algorithms for Accelerator Physics}

\author{Ryan Roussel}
\email{rroussel@slac.stanford.edu}

\author{Auralee L. Edelen}
\author{Tobias Boltz}
\author{Dylan Kennedy}
\author{Zhe Zhang}
\author{Fuhao Ji}
\author{Xiaobiao Huang}
\author{Daniel Ratner}
\affiliation{SLAC National Laboratory, Menlo Park, California 94025, USA}

\author{Andrea {Santamaria Garcia}}
\author{Chenran Xu}
\affiliation{Karlsruhe Institute of Technology, Karlsruhe, Germany}

\author{Jan Kaiser}
\author{Angel Ferran Pousa}
\affiliation{Deutsches Elektronen-Synchrotron DESY, Germany}

\author{Annika Eichler}
\affiliation{Deutsches Elektronen-Synchrotron DESY, Germany}
\affiliation{Hamburg University of Technology, 21073 Hamburg, Germany}

\author{Jannis O. Lübsen}
\affiliation{Hamburg University of Technology, 21073 Hamburg, Germany}

\author{Natalie M. Isenberg}
\author{Yuan Gao}
\affiliation{Brookhaven National Laboratory, Upton, New York 11973, USA}

\author{Nikita Kuklev}
\author{Jose  Martinez}
\author{Brahim Mustapha}
\affiliation{Argonne National Laboratory, Lemont, Illinois 60439, USA}

\author{Verena Kain}
\affiliation{European Organization for Nuclear Research, Geneva, Switzerland}

\author{Christopher Mayes}
\affiliation{SLAC National Laboratory, Menlo Park, California 94025, USA}

\author{Weijian Lin}
\affiliation{Cornell University, Ithaca, New York 14853, USA}

\author{Simone Maria Liuzzo}
\affiliation{The European Synchrotron Radiation Facility, Grenoble, France}

\author{Jason St. John}
\affiliation{Fermi National Laboratory, Batavia, Illinois 60510, USA}

\author{Matthew J. V. Streeter}
\affiliation{Queen's University Belfast, Belfast, Northern Ireland, United Kingdom}

\author{Remi Lehe}
\affiliation{Lawrence Berkeley National Laboratory, Berkeley, California 94720, USA}

\author{Willie Neiswanger}
\affiliation{Stanford University, Stanford, California 94305, USA}

\date{\today}

\begin{abstract}
Accelerator physics relies on numerical algorithms to solve optimization problems in online accelerator control and tasks such as experimental design and model calibration in simulations.
The effectiveness of optimization algorithms in discovering ideal solutions for complex challenges with limited resources often determines the problem complexity these methods can address.
The accelerator physics community has recognized the advantages of Bayesian optimization algorithms, which leverage statistical surrogate models of objective functions to effectively address complex optimization challenges, especially in the presence of noise during accelerator operation and in resource-intensive physics simulations. 
In this review article, we offer a conceptual overview of applying Bayesian optimization techniques towards solving optimization problems in accelerator physics.
We begin by providing a straightforward explanation of the essential components that make up Bayesian optimization techniques.
We then give an overview of current and previous work applying and modifying these techniques to solve accelerator physics challenges.
Finally, we explore practical implementation strategies for Bayesian optimization algorithms to maximize their performance, enabling users to effectively address complex optimization challenges in real-time beam control and accelerator design.

\end{abstract}

\maketitle

\section{Introduction} \label{sec:introduction}
Future accelerator-based experiments serving the high-energy physics, nuclear physics, and photon science communities will require a considerable increase in the capabilities of accelerator facilities to achieve the research aspirations of the next decade~\cite{gourlay_snowmass21_2022}.
Higher energy and higher brightness particle beams with more stringent requirements on reproducibility will unavoidably require complex accelerator operation stemming from an increase of nonlinear phenomena, stringent beam parameter requirements, machine protection limits, and the varied needs of different user communities.
Additionally, accelerator scientists designing future state-of-the-art accelerator facilities will need to explore and configure combinations of increasingly nonlinear and specialized accelerator elements to reach accelerator design goals, all while respecting practical constraints and minimizing construction costs.

Central to both of these challenges is the need to optimize a set of free parameters to attain a predefined objective.
Examples of this include, varying accelerator control parameters during operations to maximize performance (online tuning/optimization), identifying optimal parameters during the accelerator design process (offline simulated optimization), and matching simulated beam dynamics to experimental measurements (model calibration).
Advancements in optimization algorithms enable us to tackle more challenging optimization problems (ones with more free parameters or more complex behaviors), which in turn, improves the performance and capabilities of accelerators.

Numerical optimization algorithms have long been used to address these challenges, but often suffer from slow convergence to optimal parameter sets, are unstable in noisy environments, and can get trapped in local extrema,  making them difficult to apply in practice while limiting the complexity of optimization tasks that can be addressed.
Recently, a particular class of algorithms known as Bayesian optimization (BO) \cite{mockus1989bayesian, shahriari_taking_2016,rasmussen_gaussian_2006} has gained popularity inside the accelerator field as an efficient approach for solving both online and offline optimization problems.
These algorithms' inherent flexibility, low initialization effort, fast convergence, and robustness to noisy environments make them particularly useful for accelerator physics applications.
Multiple groups inside the accelerator physics community have investigated the advantages and disadvantages of these algorithms for solving various accelerator physics problems.
Furthermore, accelerator physics-specific modifications of basic BO components have been developed to leverage beam physics information, tailor optimization to maintain machine stability, and take advantage of high performance computational clusters.
With these developments, the study of BO techniques in the context of accelerator physics has matured to the point that these techniques are usable in regular accelerator operations and as a general high performance optimization tool in simulation.

This review article aims to inform and facilitate the wider use of BO techniques in accelerator physics by providing an easily accessible guide and reference for this class of optimization algorithms.
We begin with a discussion of the optimization challenges faced by the accelerator physics community in regards to both online control of accelerator facilities and offline optimization of simulations for beam dynamics and equipment design, which motivates the use of BO algorithms.
We then discuss basic and advanced approaches to the principal components of BO algorithms:
the Gaussian-process surrogate model most commonly used in BO; the definition of BO acquisition functions; and how the acquisition function is maximized to choose the next set of measurements.
Throughout we highlight how to incorporate beam physics information into BO algorithms in order to improve optimization performance.
Finally, we conclude with a discussion that places BO in the context of other optimization algorithms, describes best-practices for applying BO algorithms to solving optimization challenges, and future directions for research in this area. 

\section{Background and Motivation}

Optimization algorithms aim to solve the general problem

\begin{align}
	\mathbf{x}^{\ast} = & \arg \max f(\mathbf{x})\label{eq:optimization_problem_obj}\\
	& \mathrm{s.t.}\ c_i(\mathbf{x}) \leq 0 & \forall i \in[1,\ldots,m]\label{eq:optimization_problem_constraints}
\end{align}
In the above formulation, Equation~\ref{eq:optimization_problem_obj} represents the objective function, wherein we seek a parameter set $\mathbf{x}^{\ast}$ that optimizes the function $f(\mathbf{x})$ subject to the $m$ constraints specified in Equation~\ref{eq:optimization_problem_constraints}. 
These constraints may be bounds on the parameter set $\mathbf{x}$, or observables, such as safety and performance requirements. 
The formulation can be trivially transformed into a minimization problem by negating the objective function.

\textit{Iterative} optimization algorithms are a popular choice used to find solutions to Eq.~\ref{eq:optimization_problem_obj}. 
Given an initial point in parameter space, the algorithm generates a point or set of points which are evaluated using the objective and constraining functions.
Results from the evaluations are then passed back to the algorithm to generate the next point(s) to be evaluated.
The final solution is determined once the algorithm reaches a termination condition, for example, a fixed number of iterations or a satisfactory objective function value.
Selecting the right algorithm for a given optimization task is critical to success, as it directly influences the quality of the final solution, the relative speed (number of iterations) needed to identify the optimum parameter set, and resource efficiency of the applied routine (e.g., required beam time, computational resources).

The difficulty of finding a solution of a generic optimization problem is influenced by the number of optimization parameters and the complexity of the objective and constraining functions.
The so-called ``curse of dimensionality" describes the exponential growth of possible parameter states with increasing parameter space dimensionality.
As a result, optimization algorithms that perform well when optimizing a small number of parameters (such as the fitting of three beam matrix elements to quadrupole scan data) can fail to find a solution in a reasonable amount of time when applied to higher dimensional problems (such as tuning the parameters of an entire accelerator beamline).


The complexity of the objective functions also plays a role in the performance of optimization algorithms.
Objective functions that are not convex have a number of local extrema, only one of which is the global optimum.
Depending on their construction, optimization algorithms can converge to a local extremum near the initial starting parameter set, so-called \textit{local} optimization.
\textit{Global} optimization algorithms on the other hand, are designed to escape local extrema and explore the entire parameter space in search of the global optimum. 
For complex objective functions, finding the global optimum is often much more challenging \cite{horst2013global}.

\subsection{Optimization Challenges in Accelerator Physics}
\label{sec:opt_challenges_in_accelerators}

In addition to these general optimization challenges, online optimization of accelerators and offline optimization of physics simulations adds further, unique complications that need to be considered when selecting an ideal optimization algorithm.

\subsubsection{Online Accelerator Control}

\begin{figure*}
	\includegraphics[width=\linewidth]{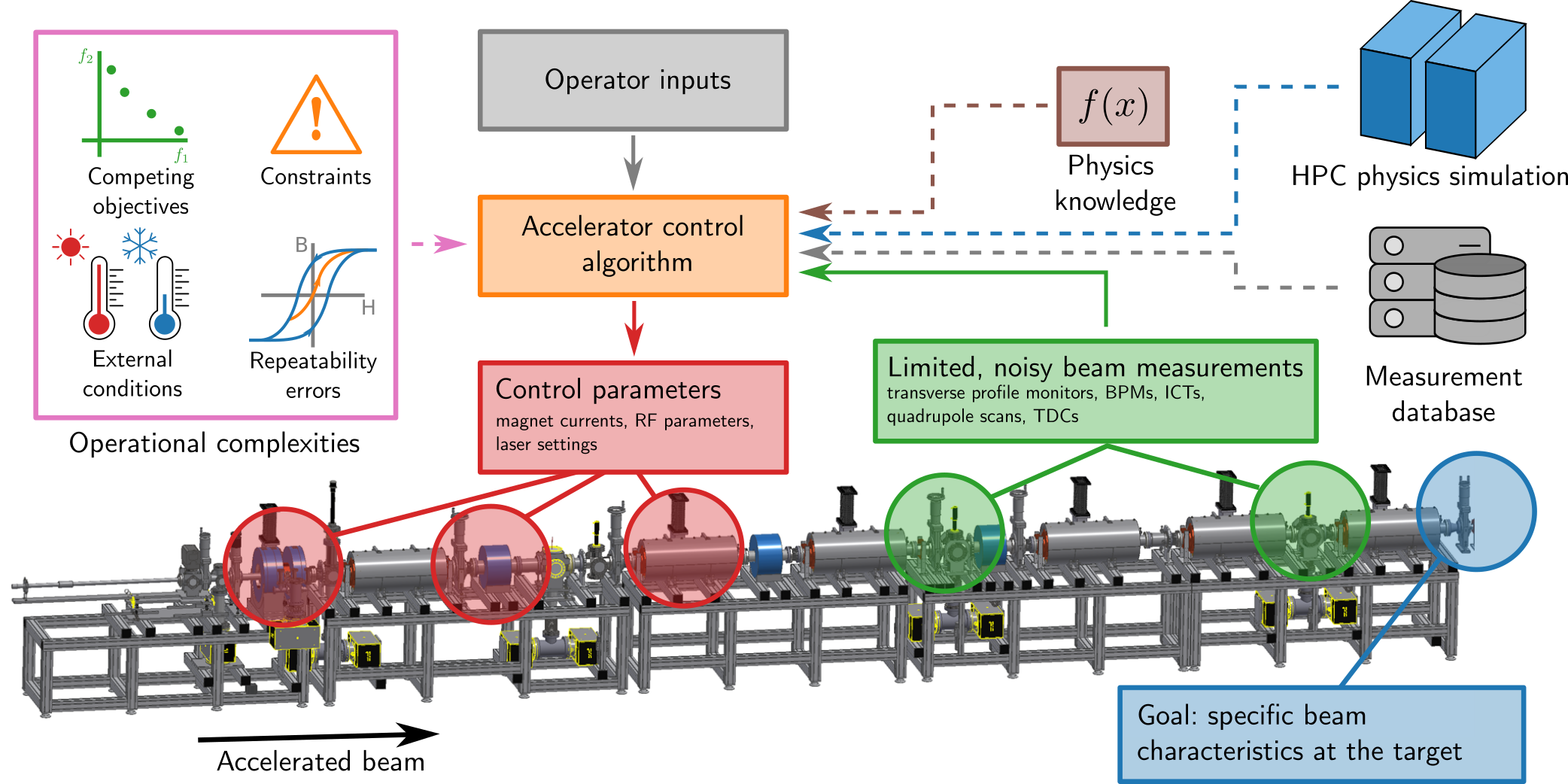}
	\caption{Overview of challenges in using optimization algorithms for online accelerator control. 
		Accelerator control algorithms make decisions about setting a wide variety of accelerator parameters in order to control beam parameters at target locations. 
		Optimal decision making takes into account limited online accelerator measurements, as well as various sources of prior knowledge about the accelerator, including previous measurements, physics simulations, and physics principals.
		Optimization must also consider complicated aspects of realistic accelerator operation including external conditions, feedback systems, safety constraints, and repeatability errors.
	}
	\label{fig:online_optimization}
\end{figure*}

Accelerator measurements are also often subject to aleatoric (random noise) or epistemic (systematic) uncertainties.
Random noise in accelerators makes it difficult for iterative algorithms to maintain stability throughout the optimization process.
This noise can also change in amplitude (e.g.\ heteroskedastic) as a function of accelerator parameters or changing environmental factors.
The amplitude of noise can also be considered as an optimization objective or constraint, given that it is often optimal to find solutions that lead to a relatively stable objective function value.
Additionally, the limited resolution of accelerator diagnostics introduces systematic uncertainties in the objective function value.
Despite this, uncertain measurements can still provide useful information to optimization algorithms if handled appropriately.
Finally, intermittent jumps (e.g. a spike in RF power, dip in beam charge, momentary fault from the machine protection system) in parameters need to be recognized and considered in automated optimization routines.

Particle accelerators are not stationary systems; they have time-dependent behavior on multiple timescales ranging from sub-milliseconds to hours. 
These behaviors include both expected time-dependent processes (such as slow loss of beam in a storage ring) or the combined effect of slow, unintended changes in the system (also known as ``drift'') that changes the relationship between settings and observed beam output. 
Drifts can come from many sources, such as changes in materials (e.g.\ loss of quantum efficiency in a photocathode) and the impact of daily and seasonal changes in temperature and humidity. 
Additionally, not all sources of drift are well-characterized or measured.

Optimizing accelerator control parameters is often framed as a \textit{multi-objective} problem, where the goal of optimization is to find a collection of potential solutions that balance trade-offs between competing objectives.
For example, many photoinjectors aim to simultaneously minimize both transverse beam emittances and bunch length of beams for high brightness applications \cite{bazarov2005multivariate}.
However, due to space charge effects, reducing the bunch length often increases the transverse beam emittance.
Multi-objective optimization identifies a set of parameter configurations that provide ideal trade-offs between objectives, known as the Pareto front (PF).
Once the PF has been identified, a single point on this PF can be selected based on objective preferences as a fixed operating point, or the entire front can be utilized to provide multiple operating modes for different applications.

Accelerators often operate in tightly constrained parameter spaces that limit beam losses that contribute to accelerator downtime, radiation generation, and hardware degradation.
This is especially important for high-power beams, since even the lower-density edges of the beam distribution (or ``halo'') can damage equipment if the trajectory is not carefully controlled.
These limits are often unknown prior to performing optimization, so algorithms must learn valid and invalid regions of parameter space that satisfy the constraints on-the-fly during optimization.
On the other hand, there are cases where operational constraints are not as strict, such as beam losses in lower power facilities or non-safety related beam quality constraints (maximum beam emittance or energy spread).
In these cases, occasional violations of the constraints can be tolerated if they lead to increased convergence speed to optimal values.
Algorithm design for online accelerator operations needs to balance conservative adjustments of accelerator parameters to avoid constraint violations with the need to explore the input space to find optimal solutions.

The type of operating conditions for a particular accelerator also impacts how challenging it is to arrive at an optimal configuration. 
Some particle accelerators deliver highly customized beams to their users, which requires a new combination of accelerator settings for each request, however often. 
Large changes in machine setup introduce additional challenges, such as the need to deal with path-dependent processes like magnetic hysteresis and mechanical backlash. 
Additionally, rapid changes to accelerator parameters can lead to instabilities in the machine due to interacting feedback algorithms in multiple accelerator sub-systems. 
The degree to which these processes need to be considered depends in part on whether the accelerator must undergo somewhat global optimization frequently, as opposed to keeping a single configuration stable for long periods of time. 


\subsubsection{Simulation-Based Optimization of Accelerator Systems}
Optimization algorithms are also used in simulation to design new accelerator components and facilities, as well as calibrate physics models to experimental measurements. 
Simulated optimization shares some of the same challenges as online optimization, including satisfying constraints, limited evaluations, and  balancing the trade-offs between multiple competing objectives.

Detailed physics simulations are often used in the design of new particle accelerators and new experimental setups.  
To include the full detail of nonlinear beam dynamics or collective effects, computationally intensive, high-fidelity particle-in-cell simulations are often used to accurately make predictions of real-world beam dynamics.
However, running these simulations consumes a significant amount of computational resources and run time.
Thus, using algorithms that reduce the need for high-fidelity simulations is essential to keep computational costs to a minimum.

\begin{figure}
	\includegraphics[width=\linewidth]{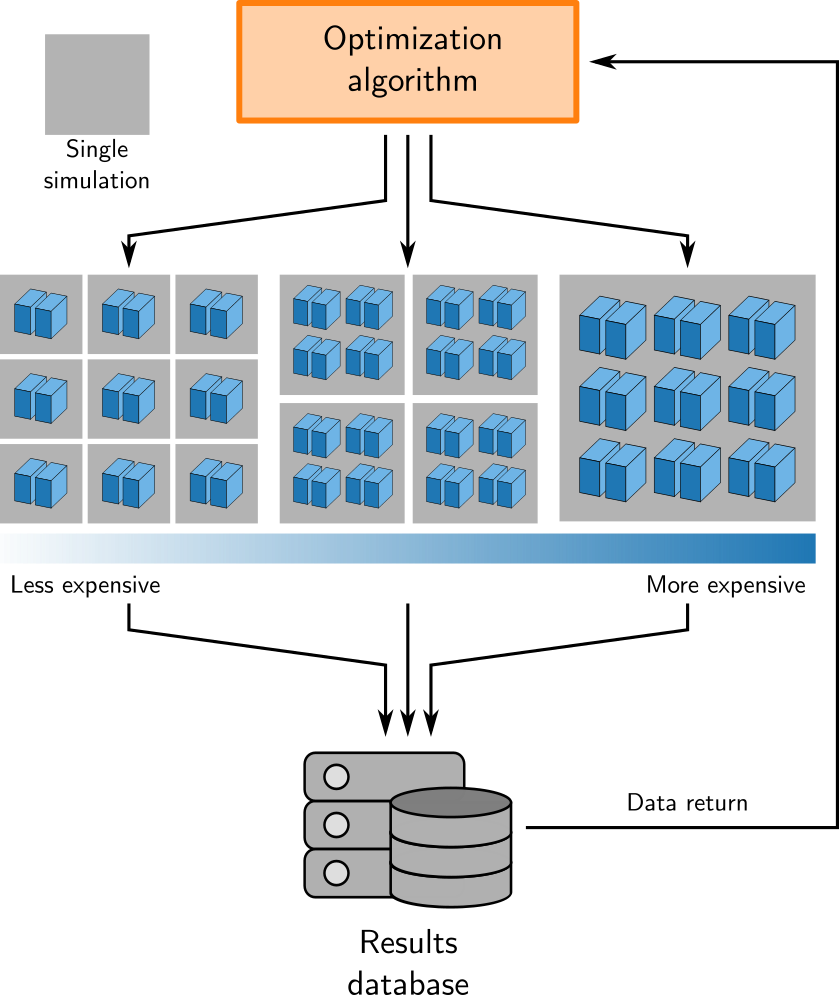}
	\caption{Overview of optimization challenges in accelerator physics simulations. Ideal algorithms aim to minimize the computational cost of performing optimization by orchestrating parallel simulation evaluations at multiple fidelities ranging from analytical models to high fidelity (computationally expensive) simulations. Correlations between simulation predictions at different fidelities can be leveraged to reduce the number of high fidelity simulation evaluations needed to find an ideal solution at the highest fidelity level.}
	\label{fig:simulated_optimization}
\end{figure}

To speed up optimization, multiple simulations can be performed in parallel on high performance computing clusters.
Additionally, reducing the cost of simulations by making approximations or neglecting higher order effects can speed up the optimization progress, but at the cost of predictive accuracy.
In an ideal scenario, multiple, inexpensive evaluations of an approximate model would be used to identify promising regions of parameter space before evaluating expensive, high-fidelity simulations using those parameters.
Accelerator scientists have typically implemented this strategy by pre-selecting a suitable balance between simulation precision and computational cost. However, recent advances in optimization algorithms [citation] enable an automated approach to this process, where simulations of different fidelity can be combined into a single optimization to reduce the overall run time and computational cost, as shown in Fig.~\ref{fig:simulated_optimization}.


\subsection{Optimization Algorithm Selection}
\label{sec:opt_algorithm_selection}
Selecting an optimization algorithm for a specific problem aims to minimize the overall cost needed to reach a given optimization goal.
Costs associated with optimization can be characterized in a variety of ways, for example: beam time at an accelerator, computational assets at a computing cluster, personnel time resources, or financial expenditure.
These costs depend on four factors that make up the optimization process, including (1) the number of steps required to reach an optimization goal, (2) the cost of evaluating objectives (and potentially constraints), (3) the cost of decision-making inside the optimization algorithm (i.e. selecting the next point in parameter space to evaluate), and (4) the initialization cost associated with setting up the optimization algorithm.

Choosing the correct algorithm for solving an optimization problem requires balancing the trade-offs between the different costs associated with performing optimization.
To observe these effects we consider a toy model of the total cost of solving a single optimization problem.
We start by assuming that the steps of performing iterative optimization are done sequentially with a single evaluation of the objective function at a time.
We also assume that an optimization algorithm $A$, takes $N(A)$ steps to find a solution that meets a predefined optimization goal. 
Evaluating the objective and constraint functions has a constant cost $E$ and the algorithm makes decisions (chooses the next point to observe) with a constant cost $D(A)$.
Finally, preparing the algorithm to perform optimization has a one-time initialization cost $I(A)$.
With these assumptions, the total cost $T$ of reaching the predefined optimization goal is given by
\begin{equation}
	T = N(A) [E + D(A)] + I(A).
\end{equation}

From this formula we can observe how making trade-offs between different aspects of performing optimization can be leveraged to reduce the overall cost of finding an optimal solution.
For example, if the evaluation cost is large, we can select more ``sample efficient" optimization algorithms that incur a larger decision making cost in order to reduce the number of iterations needed to find a solution.
On the other hand, if the evaluation cost is relatively low, an inexpensive optimization algorithm can be used to reduce overall cost, even if that means an increase in the number of iterations needed to find a solution.
However, if poor decision making can for instance, lead to safety violations or negatively effect machine stability, it makes sense to use additional resources to make good decisions, regardless of the evaluation cost.
Finally, if more initialization effort leads to faster convergence speed then it makes sense to accept this initialization cost, especially in contexts where optimization of the same problem is repeated multiple times or if the evaluation costs are large.
A careful consideration of all of these factors will minimize the total cost of solving a given optimization problem.

\subsection{Bayesian Optimization}
\begin{figure*}
	\includegraphics[width=\linewidth]{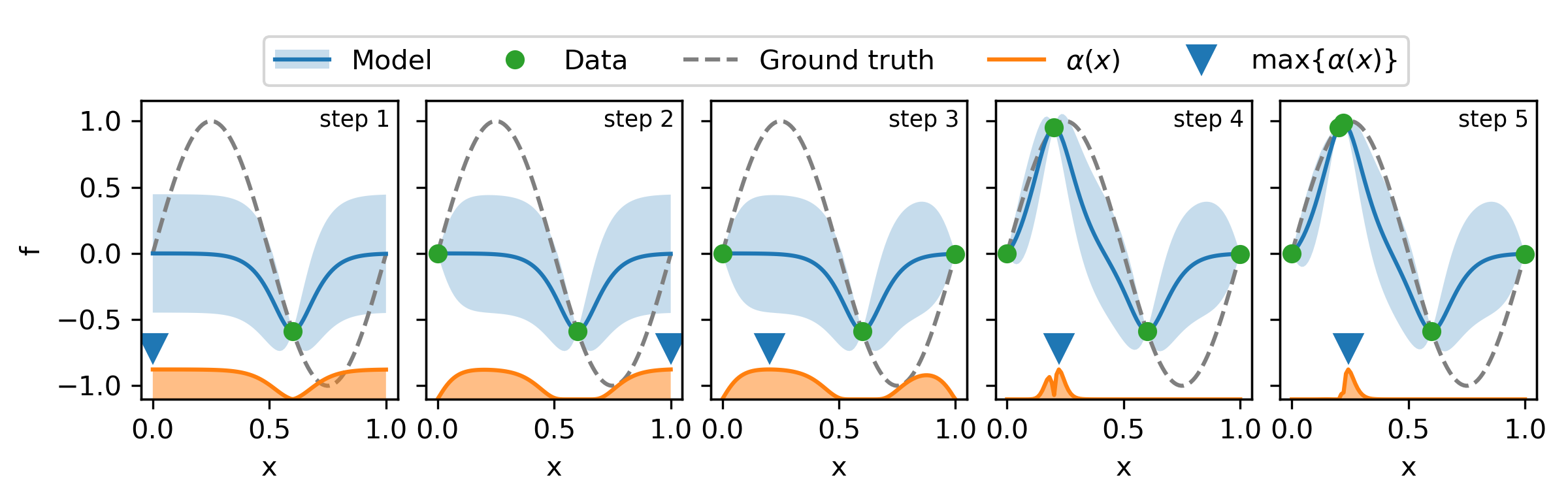}
	\caption{Illustration of the Bayesian optimization process to find the maximum of a simple function.
		A Gaussian process (GP) model makes predictions of the function value (solid blue line) along with associated uncertainties (blue shading) based on previously collected data.
		An acquisition function then uses the GP model to predict the ``value'' of making potential future measurements, balancing both exploration and exploitation.
		The next observation is chosen by maximizing the acquisition function in parameter space.
		This process is repeated iteratively until optimization goals have been reached.
		\label{fig:ucb_cartoon}}
\end{figure*}

Bayesian optimization (BO) is an iterative, model-based optimization algorithm that is particularly well-suited for sample-efficient optimization of noisy, expensive-to-evaluate functions.
In general, BO consists of three steps, as illustrated in Figure \ref{fig:ucb_cartoon} and is summarized in Algorithm~\ref{algo:bo}.
The first is the construction of a statistical surrogate model of the objective and constraining functions based on measured data, often using Gaussian process (GP) modeling \cite{rasmussen_gaussian_2006}.
The second step is the definition of an acquisition function based on the GP model, which defines the relative ``value" of potential future measurements in input space in order to achieve optimization goals.
The final step solves for the point (or set of points) that maximize the acquisition function and are thus predicted to provide the most value towards optimization goals.
Points that are selected in the last step are then passed to the objective and constraint function(s) to be evaluated; the results of which are then passed back to the algorithm to be incorporated into the model data set.
This process repeats until an optimization criteria is met.

An additional benefit of BO is that the model created and trained during the optimization process can also be used outside of the context of optimization.
For example, the model can provide information about objective function sensitivities to accelerator parameters, be integrated as a fast-executing surrogate into other models of the accelerator, or be used to identify unknown parameters of the beamline, such as element misalignments or hysteresis effects.
Finally, as a result of the BO sampling process, these models are often most accurate in regions of parameter space that are of the highest interest, namely regions of parameter space that are near optimal parameter sets.

\begin{figure}
	\begin{minipage}{\linewidth}
		\begin{algorithm}[H] \small
			\caption{Bayesian Optimization} \label{algo:bo}
			\begin{algorithmic}[1]
				\REQUIRE Objective function $f$, observation dataset $\mathcal{D_N}$, GP prior $M=\mathcal{GP}(\mu(\textbf{x}), k(\textbf{x}, \textbf{x}'))$, acquisition function $\mathcal{A}(\cdot)$.
				\FOR {$t=1,2,\ldots$}
				\STATE Decide a new sample point $\textbf{x}_{new}=\arg \max_{\textbf{x}} \alpha(\textbf{x})$.
				\STATE Query the objective $y_{new}=f(\textbf{x}_{new})+\epsilon$.
				\STATE Update $\mathcal{D_N}$ and the GP model.
				\ENDFOR
			\end{algorithmic}
		\end{algorithm}
	\end{minipage}
\end{figure}

\subsection{Demonstrations of BO in Accelerator Physics}
Bayesian optimization has already been used to solve a wide variety of optimization problems in accelerator physics. These demonstrations include:
\begin{itemize}
	\item Single-objective, online and offline optimization of accelerator parameters, e.g. of magnetic optics, RF parameters, in conventional linear \cite{mcintire16BOFEL, duris_bayesian_2020, xu_bayesian_2023, miskovich_online_2022, iwai2023spectral, morita2023accelerator, awal2023optimization} and circular \cite{gao_bayesian_2022, salehi2024bayesian} accelerators, as well as novel accelerator concepts \cite{shalloo_automation_2020,jalas_bayesian_2021,ye_fast_2022,loughran_automated_2023,jalas_tuning_2023, sha2023bayesian}.
	\item Time-dependent optimization to maintain optimal tuning configurations in problems subject to drift \cite{kuklev_online_2022, xu_bayesian_2023, kuklev_robust_2023} (Sec.~\ref{subsec:time_dependence}).
	\item Online optimization that leverages prior physics knowledge or simulations \cite{duris_bayesian_2020, xu_neural_2022} (Sec.~\ref{subsec:custom_kernels},~\ref{subsec:prior_means}). 
	\item Online optimization subject to repeatability errors (hysteresis, motor backlash) \cite{roussel_differentiable_2022} (Sec.~\ref{subsec:hysteresis}).
	\item Autonomous characterization of objective functions in experiment \cite{roussel_turn-key_2021} (Sec.~\ref{subsec:bayesian_exploration}).
	\item Optimization with unknown constraints \cite{Kirschner_2019, Kirschner_2022, Luebsen_2023} (Sec.~\ref{subsec:constraints}).
	\item Multi-objective optimization to discover ideal trade-offs between competing objectives in experiments \cite{ji_multi_2022,jalas_tuning_2023} and simulations \cite{roussel_multiobjective_2021,irshad_multi-objective_2023, huang2021multi} (Sec.~\ref{subsec:mobo}).
	\item Bayesian algorithmic execution, e.g. optimization of beam emittance using virtual quadrupole scans \cite{miskovich2024multipoint} (Sec.~\ref{subsec:virtual_objectives}).
	\item Multi-fidelity optimization, e.g. of beam dynamics and plasma wakefields in simulations \cite{irshad_multi-objective_2023,ferran_pousa_bayesian_2023} (Sec.~\ref{subsec:multi_fidelity}).
\end{itemize}

In the following sections we describe BO techniques in detail; first by introducing common approaches and methods for each step.
We then describe advanced modifications of basic techniques that have been shown to be advantageous towards solving accelerator physics problems.

\section{Gaussian Process Modeling} \label{sec:gp_modelling}
Bayesian optimization uses a computational surrogate model of the objective function in order to inform the selection of new measurement points in input space.
In practice, the surrogate model should use data collected during optimization to make predictions of the objective function value as well as provide an estimate of corresponding uncertainties with those predictions.
While any surrogate model with these properties could potentially be used in this context, models known as ``Gaussian Processes" (GPs) \cite{rasmussen_gaussian_2006} are often used.

\subsection{Bayesian Inference}
Before starting a discussion of models used in BO, it makes sense to first develop a conceptual understanding of \textit{Bayesian statistics}.
A Bayesian interpretation of probability expresses a degree of belief in an event, or a probability distribution, based on prior knowledge of that event.
This is different than a frequentest view of probability which reflects the measured outcomes of many trials.
Bayesian statistics uses \textit{Bayes' rule} to predict the conditional likelihood of an event $A$ occurring given an event $B$ happened as
\begin{equation}
	p(A\ |\ B) = \frac{p(B\ |\ A)p(A)}{p(B)}
\end{equation}
where $p(B\ |\ A)$ is known as the \textit{likelihood function}, $p(A)$ is the \textit{prior} probability distribution, $p(B)$ is the \textit{marginal likelihood} or the \textit{evidence}, and $p(A\ |\ B)$ is the \textit{posterior} probability.

\begin{figure*}
	\includegraphics[width=\linewidth]{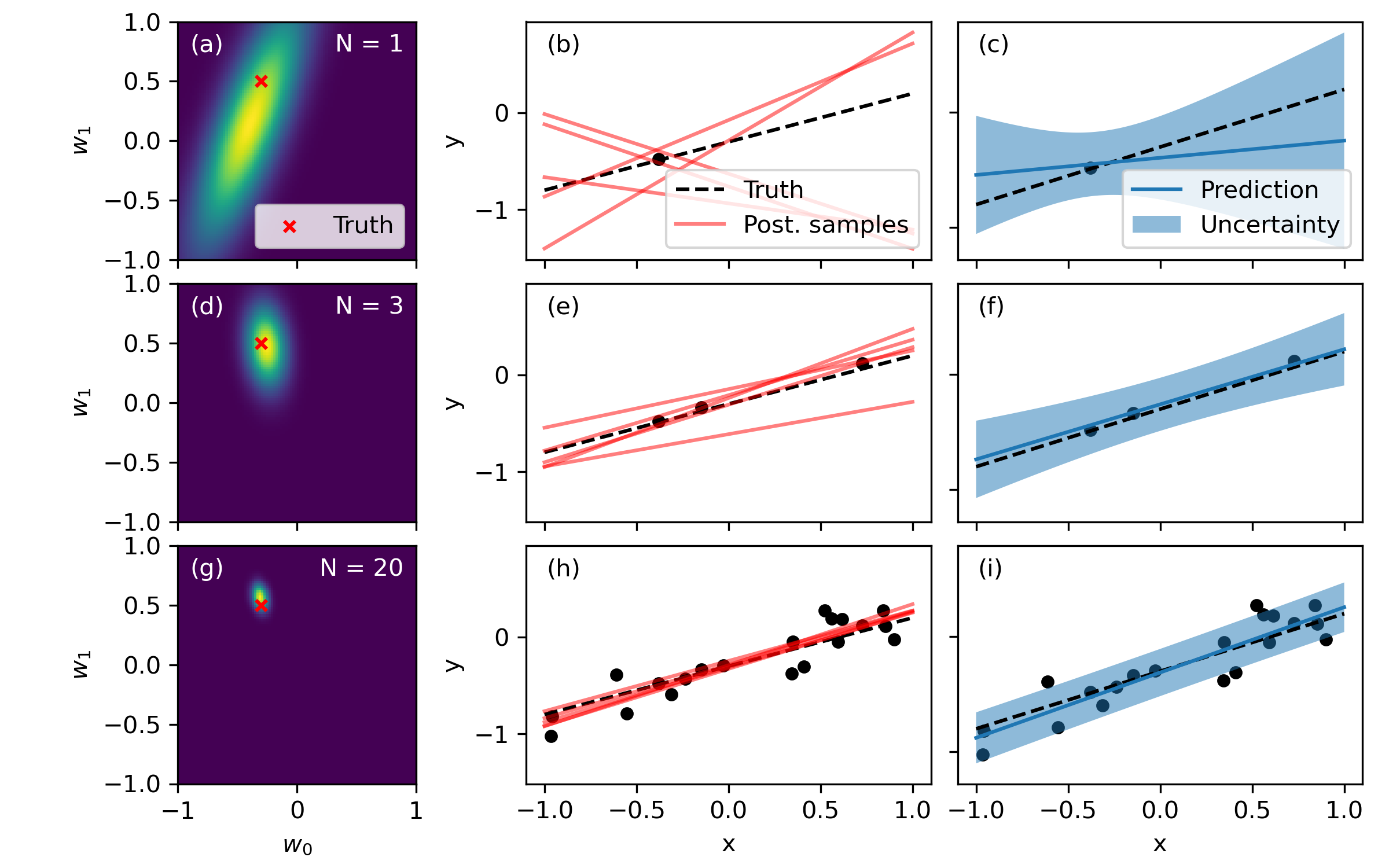}
	\caption{Illustration of Bayesian regression using a linear model $f(x) = w_1 x + w_0$. (a,d,g) Posterior probability density of the linear weights $\{w_1,w_0\}$ conditioned on $N$ observations of the function $y=f(x) + \epsilon$. (b,e,h) Model predictions using random samples of $\{w_1,w_0\}$ drawn from the posterior probability distribution. (c,f,i) Predictive mean (solid line) and 90\% uncertainty intervals (shading) of the posterior model. Red cross and black dashes denote true parameters and values of the function $f(x)$ respectively. Reproduced with permission from \cite{martin_krasser_2020_4318528}.}\label{fig:bayes_regression}
\end{figure*}

To make the interpretation of Bayes' rule more concrete, we examine fitting a linear model $f(x) = w_0x + w_1$ to experimental measurements corrupted by noise $y = f(x) + \epsilon$ as is shown in Fig.\ref{fig:bayes_regression}.
The goal of this analysis is to determine the likelihood of the two model parameters $\{w_0, w_1\}$ given a collection of experimental measurements $\mathcal{D} = \{\mathbf{x}, \mathbf{y}\}$.
Using Bayes' rule, the posterior probability distribution of these parameters is given by
\begin{equation}
	p(w_0,w_1|\mathcal{D}) = \frac{p(\mathcal{D}|w_0, w_1)p(w_0,w_1)}{p(\mathcal{D})}.
\end{equation}
where the likelihood $p(\mathcal{D}|w_0,w_1)$ captures how well the linear model with the given parameter values $\{w_0,w_1\}$ represents the measured data, and $p(w_0,w_1)$ represents the prior probability distribution of the weights.
In this case, the prior distribution of the weights is a multivariate Gaussian distribution centered at the origin.
This prior distribution is equivalent to adding a regularization term to least-squares curve fitting, which aims to prevent over-fitting by penalizing large parameter values.

After a single observation, the posterior probability distribution of the weights is shown in Fig.~\ref{fig:bayes_regression}(a).
Based on the single measured data point Bayes' rule predicts a positive correlation between the y-intercept ($w_0$) and the slope ($w_1$).
This is evident in function samples drawn from the posterior distribution shown in Fig.~\ref{fig:bayes_regression}(b).
These samples can be collected into a distribution that predicts the mean value of the function and associated uncertainty as is shown in Fig.~\ref{fig:bayes_regression}(c).

As additional measurements are introduced into the model, Fig.~\ref{fig:bayes_regression}(d-i), the likelihood and posterior probability distributions become sharper since a smaller range of parameters leads to accurate models of the experimental data.
How rapidly the posterior probability distribution shrinks according to new evidence depends on the relative ``strengths" of the prior and likelihood distributions.
A strong prior probability distribution on the weights (one that is highly peaked in a small local region) will result in a similar posterior distribution unless significant experimental evidence that is contrary to the prior is incorporated into Bayes' rule via the likelihood.
On the other hand, if the prior distribution is relatively weak (ie. nearly uniform across parameter space) then it has relatively little impact on the posterior distribution.

The process of determining the posterior probability distribution of model parameters based on observed data and Bayes' rule is referred to as \textit{Bayesian inference}.
In most cases, determining the exact posterior probability distribution for the entire parameter space requires performing integrals that are computationally intractable to compute, specifically when evaluating the evidence term in Bayes' rule $p(\mathcal{D})$.
Rather than directly assessing the posterior probability distribution, a variety of alternative analytical techniques are used to perform inference.
First is \textit{maximum likelihood estimation} (MLE), which estimates point-like values of the model parameters $\theta$ by solving for the point $\theta^*$ that maximizes the likelihood term (which ignores any priors on the parameters)
\begin{equation}
	\theta^*_{\mathrm{MLE}} = \arg \max_\theta p(\mathcal{D}|\theta).
\end{equation}
If the likelihood takes the form of a Normal distribution, this is equivalent to performing mean squared error curve fitting.

The second analysis method is \textit{maximum a posteriori} (MAP), which also determines point-like values of the parameters $\theta$, this time by maximizing a quantity that is the mode of the posterior distribution
\begin{equation}
	\theta^*_{\mathrm{MAP}} = \arg \max_\theta p(\mathcal{D}|\theta)p(\theta).
\end{equation}
which incorporates the prior without having to compute the full posterior probability distribution.
Finally, we can also determine an approximate posterior probability distribution of $\theta$ using \textit{variational inference}, which uses optimization to fit a computationally tractable distribution to values of the exact posterior distribution in order to minimize the evidence lower bound (ELBO) in terms of the Kullback-Leibler divergence (see \cite{hoffman2013stochastic} for details).
Depending on the application, any of these three methods can be used to estimate posterior parameter values using Bayesian inference, albeit with varying computational costs required to solve the respective optimization problems associated with performing each type of inference.

\subsection{Gaussian Process Modeling Basics}
Gaussian process models \cite{rasmussen_gaussian_2006} are non-parametric models that use Bayes' rule to describe unknown functions by leveraging high level functional behavior to establish correlations between function values at points in objective space.
As opposed to parametric models, which use Bayes' rule to identify probability distributions of model parameters, GP models use Bayes' rule to predict probability distributions of function values at arbitrary locations in parameter space using measured data. 

We start by assuming that the output $y$ of a function $f$ at input parameter $\mathbf{x}$ is given by 
\begin{equation}
	y = f(\mathbf{x}) + \epsilon
\end{equation}
where corrupting noise is given by $\epsilon \sim \mathcal{N}(0, \sigma_\epsilon^2)$.
A GP model is a distribution of possible functions
\begin{equation}
	f(\mathbf{x})\sim \mathcal{GP}(m(\mathbf{x}), k(\mathbf{x},\mathbf{x'}))
\end{equation}
where $m(\mathbf{x}) = \mathbb{E}[f(\mathbf{x})]$ is referred to as the \textit{prior mean function}, and $k(\mathbf{x},\mathbf{x}') = \mathbb{E}[(f(\mathbf{x}) - m(\mathbf{x}))(f(\mathbf{x}') - m(\mathbf{x}'))]$ is commonly called the \textit{covariance kernel function}.
Finally, the probability distribution of the observable $y$ is given by our assumed \textit{likelihood}, which in this case is a Normal distribution $p(y|f(\mathbf{x})) = \mathcal{N}(f(\mathbf{x}), \sigma_\epsilon^2)$.
To simplify calculations the prior mean function is often specified to be $m(\mathbf{x})=0$, although a fixed non-zero prior mean can also be learned from the data.

Given a set of $n$ collected data samples $\mathcal{D} = \{X, \mathbf{y}\}$, we can make predictions for the probability distribution of the function value evaluated at $n_*$ test points 
using Bayesian inference.
The resulting posterior distribution $p(\mathbf{y_*}| X_*, \mathcal{D}) = \mathcal{N}(\bm{\mu_*}, \bm{\sigma_*}^2)$ with the mean and variance given by \footnote{The posterior discussed here is often referred to as the \textit{posterior predictive}. The true posterior of the GP model is with respect to function values and is written as $p(\mathbf{f_*}| \mathbf{X_*}, \mathcal{D})$.}
\begin{align}
	\bm{\mu_*} &= K(X_*, X)[K(X,X) + \sigma_\epsilon^2I]^{-1}\mathbf{y}\label{eq:posterior_mean}\\
	\bm{\sigma_*}^2 &= K(X_*,X_*) -\label{eq:posterior_variance_1}\\ &\hspace{4mm}K(X_*, X)[K(X,X) + \sigma_\epsilon^2I]^{-1}K(X_*,X)^T \nonumber
\end{align}
where $K(X,X)$ is an $n\times n$ covariance matrix between each data set element locations, $K(X_*,X)$ is an $n_*\times n$ covariance matrix between test points and data set element locations, $K(X_*, X_*)$ is an $n_*\times n_*$ covariance matrix between test point locations, and $I$ is the identity matrix.

\begin{figure*}
	\includegraphics[width=\linewidth]{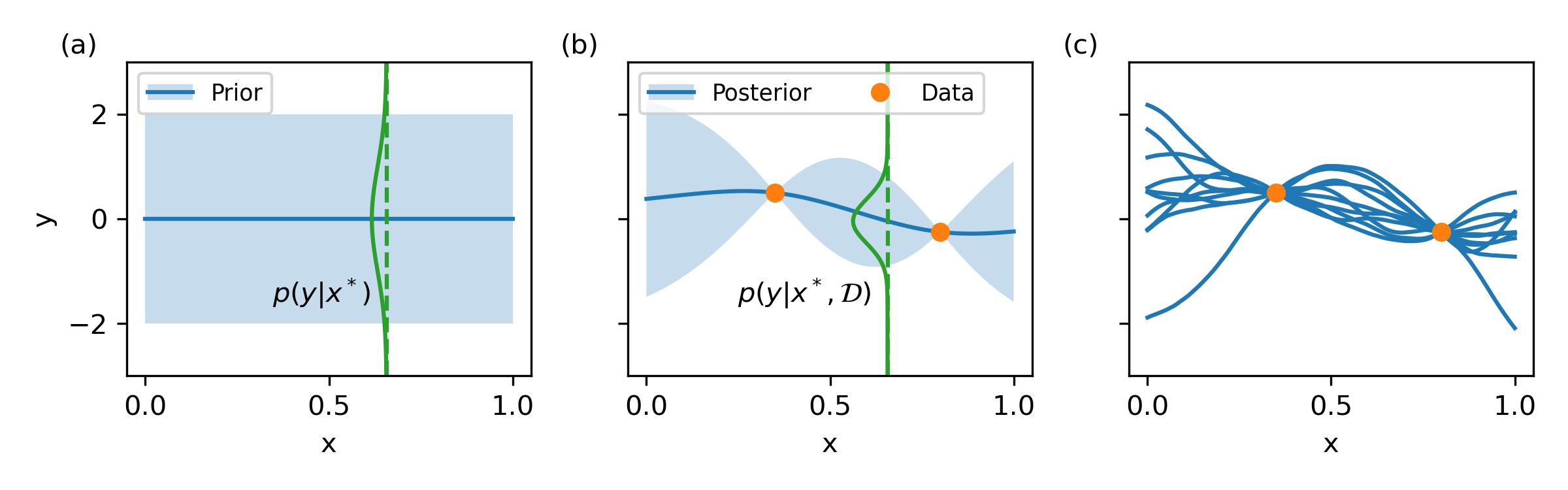}
	\caption{Illustration of GP model predictions. (a) Prior model prediction of the function mean (solid blue line) and confidence interval (blue shading) at a set of test points in parameter space. The probability of the output value $y$ at any given test point $x^*$ is a Normal distribution. (b) The posterior GP model also predicts Normal probability distributions at each test point, conditioned on the data set $\mathcal{D}$.   (c) Individual function samples can also be drawn from the posterior GP model and can be used for Monte Carlo computations of function quantities.}\label{fig:gp_cross_section}
\end{figure*}

An example of GP predictions is shown in Fig. \ref{fig:gp_cross_section}, assuming that the noise parameter $\sigma_\epsilon=0$.
Figure~\ref{fig:gp_cross_section}(a) shows the prior mean and confidence bounds (equal to $2\sigma$ above and below the mean) of the observable $y$ for a set of 100 test points in the domain $x^* \in [0,1]$.
At an arbitrary point in parameter space, the GP prior distribution $p(y|x^*)$ is a Normal distribution with a mean of zero and a unit variance.
By adding a data set $\mathcal{D}$ to the GP, the model predictions are updated to form the posterior predictive distribution as shown in Fig.~\ref{fig:gp_cross_section}(b).
Posterior predictions at a single test point also take the form of Normal distributions with predictive means and variances conditioned on the data set according to Eq.~\ref{eq:posterior_mean} and \ref{eq:posterior_variance_1}.
We can also draw individual function samples at points in parameter space from the posterior distribution, as shown in Fig.~\ref{fig:gp_cross_section}(c).
These function samples are generated by drawing multiple random values from the Normal distribution at every point in input space.

Conceptually, GP models use Bayes' rule to derive a posterior probability distribution of the function value $f(\mathbf{x})$ conditioned on the observed data set and covariances in function values between observed data and test points.
These covariances are defined by the kernel function $k(\mathbf{x},\mathbf{x'})$ and a likelihood function (which describes probabilities due to measurement noise).
A physical analog of GP modeling is a vibrating string with a collection of fixed nodes along the string length.
The possible locations of the string at any point along its length is constrained by where the nodes are located on the $x-y$ plane (observed data) and the elasticity of the string (kernel function).
For a given string we can be quite confident where the string is in space close to fixed nodes.
However, far away from any nodes the string position possibilities can vary widely.
Increases in the elasticity of the string creates more uncertainty in both of these cases owing to its' ability to stretch; this corresponds to weaker covariances between function values. 

\subsubsection{Kernel function definition} \label{par:kernels}
By defining the covariances of function values between different locations in parameter space, the kernel function dictates the overall functional behavior of the predictive model. 
Selection of a particular kernel function is usually based on prior knowledge of the real function's behavior in parameter space and is critical to creating accurate models with limited amounts of data.
Kernel functions are often contain \textit{hyperparameters}, which alter the high level functional behavior of the GP posterior, and can be specified prior to modeling or inferred from training data. 
Kernels are generally divided into two categories, stationary and non-stationary.

Stationary kernels are invariant under translations of the input space $k(\textbf{x},\textbf{x}') = k_S (||\textbf{x} - \textbf{x}'||)$, which means it only depends on the relative positions of its two inputs \textbf{x} and \textbf{x}', and not on their absolute positions. 
This feature makes stationary kernels a popular choice for modeling arbitrary functions when limited prior information is present.

One of the most basic stationary kernels is the Radial Basis Function kernel (RBF).
The RBF kernel is defined as:
\begin{equation}
	k_\text{RBF} (\textbf{x},\textbf{x}') = \exp \left(- \frac{||\textbf{x} - \textbf{x}'||^2}{2l^2} \right)
\end{equation}
where $l$ is the length scale hyperparameter of the kernel.
\begin{figure}
	\includegraphics[width=\linewidth]{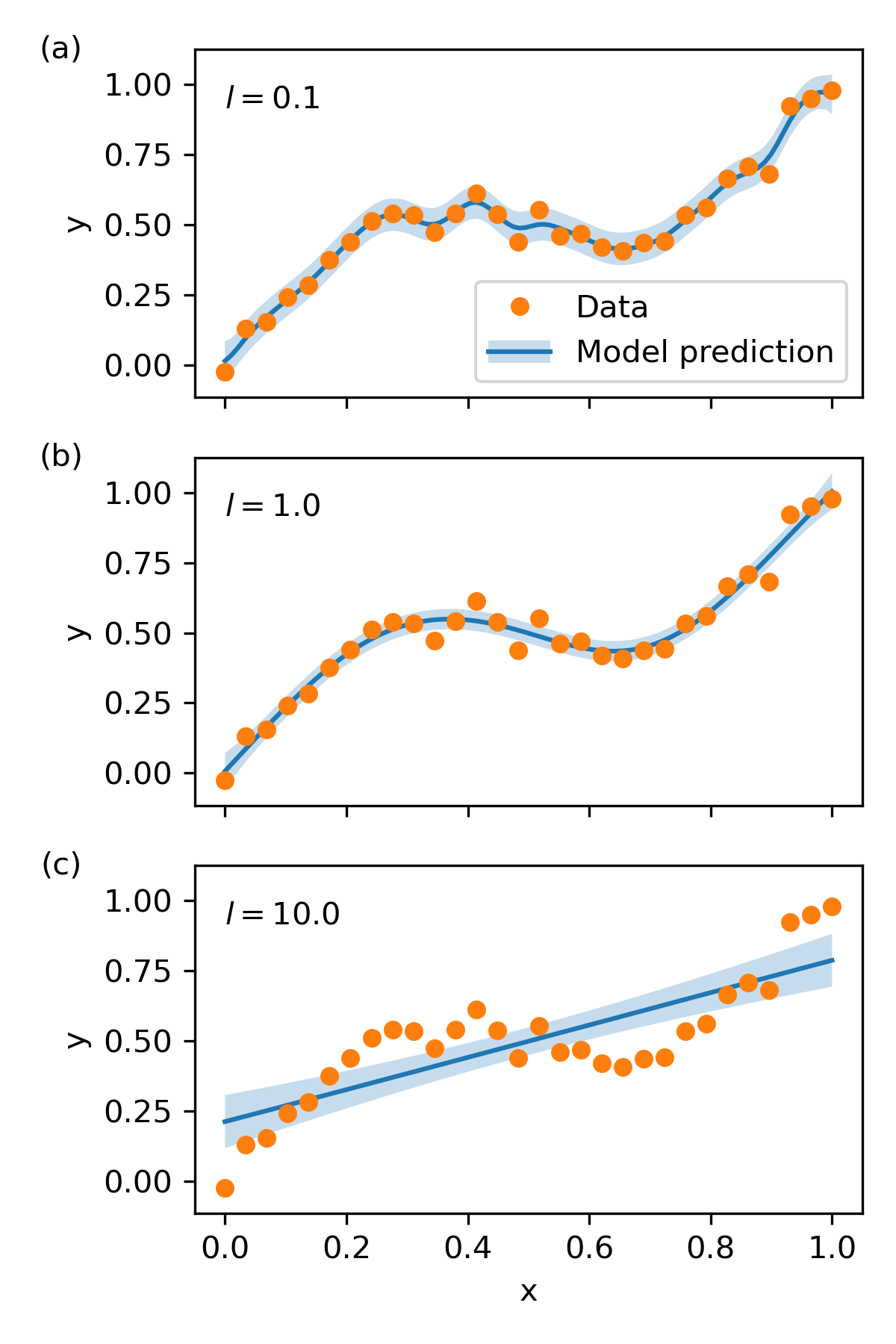}
	\caption{Visualization of how the length scale hyperparameter $l$ effects GP modeling. Three GP models are trained on the same data set using a Matérn kernel with fixed length scales of (a) 0.1, (b) 1.0, and (c) 10.0. Remaining hyperparameters are trained by maximizing the marginal log-likelihood. \label{fig:lengthscale_example}}
\end{figure}
While the RBF kernel's simplicity makes it easy to use and adapt to specific purposes (see Sec.~\ref{subsec:custom_kernels}), it results in predictions that are infinitely differentiable, which are generally too smooth for describing realistic functions.

A more generalized version of the RBF kernel is the Matérn kernel~\cite{rasmussen_gaussian_2006}. 
The Matérn kernel is defined as:

\begin{equation}
	k_\text{MA}(\textbf{x},\textbf{x}') = \frac{2^{1-\nu}}{\Gamma(\nu)} \left( \sqrt{2\nu} \frac{d}{l} \right)^\nu K_\nu \left( \sqrt{2\nu} \frac{d}{l} \right)
\end{equation}

Here, $d = ||\textbf{x} - \textbf{x}'||$ represents the Euclidean distance between inputs, $\Gamma$ is the gamma function, and $K_\nu$ is the modified Bessel function of the second kind. 
The length scale of the kernel is denoted by $l$, and $\nu$ controls the smoothness of the resulting function.
As $\nu \rightarrow \infty$, the Matérn kernel converges to the RBF kernel. 

Commonly used values for $\nu$ are $\nu = 1.5$ for once differentiable functions and $\nu = 2.5$ for twice differentiable functions. 
Limiting the differentiability enables GP models with Matérn kernels to accurately predict realistic physical processes. 
As a result, the Matérn kernel with $\nu = 2.5$ is often employed as a starting point for modeling physical functions in the absence of prior information.

For modeling functions that are expected to be periodic, a periodic kernel can be used to increase model accuracy containing small data sets. 
A periodic kernel, also called a Exp-Sine-Squared kernel, is defined as:
\begin{equation}
	k_\text{PER} (\textbf{x},\textbf{x}') = \exp \left(- \frac{2 \sin^2 (||\textbf{x} - \textbf{x}'|| / p)}{l^2} \right)
\end{equation}
where $l$ is the length scale of the kernel, and $p$ is the periodicity of the kernel.

Hyperparameters of these kernels control high-level model behavior by modifying the covariance between function values at different points in parameter space.
For example, Fig.~\ref{fig:lengthscale_example} shows how the length scale hyperparameter of a stationary kernel effects GP predictions.
Models that contain kernel functions with short length scales vary rapidly to precisely match the training data.
As the length scale increases the smoothness of the model prediction increases, capturing more of the general trend of the training data with a reduction in accuracy.
Selecting hyperparameter values depends on improving the accuracy of the GP model while reducing the complexity of the model, thus preventing over fitting of the data (see the next section for a detailed discussion).
Other hyperparameters, such as the period in the periodic kernel and the offset in the polynomial kernel, have similar macro-scale effects on model behavior, and thus significantly impact the accuracy of model interpolation and extrapolation.

One straightforward modification of stationary kernels often used in practice is replacing the scalar length scale hyperparameter with a vector of independent length scales corresponding to each optimization parameter.
In this case, the RBF kernel for example can be specified by 
\begin{equation}
	\label{eq:factor_analysis_distance}
	k(\mathbf{x}, \mathbf{x}') = \exp\Big(-\frac{1}{2}(\mathbf{x} - \mathbf{x}')^TM(\mathbf{x} - \mathbf{x}')\Big)
\end{equation}
with $M=\mathrm{diag}(\mathbf{l})^{-2}$ where $\mathbf{l}$ is a vector of positive real values.
This technique is often referred to as \textit{automatic relevance determination} \cite{neal2012bayesian} and can be used to identify the sensitivity of the objective function to each optimization parameter. 
A long length scale implies weak dependence of the objective function on a particular parameter while a small length scale implies strong dependence.

The flexibility of this approach can be expanded even further by specifying a full positive semi-definite matrix $M=\Lambda^T\Lambda + \mathrm{diag}(\mathbf{l})^{-2}$, where $\Lambda$ is an upper triangular matrix, often referred to as \textit{factor analysis distance} due to the analogy with factor analysis methods used to find low rank decomposition of the data along arbitrary axes in parameter space.
This parameterization is used less often in practice due to the large data sets necessary to learn the covariances on-the-fly during optimization.
It can however be useful in cases where the decomposition can be determined prior to conducting optimization and the low rank behavior is not aligned with individual parameter axes, as is often the case when tuning quadrupole parameters (see Sec.~\ref{subsec:custom_kernels}).
Automatic relevance determination and factor analysis distance methods allow the GP model to represent low dimensional structure within high dimensional optimization spaces, increasing model accuracy with fewer data points.

Non-stationary kernels depend explicitly on the locations of the two inputs \textbf{x} and $\textbf{x}'$. 
Using non-stationary kernels can provide more accurate predictions with fewer data points, at the cost of reduced model flexibility.
A commonly used non-stationary kernel is the polynomial kernel~\cite{genton_classes_2002}. 
A polynomial kernel of degree $p$ is defined as:
\begin{equation}
	k_\text{POL} (\textbf{x},\textbf{x}') = (\textbf{x}^T \textbf{x}' + c)^p
\end{equation}
where $c \geq 0$ is a constant offset parameter.
Using a polynomial kernel in a GP model is equivalent to performing Bayesian regression of data using the same-order polynomial.
Functional samples drawn from the GP model are then also polynomial functions of the same order as the kernel.

\subsubsection{Determining model hyperparameters}
The hyperparameters of the GP kernel can be learned from training data gathered during optimization or specified \textit{a priori} using prior knowledge of the objective function.
A common strategy for learning the hyperparameters from experimental data is maximizing the \textit{marginal log-likelihood} (MLL) of the GP model with respect to the hyperparameter values.
While in most cases calculating the marginal likelihood requires performing analytically intractable integrals, the marginal likelihood of GP models with Gaussian likelihoods can be calculated analytically, and is given by
\begin{equation}
	\log p(\mathbf{y}| X, \mathbf{\theta}) = -\frac{1}{2}\mathbf{y}^TK_y^{-1}\mathbf{y} - \frac{1}{2}\log |K_y| - \frac{n}{2}\log 2\pi.
\end{equation}
where $\theta$ is the set of GP model hyperparameters contained in $K_y=K(X,X) + \sigma_\epsilon^2I$.
The MLL has three terms, each having an interpretable role. 
The first term, which is the only term that contains training data, is the data fit term which is maximized when model predictions accurately predict experimental data.
The second term describes model complexity and is maximized given the simplest model, ie., models whose kernel matrices have determinants close to zero.
The final term is a normalization constant based on the number of training points in the data set.
Maximizing the MLL naturally regularizes fitting of the GP, resulting in model hyperparameters that create the simplest model which accurately reproduces the training data.
For relatively small data sets ($<$ 300 data samples), maximizing the MLL takes a few seconds on most modern CPUs, making it feasible to perform this process during each iteration of BO (see Sec.~\ref{subsec:computational_benchmarks} for details).

Alternatively, fixed individual hyperparameter values can be specified before modeling occurs, based on prior knowledge of the function, either from previous sets of data or physics knowledge.
While fixing hyperparameter values circumvents the need for retraining the model at each optimization step during BO, this limits the ability of BO to adapt to novel functional behavior that is not well characterized by the fixed hyperparameter values.

Since maximizing the MLL is itself an optimization problem, this process suffers from the same complexities and challenges associated with solving general optimization problems in practice.
A wide variety of numerical optimization algorithms can be used for this purpose, given that the number of hyperparameters that are included inside the GP model is generally small ($< 5-10$).
Current state-of-the-art software packages developed for GP modeling (see Sec.~\ref{subsec:implementations}) employ two strategies to maximize speed and robustness when optimizing the MLL.

The first strategy uses of so-called \textit{differentiable} calculations, which allow cheap computation of the MLL gradient with respect to the hyperparameters. 
This enables the use of gradient-based optimization algorithms that scale well towards performing optimization given a large number of hyperparameters.
Since gradient descent optimization algorithms often converge to local extrema, optimization can be repeated in parallel, starting with randomly chosen initial points in hyperparameter space to improve the chances of finding a global extrema.

The second strategy used to improve MLL maximization robustness is training data normalization and standardization.
As is common in other machine learning disciplines, it is recommended that training data sets are transformed such that they are near unity value when passed to the model, thus preserving unit scale gradients with respect to hyperparameters.
For GP modeling, it is common to normalize input data into the unit domain $[0,1]$ and standardize the outcome data such that it has a mean of zero and a standard deviation of one (to match the default zero prior mean and unit standard deviation in most GP modeling frameworks). 

These two strategies make maximizing the MLL fairly robust in practice, such that monitoring and customizing the fitting of model hyperparameters in BO is only necessary in specialized cases.

\subsubsection{Observation noise and heteroskedasticity}
In most cases when performing online optimization of accelerator parameters, measurements of the objective function are corrupted by noise and/or systematic uncertainties.
Through the use of Bayesian inference, GP models explicitly support a notion of measurement uncertainty when making predictions.
Furthermore, depending on the application, GP models can be tailored to account for measurement uncertainty in a variety of ways based on measurements or prior physics information.

\begin{figure}
	\includegraphics[width=\linewidth]{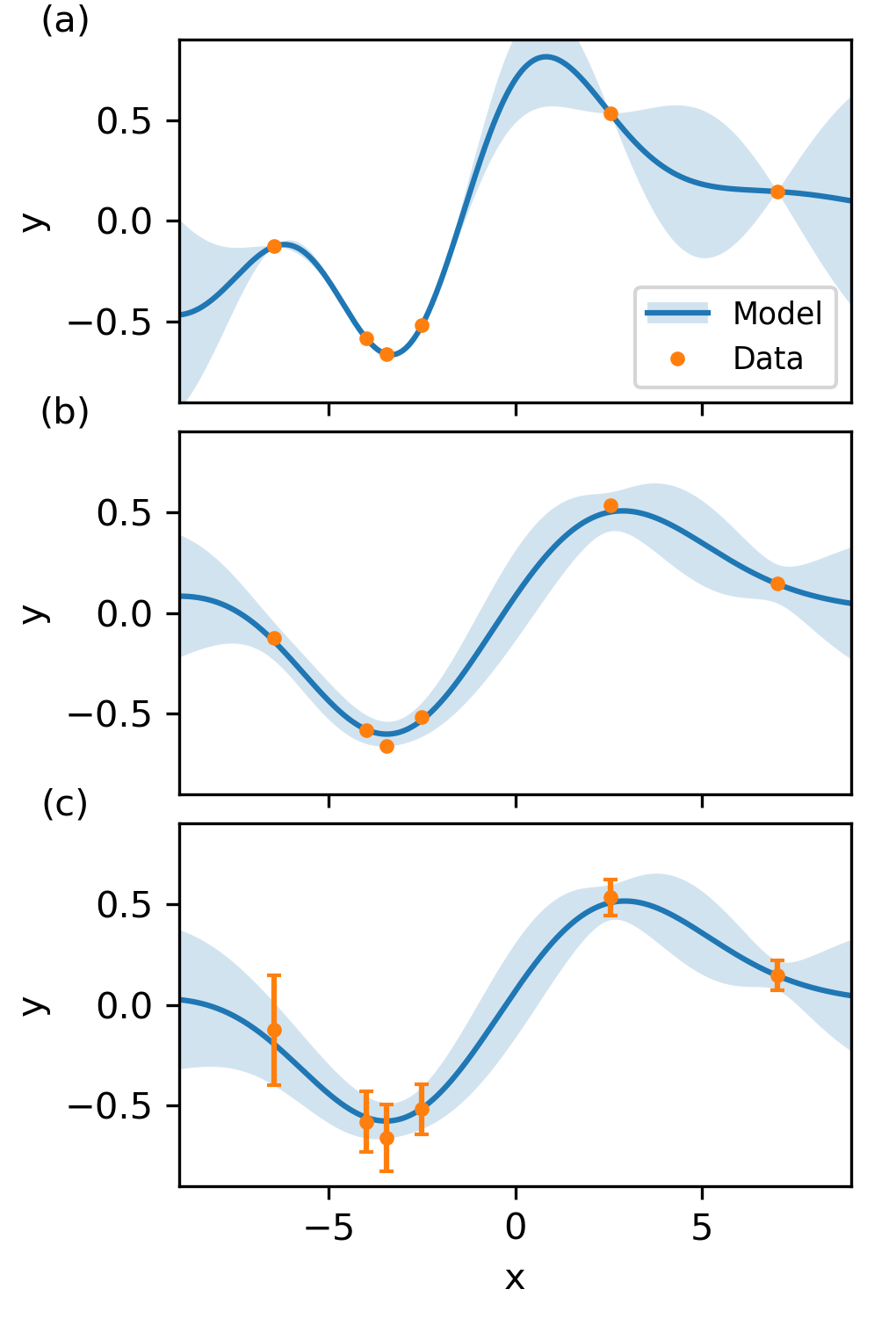}
	\caption{Examples of GP modeling with varying treatment of measurement noise. 
		(a) Shows a GP model containing zero noise, forcing the GP prediction to fit experimental data exactly. (b) Shows a GP model trained on the same data with a fixed (homoskedastic) noise parameter. (c) Illustrates a GP model incorporating heteroskedastic noise, where the data variance for each point is explicitly specified.
		\label{fig:GPR_noise_example}}
\end{figure}

The most straightforward method for representing measurement uncertainty uses a noise hyperparameter $\sigma_\epsilon$ that is incorporated into Eq. \ref{eq:posterior_mean} and \ref{eq:posterior_variance_1} by assuming a fixed Gaussian model of the uncertainty for all measurements.
This \textit{homoskedastic} uncertainty assumption adds $\sigma_\epsilon^2$ terms to the diagonal elements of the kernel matrix, in what is sometimes referred to as Tikhonov regularization or ridge regression \cite{golub1999tikhonov}.
In cases where no noise is present, as in deterministic simulations, this parameter can be set to zero, resulting in GP models that predict exact values at training data locations, as shown in Fig.~\ref{fig:GPR_noise_example}(a).
In the case of experimental measurements containing noise, the noise hyperparameter can be determined during optimization alongside other model hyperparameters by maximizing the MLL. 
This process serves to regularize the GP model, mitigating high-frequency behavior and treating it as uncertainty at measurement locations, as exemplified in Fig.~\ref{fig:GPR_noise_example}(b).

In some situations, observation uncertainty is known beforehand either from systematic uncertainties or stochastic noise.
This uncertainty can be different for each measurement, or \textit{heteroskedatic} in nature.
If the observation uncertainty can be estimated, e.g. by taking repeat measurements to estimate stochastic noise, or by specifying systematic measurement uncertainty, this information can be included for each point individually in a heteroskedastic model.
In this case, different values of $\sigma_{\epsilon,i}^2$ can be added to the diagonal elements of the covariance matrix for each data point $y_i$.
This allows for individual measurement uncertainty to be accounted for explicitly in the GP model, as illustrated in \cref{fig:GPR_noise_example}(c).
An alternative approach is to use a second GP to model the variance (or log-variance) over the parameter space and use this model to provide the weighting of the GP of the observations.

\subsubsection{Computational costs}
If the likelihood of the GP model is a Normal distribution then calculating the posterior distribution is analytical via matrix computations.
However, for more complex models the posterior cannot be obtained analytically and may require the use of a sampling algorithm such as Markov Chain Monte Carlo (MCMC) \cite{geyer1992practical} to estimate the posterior, which are known to be more computationally intensive. 

Calculation of the GP posterior can become a significant bottleneck given a large dataset of training points.
The computational cost of evaluating GP model predictions is primarily due to the matrix inversion operations in Eq. \ref{eq:posterior_mean} and Eq. \ref{eq:posterior_variance_1} which has a computational cost of $O(n^3)$ where $n$ is the number of training points (note that it is independent of the dimensionality of $\mathbf{x}$).
As a result, the decision making cost can increase substantially as the number of optimization iterations increases due to the need to train and evaluate GP models (bench-marking of these computational costs on modern hardware architectures can be found in Sec.~\ref{subsec:computational_benchmarks}).
Thus, when using BO, it is advantageous to find strategies that reduce the number of training data points needed to make an accurate model of the objective function.
This is where advanced modeling techniques come into play.

\subsection{Advanced Gaussian Process Modeling Techniques}
The goal of advanced modeling techniques is to encode prior information into the GP model such that it makes more accurate predictions with smaller sets of data.
This prior information can come from a number of different sources, including (but not limited to) physics knowledge, historical data sets, and/or prior optimization runs.
Improving the predictive accuracy of GP models prior to starting optimization allows BO to select more optimal points to measure at each step, reducing the average number of iterations needed to reach convergence.
Furthermore, reducing data requirements for accurate modeling reduces the computational cost of evaluating the GP model, enabling faster decision making.
Here we describe advanced techniques that can be used to improve model accuracy with smaller data sets.

\subsubsection{Kernel customization}
\label{subsec:custom_kernels}
\paragraph{Combining kernels} For more expressive behavior, multiple kernels can be combined into a single kernel through addition, multiplication, and tensor products. 
This combines the high level functional behavior of expected phenomena into a single model. 
For example, when modeling beam size squared as a function of beamline parameters, the second-order dependence of beam size on quadrupole strength can be captured by a polynomial kernel, while a more general Matérn kernel can be used for other beamline parameters whose effect on beam size is less well known..

\paragraph{Hyperparameter priors} 
Kernel functions can also be customized by specifying prior distributions for kernel hyperparameters.
Incorporating priors into the hyperparameter training process, biases MLL hyperparameter training towards certain values according to the prior distribution.
This provides a convenient middle ground between fixing hyperparameter values during optimization and training from scratch.
For example, we can be relatively confident that linear beamline elements (such as quadrupole magnets) have a minimal effect on beam emittance.
In this case, we can specify prior probabilities on the GP model length scales with respect to linear element parameters that encode this independence information into the GP model of the emittance, similar to what is done in the Sparse Axis Aligned Subspaces (SAAS) kernel \cite{eriksson2021high}.
Incorporating this assumption into the kernel rapidly speeds up convergence in objective functions that are only strongly dependent on a small subset of optimization parameters by reducing the effective dimensionality of the problem.
However, if non-linearities exist in these magnetic elements that do effect the beam emittance, the notion of independence in the GP model can be updated provided experimental evidence, instead of being ignored completely.



\begin{figure}
	\includegraphics[width=\linewidth]{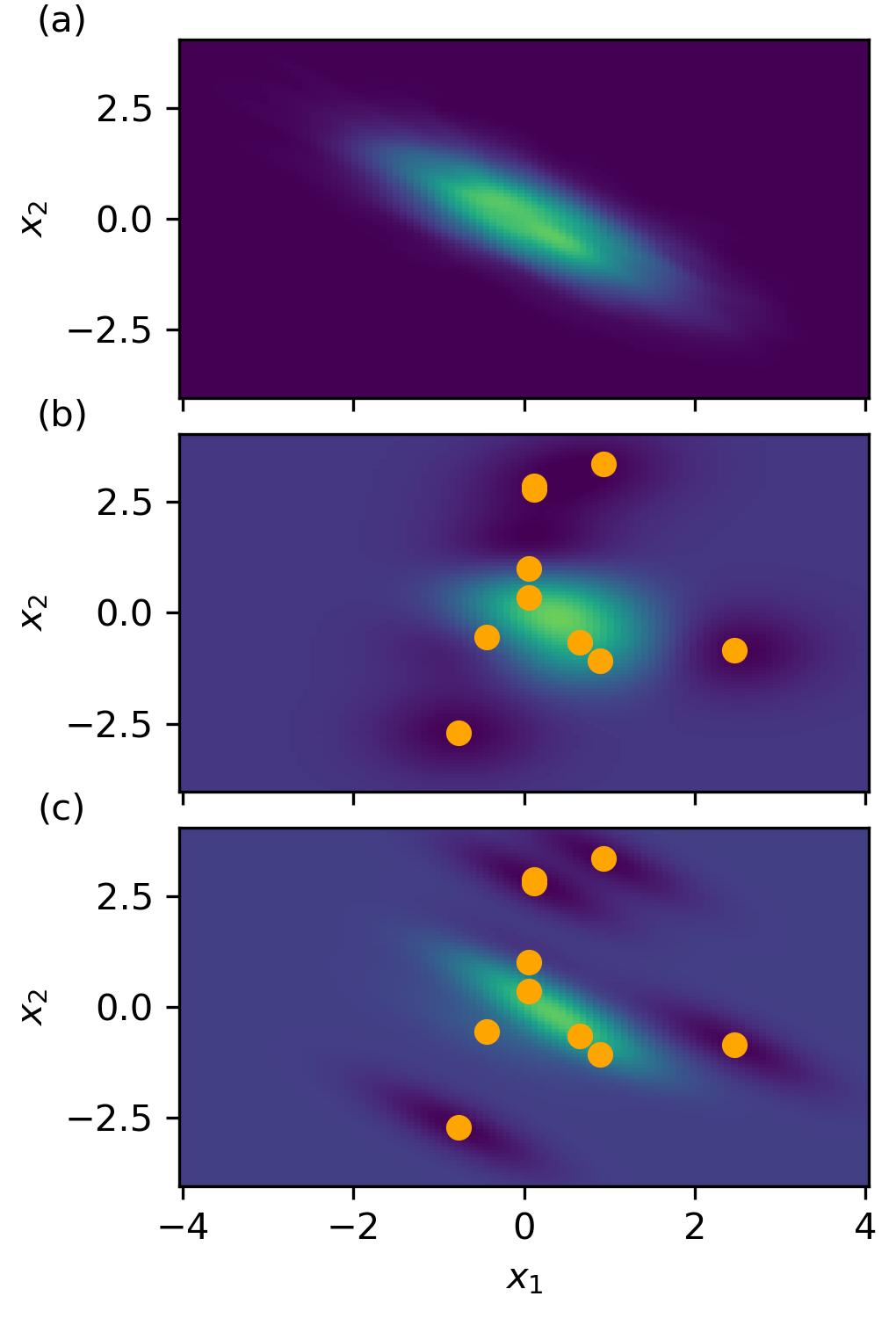}
	\caption{Illustration of the improvement in sample efficiency that can be gained by including expected correlations, such as those that arise from adjacent quadrupoles, into the GP kernel design. Here, a 2D function has input correlations that are similar to what one might observe between adjacent quadrupoles (a). For a given set of training data points (shown in orange), the GP that is provided with the correct correlated kernel (c) is able to learn a substantially more accurate model than the one that is provided with an  uncorrelated kernel (b). In the context of BO, learning a more accurate model with fewer data training points translates to faster convergence in optimization.
		\label{fig:corr_kernel}}
	
\end{figure}

\paragraph{Kernel estimation from Hessian}
\label{subsec:physics_kernel}
Another way of encoding prior information of an objective function into the kernel can be specifying fixed kernel function hyperparameters that express expected functional correlations in a local region around the expected optimal point.
For example, objective functions that depend on quadrupole strengths often contain cross-correlation structure, similar to what is shown in Fig.~\ref{fig:corr_kernel}(a), between adjacent quadrupoles due to the focusing-defocusing nature of first order beam dynamics.
These cross correlations are difficult to learn on-the-fly without making a large number of measurements, as shown in Fig.~\ref{fig:corr_kernel}(b).
Adding information about cross correlated structure in the objective function can significantly increase the accuracy of the GP model in high dimensional spaces without having to make a large number of measurements.
An efficient method for doing this is to compute the Hessian matrix of the objective function near the predicted optimal point in parameter space $\mathbf{x}^*$ \cite{hanuka_physics_2021}.
This can then be used as the factor analysis distance metric described in Eq.~\ref{eq:factor_analysis_distance}, where $M = H_f(\mathbf{x}^*)$.
As shown in Fig.~\ref{fig:corr_kernel}(c), incorporating the Hessian matrix into the RBF kernel improves the accuracy of the GP model with fewer training data points.
Identifying this low dimensional structure in the objective leads to faster convergence speeds during optimization, especially in high-dimensional optimization problems \cite{duris_bayesian_2020}.

\paragraph{Deep kernel learning}
Neural networks (NN) can also be used as drop-in replacements for kernel functions in what is often referred to as Deep Kernel Learning (DKL) \cite{wilson2016deep}.
Neural networks can be incredibly powerful when modeling complex features in accelerator physics measurements such as images and signals.
However, they require large or information-rich training data sets to accurately predict functional covariances between points in parameter space.
As a result, learning kernel functions specified by NN on-the-fly during optimization is impractical for cases where measurements are expensive.
Furthermore, for most optimization cases in accelerator physics the objective functions are relatively smooth, thus much simpler kernel functions can reasonably predict accurate covariances without the cost of training a NN representation.
As a result, there has been limited work in accelerator physics towards investigating the use of NN models in kernel functions for the purpose of conducting optimization.
On the other hand, NN models have been investigated for use in other aspects of GP modeling, see Sec.~\ref{subsec:prior_means} for details.



\subsubsection{Non-constant prior means}
\label{subsec:prior_means}
An alternative approach for incorporating prior information into the GP model is to specify an explicit, non-zero prior mean function.
In this case, instead of embedding prior information about the high-level functional characteristics of the objective function, we provide an explicit guess as to what the objective function value is at every point in parameter space.
This is especially useful if we have detailed knowledge of the objective function from beam dynamics calculations, historical data sets, and/or previous optimization runs.


\begin{figure}
	\includegraphics[width=\linewidth]{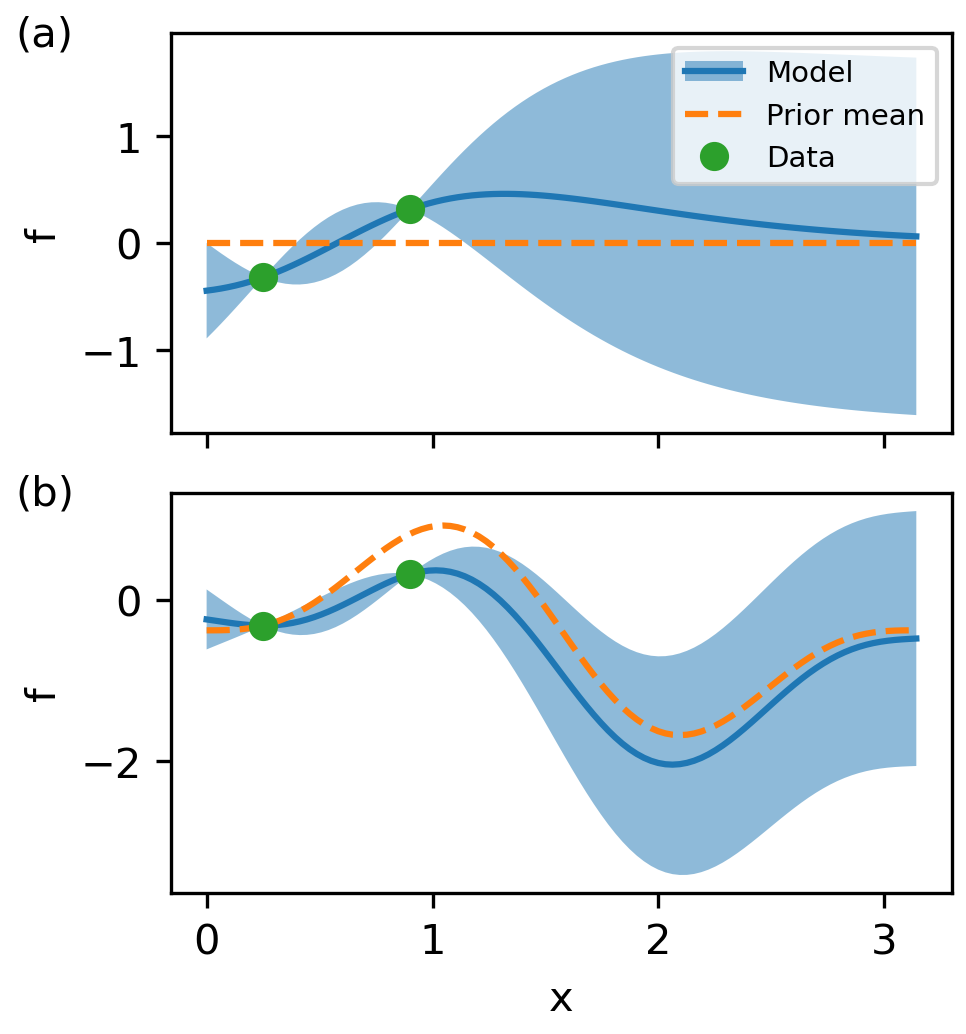}
	\caption{Illustration of non-zero prior mean. In the absence of local data, the mean of the posterior distributions reverts to (a) zero or (b) the non-zero prior mean. The variance remains unchanged.}
	\label{fig:prior_mean_example}
\end{figure}

Incorporating a non-zero prior mean function $m(\mathbf{x})$ into a GP model re-incorporates an extra term ignored in Eq.~\ref{eq:posterior_mean}, producing posterior mean function values at the test points $X_*$
\begin{equation}
	\bm{\mu_*} = \mathbf{m}(X_*) + K(X_*, X)K_y^{-1}(\mathbf{y} - \mathbf{m}(X))
	\label{eq:posterior_mean_for_non_zero_prior}
\end{equation}
with $K_y = K(X,X) + \sigma_\epsilon^2I$. For test points that are far away from previous measurements in parameter space ($K(X_*, X) \rightarrow 0$), the posterior mean function values $\mathbf{f_*}$ are equal to the prior mean values at the test points $\mathbf{m}(X_*)$.
This effect is illustrated in Fig.~\ref{fig:prior_mean_example}, where the mean of the posterior distribution reverts back to the prior mean as the distance between test points and training data increases.
Thus, if the prior mean function accurately predicts the objective function, the GP model can make similarly accurate predictions of the objective without any data.
Conversely, if portions of the prior mean incorrectly make predictions, the posterior predictions of the GP model will reflect updated values from training data. 
In this way, the GP can be interpreted to only model the difference to the prior mean function $m(\mathbf{x})$ instead of the full objective.

A custom mean function can be parameterized in a number of different ways, with the only requirement being that it maps parameter values to function values.
For example, prior mean functions can be analytic functions, surrogate models, or even full particle dynamics simulations.
However, for the best performance in the context of BO, prior mean functions should be differentiable (i.e.\ support backwards automatic differentiation) and relatively inexpensive to evaluate.
These attributes allow the acquisition function to be quickly optimized to find the next measurement point (see Sec.~\ref{sec:acq_opt} for details).

For the purposes of online optimization in accelerators, neural network (NN) surrogate models \cite{nn_prior_experiment} or fast differentiable beam dynamics simulations \cite{kaiser_cheetah_2024}  have been identified as promising prior mean functions.
Neural networks of sufficient complexity are known as universal function approximators \cite{hornik1989multilayer} and typically execute much faster than conventional physics simulations \cite{edelen2020machine}.
Furthermore, NN models are generally differentiable due to training requirements, making them ideal for representing prior mean functions.
Most importantly, unlike GP models which scale poorly with data set size, NN surrogate model execution time is independent of the size of the data set used to train the surrogate.
This allows large amounts of historical measurement and simulation data to be incorporated into the GP model without hurting online performance.
Alternatively, fast-executing beam dynamics simulations that are differentiable can be used to model the prior mean functions.

Incorporating prior mean functions inside GP models has a substantial impact on BO convergence speed.
If the prior mean exactly matches the true objective function, convergence can happen immediately depending on the ratio of the objective function's ideal value with respect to the prior model uncertainty and which acquisition function is chosen to perform BO.
Offline studies using simulated objectives similar to those of the LCLS injector demonstrates that prior mean functions which have positive correlations with the true objective function also improve convergence speed of BO to optimal values \cite{xu_neural_2022}.

\begin{figure}
	\includegraphics[width=\linewidth]{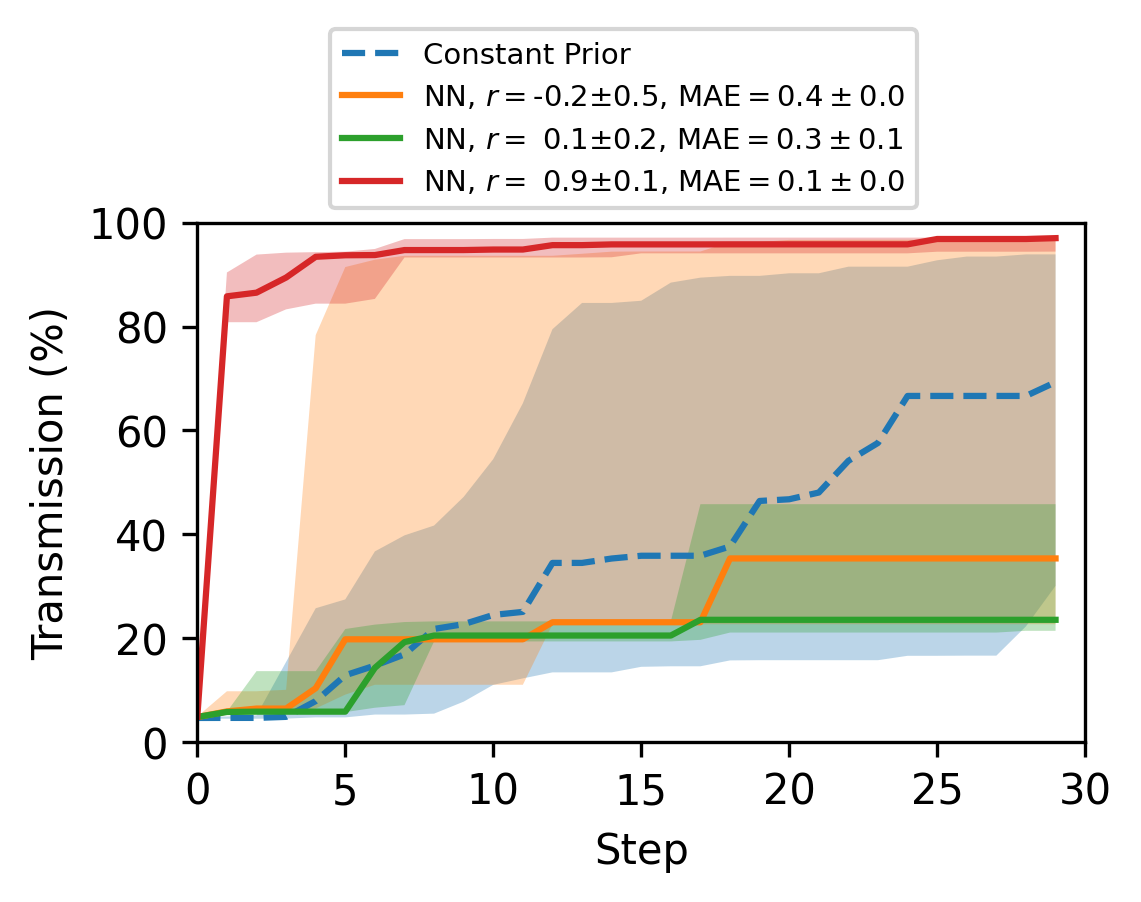}
	\caption{Transmission optimization at ATLAS subsection using different prior mean functions. Solid and dashed lines depict the medians and the shaded areas the corresponding 90\% confidence levels across 10 to 20 runs. Reproduced from~\cite{nn_prior_experiment}.}
	\label{fig:ATLAScorr}
\end{figure}

This benefit was also observed experimentally at the ATLAS accelerator \cite{nn_prior_experiment} at Argonne National Laboratory.
In this example, BO was used to tune the strengths of 5 quadrupole magnets to maximize the beam transmission through a beamline.
They repeated the optimization multiple times starting with the same initial point using a constant prior mean function and 3 different NN surrogate models based on previous experimental data such that the models have varying correlations with the true objective function.
Figure~\ref{fig:ATLAScorr} shows that using a prior mean that has a high correlation with the ground truth resulted in better BO performance than standard GP models with constant priors.
However, if the prior model is poorly correlated with the true objective function then BO performance can suffer. 
Although measures can be taken to mitigate these effects~\cite{nn_prior_experiment}, the quality of the prior mean is critical to improving the performance of BO when using a non-zero prior mean.
Additional work at the LCLS photoinjector has demonstrated similar benefits to using NN surrogate models as priors in GP modeling up to 9 free optimization parameters~\cite{nn_prior_experiment}.

\subsubsection{Modeling in transformed spaces}
\begin{figure}
	\includegraphics[width=\linewidth]{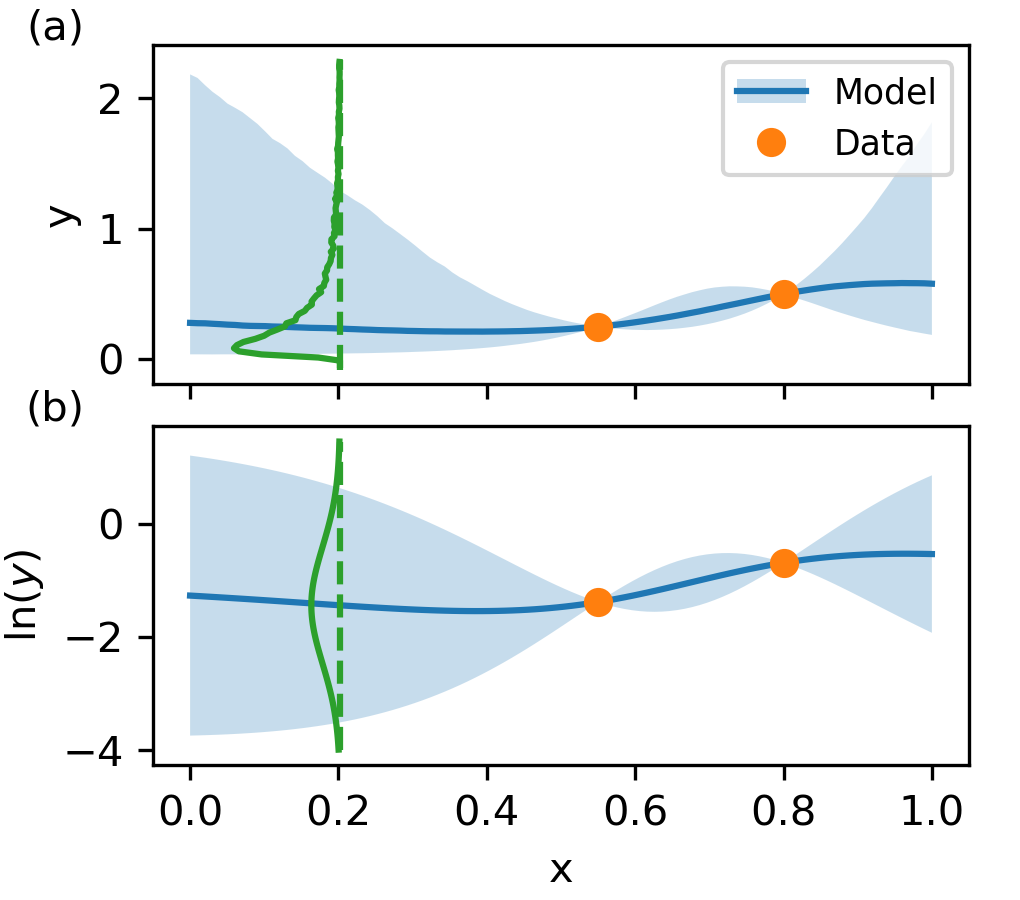}
	\caption{Example of using Log-transformations in GP modeling for strictly positive output values. Data in real space (a) is transformed to log space before fitting a GP model (b). 
		Samples drawn from the GP model in log-space are then transformed back into real space to make GP predictions.
		The resulting likelihood in real space is then a Log-Normal distribution which is strictly positive.
		\label{fig:log_transform}}
\end{figure}

In some cases it is advantageous to transform input or output data into an intermediate space before training hyperparameters and making model predictions.
This strategy is useful when modeling objectives according to known physical principles or constraints.
For example, a number of objectives in accelerator physics are strictly positive, such as beam size and emittance.
In order to restrict the range of GP predictions to positive values, data can be transformed into log space before fitting the GP model, as is shown in \ref{fig:log_transform}.
Samples drawn from the GP model in log-space are then transformed back into real space, resulting in model predictions that follow a Log-Normal distribution, respecting the requirement that model predictions are strictly positive.

The \textit{Bilog} transform \cite{eriksson2021scalable} can improve modeling accuracy near zero by magnifying output values near zero and damping output values far away from zero.
Applying this transform to data that is used to model constraint functions helps improve constrained optimization (see Sec.~\ref{subsec:constraints}), since constraints are generally defined with respect to zero, see Eq.~\ref{eq:optimization_problem_constraints}.
The \textit{Gaussian copula} \cite{wilson2010copula} has also been proposed as a useful transformation for magnifying objective function values that are on the edges of the observed range -- namely maximum or minimum values.  

While in principle a variety of transformations can be used to achieve different effects, their use can degrade computational performance.
For many acquisition functions, GP model predictions should be mapped back into real space in order to calculate the acquisition function value. 
Most transformations do not provide an analytical mapping of posterior predictive means and variances back to un-transformed space, meaning that samples must be drawn from the GP posterior and then transformed back into real space before being passed to the acquisition function.
This can increase the computational cost of making GP model predictions and prohibit the use of analytical acquisition functions, which are generally faster than sample-based acquisition functions.

\subsubsection{Time-dependent modeling}
\label{subsec:time_dependence}
In practical applications, the accelerator systems under consideration are often affected by factors beyond those represented in the input space, such as incoming beam parameters or drifts in auxiliary equipment due to external factors. 
While these factors cannot be controlled and changed explicitly by the optimization process, they can be incorporated into the GP model to improve the model predictive power and thus the convergence speed of BO. 
Using above-described approach of kernel multiplication, the standard model can be extended with time and other contextual dimensions to represent a modified system
\begin{equation}
	\textbf{y} = f(\mathbf{x},\mathbf{\phi}) + \varepsilon,
\end{equation}
where $\phi$ represents the contextual parameters. 
A classic use of such GP models has been in time-series predictions (i.e. stock prices), where $\phi$ contains the time dimension $t$. 
This method is referred to as adaptive BO (ABO)~\cite{nyikosa_bayesian_2018} or contextual BO~\cite{krause_contextual_2011}, and can be used to compensate for  changes in the accelerator as long as they can be correlated to an observable. 
Note that to use such extended GP models in Bayesian optimization, all contextual parameters will need to be specified explicitly for the next point(s) and then held fixed during acquisition function optimization. 

Incorporating additional dimensions into the GP model will increase the amount of data required for a good fit and thus slow down initial BO convergence (but not as much as a regular input parameters). 
ABO should only be used if the impact of contextual variables is significant relative to the noise in the objectives. 
Otherwise, it is advisable to use standard BO which will incorporate small drifts into the fitted noise parameter. 
For the most common case of time-adaptive BO, there are several choices of auxiliary variables that can be used - only time (which correlates to all drift sources, but potentially has a complicated relationship that cannot be represented well by GP), time and specific drift sources, or only specific drift sources. 
Where possible, specific sources should be used to simplify the model. 
For example, if it is known that only room air and cooling water temperatures contribute to time-dependent drifts in RF cavities, it is best to only include temperature values as contextual variables instead of time. 
However, using time permits ABO to be very flexible with little to no tuning, or when drift sources are distributed over too many dimensions.

\begin{figure}
	\includegraphics[width=\linewidth]{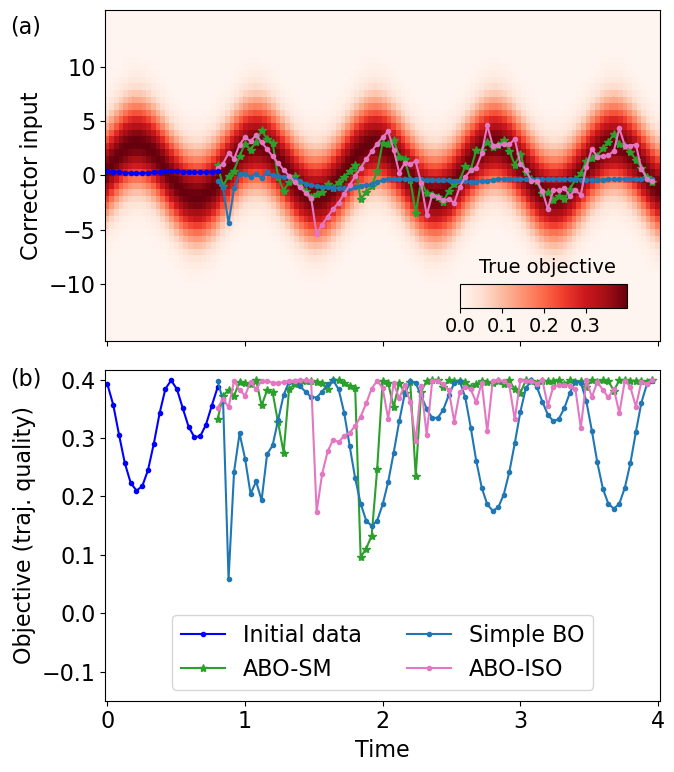}
	\caption{Simulated application of standard and time-aware BO in a drifting trajectory stabilization problem. BO settles on the mean value of the oscillations. ABO-ISO (isotropic) follows the changes but lags them because it only uses isotropic (local) kernel. ABO-SM (spectral mixture) captures long-range correlations and eventually correctly predicts necessary future changes in-phase. By default, ABO-SM continues to explore around maximum value for optimization, producing a small step jitter. It can be eliminated by using posterior mean as the acquisition function at cost of convergence speed.}
	\label{fig:abosim}
\end{figure}

Regardless of the choice of contextual variables, to achieve good model fit with standard local kernels like RBF the perturbations must be slow enough such that BO loop can sample the highest-frequency features of the drift with at least a few points, by analogy to Nyquist's theorem. 
This requirement can be relaxed somewhat if a custom kernel is used that can account for long-range structure in the data. 
Experimentally, many drift signals are either directly correlated to something or have periodic structures. 
To avoid excessive hyperparameter tuning, special kernels like the Spectral Mixture kernel~\cite{wilson_gaussian_2013} can be used for cases where the exact number and periodicity of oscillatory signals are not known, but require more data to achieve a good initial fit. 
A more advanced strategy is to use the rate-of-change of GP model parameters, such as length scales, with time to choose the most appropriate kernel so as to ensure an acceptable trade-off between worst-case performance and convergence. 
In Fig. \ref{fig:abosim} an example application to a linac trajectory stabilization problem is demonstrated. 
Performance of the more advanced methods strongly depends on drift magnitude, sampling rate, and measurement noise levels - it is suggested to perform similar simulations to evaluate suitability of contextual methods to specific tasks. 

Both time and generic ABO has been demonstrated experimentally at the APS linac ~\cite{kuklev_online_2022} and in the KARA storage ring ~\cite{xu_bayesian_2023}. 
Extensions of ABO to constrained problems with robustness requirements and dynamic kernel selection have also recently been tested at APS ~\cite{kuklev_robust_2023}.

\subsubsection{Multi-fidelity modeling}
\label{subsec:multi_fidelity_modeling}

In some cases, it may be possible to obtain data about the behavior of the accelerator from different sources of information. 
For instance, we may have access to data from both experimental measurements and numerical simulations, or from numerical simulations using different computational models and/or different resolutions.
Additionally, we may be able to make experimental measurements that trade varying levels of detail and accuracy for evaluation speed or cost.
In these cases, it is desirable for the GP model to be able to learn from these different sources of data, while keeping track of their respective origin and evaluating their respective trustfulness.

In this context, the source of the data is encoded by assigning a \emph{fidelity} value $s$ to each data point in the data set. 
Depending on the context, this fidelity parameter may take discrete values or continuous values. 
For instance, when combining experimental and simulation data (discrete fidelity), one may assign $s=0$ to data points coming from simulations, and $s=1$ to data points from experimental measurements. 
When combining simulations at different resolutions (continuous fidelity), one may assign $s=0$ to data points from low-resolution simulations, $s=1$ to data points from high-resolution simulations, and an intermediate value of $s$ to simulations at intermediate resolutions.

A GP can then be trained on this combined data set, taking $\boldsymbol{x}$ \emph{and} $s$ as input and predicting the associated $\boldsymbol{y}$. 
In the case where $s$ takes continuous values, the fidelity dimension is simply treated as another input dimension to the GP \cite{kandasamy_multi-fidelity_2017}, with its associated kernel and hyperparameters. 
It is common to choose the kernel for $s$ and $\boldsymbol{x}$ to be separable (see \ref{subsec:custom_kernels}), and to use a stationary kernel for $s$ \cite{kandasamy_multi-fidelity_2017}:
\[ k((s,\boldsymbol{x}), (s',\boldsymbol{x}')) = \tilde{\kappa}(||s-s'||) \kappa(\boldsymbol{x},\boldsymbol{x'})\]
In this case, the lengthscale hyperparameter $l$ for the kernel $\tilde{\kappa}$ quantifies the extent to which similar fidelities give similar results. 
For instance, a large lengthscale $l$ would cause the GP to predict similar output $\boldsymbol{y}$ even for relatively different values of the fidelity $s$. 
As usual, during training, this hyperparameter is often learned on-the-fly using hyperparameter tuning, and thus the GP automatically learns how much the prediction $\boldsymbol{y}$ varies with the fidelity $s$. 
Or, in other words, the GP learns to which extent the low-fidelity data can be relied on when trying to predict high-fidelity data.

In the case where the fidelity $s$ is discrete, one type of multi-fidelity GP is the multi-task GP \cite{bonilla_multi-task_2007}, where the kernel is expressed in a similar manner:
\[ k((s,\boldsymbol{x}), (s',\boldsymbol{x}')) = \tilde{\kappa}_{s,s'} \kappa(\boldsymbol{x},\boldsymbol{x'})\]
where $\tilde{\kappa}_{s,s'}$ is a positive semi-definite \emph{matrix} (given that $s$ and $s'$ take discrete values). The values of the entries of this matrix are obtained by hyperparameter tuning, and they, again, quantify the extent to which low-fidelity and high-fidelity data are related.

\begin{figure}
	\includegraphics[width=\linewidth]{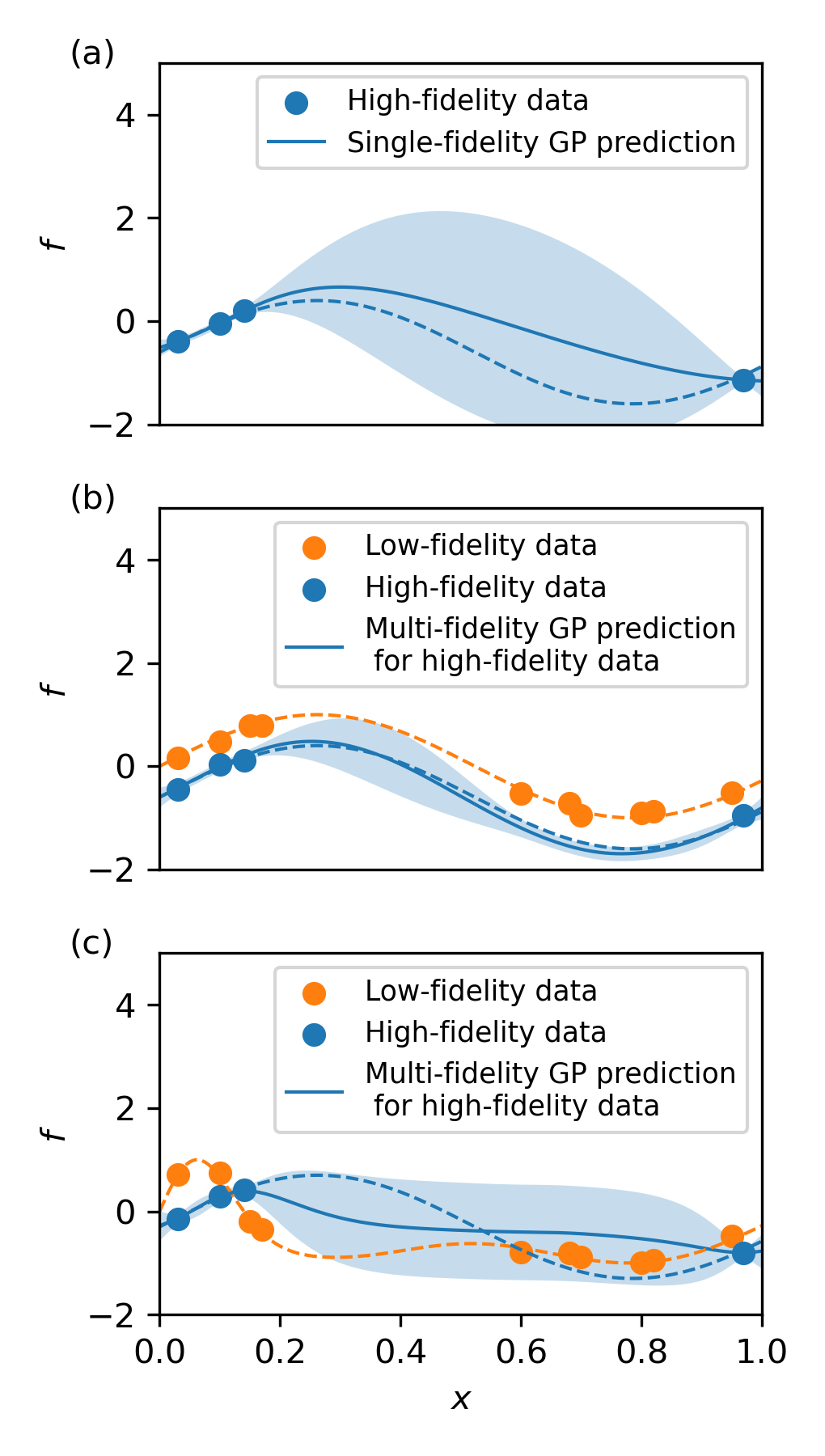}
	\caption{Illustration of the prediction of a multi-fidelity Gaussian Process, by comparing (a) a single-fidelity Gaussian Process trained only on high-fidelity data, and (b-c) a multi-fidelity Gaussian Process trained on both high-fidelity and low-fidelity data, in the case where
		(b) high-fidelity and low-fidelity data are highly correlated, as well as (c) high-fidelity data and low-fidelity data are largely uncorrelated. In this particular example, the multi-fidelity GP is a multi-task GP \cite{bonilla_multi-task_2007}, as implemented in the library BoTorch. Dashed lines denote ground truth values of the low and high fidelity functions.
		\label{fig:Multifidelity_GP}}
\end{figure}

This is illustrated with an example in Fig.~\ref{fig:Multifidelity_GP}, where low and high fidelity versions of an objective function can be evaluated with corresponding evaluation costs.
Fig.~\ref{fig:Multifidelity_GP}(a) shows a conventional GP model that predicts the output of the high-fidelity objective trained solely on a small, high-fidelity data set.
Instead of continuing to evaluate the expensive high-fidelity objective function, a multi-fidelity modeling approach incorporates inexpensive, low-fidelity data into the model of the high-fidelity objective.
If the low-fidelity data serves as a good approximation of (ie. is highly correlated with) the high-fidelity objective function, adding this data will reduce the uncertainty of the multi-fidelity model and increase its accuracy, as shown in Fig.~\ref{fig:Multifidelity_GP}(b).
However, if the low fidelity data is largely uncorrelated with the high fidelity data, as in Fig.~\ref{fig:Multifidelity_GP}(c), the model prediction of the high-fidelity objective function is weakly influenced by the low fidelity data.


Multi-fidelity GPs have been used in the context of simulation-based design optimization for laser-plasma accelerators \cite{ferran_pousa_bayesian_2023,irshad_multi-objective_2023}. 
In these instances, the fidelity was either continuous and corresponded to the resolution of the simulation grid \cite{irshad_multi-objective_2023}, or discrete and corresponded to different simulation codes making different approximations \cite{ferran_pousa_bayesian_2023}.
In both cases, the ability of multi-fidelity GP to partially rely on cheap, low-fidelity simulations (either low-resolution simulations, or approximated simulation codes) significantly reduced the cost of performing optimization.
These examples are discussed in more detail in Sec.~\ref{subsec:multi_fidelity}.


\subsubsection{Embedding complex modeling processes}
\label{subsec:hysteresis}
As fast executing surrogate models, GPs can be used as a drop-in replacement for other numerical models when creating multi-component models of complex systems.
This can add flexibility, a robust treatment of uncertainty, and data-efficiency to arbitrary models of accelerator physics.
These hybrid models, in turn, can increase the interpretability of GP modeling, and in some cases expand the applicability of GP modeling to new domains.

For example, basic GP modeling is insufficient when describing systems that exhibit path-dependent physical processes, most notably, mechanical and magnetic hysteresis.
Hysteresis is a path dependent process such that beam properties depend not only on the current state of the machine but also on the historical path taken to get to the current state.
This creates repeatability issues when optimizing accelerator parameters in magnetic and mechanical systems. 

\begin{figure}
	\includegraphics[width=\linewidth]{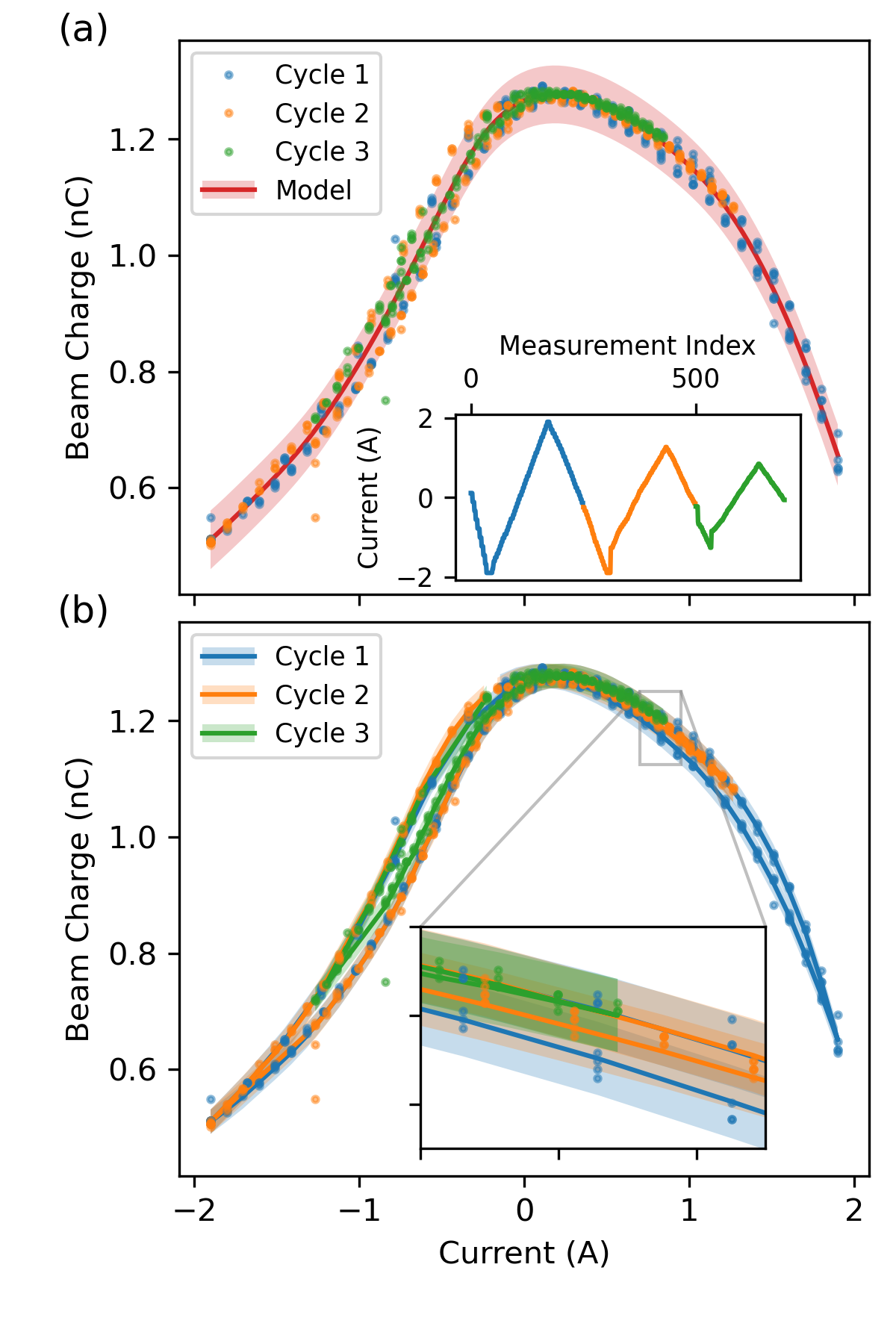}
	\caption{Demonstration of combining GP models with a differentiable physics model of magnetic hysteresis. (a) Measured beam charge after passing through a linac section at APS is plotted over three cycles of varying the current in an upstream quadrupole. Transmitted beam charge measurements are not repeatable due to hysteresis effects in the upstream quadrupole. (b) GP modeling with differentiable hysteresis model included accurately predicts beam charge over multiple hysteresis cycles with improved (reduced) uncertainty predictions. Reproduced from \cite{roussel_differentiable_2022}.
		\label{fig:hysteresis}}
\end{figure}

Basic GP modeling, see Fig. \ref{fig:hysteresis}(a), interprets this error as stochastic noise, reducing the accuracy of model predictions, underestimating measurement uncertainty in some regions of parameter space, and overestimating uncertainty in others.
However, if a GP model is combined with a numerical model of hysteresis, the hybrid model can make predictions with higher accuracy and better calibrated uncertainty estimates.
In Fig.~\ref{fig:hysteresis}(b), a Preisach hysteresis model \cite{preisach_uber_1935} is used to map the control parameter (magnet current) to magnetic field, while the GP model represents the mapping of magnetic field to beam dynamics \cite{roussel_differentiable_2022}.
Both hysteresis model parameters and the GP hyperparameters are trained simultaneously on the data using MLL.
The resulting hybrid model has a higher predictive accuracy and provides uncertainty estimates that are well calibrated to stochastic experimental noise.
As a result, the hybrid model improved optimization convergence, mitigating the detrimental effects of hysteresis on optimization when using basic GP models.

A key factor that enabled the simultaneous training of both hysteresis model parameters and GP hyperparameters was that calculations in both models were differentiable, allowing the use of gradient descent to maximize the MLL.
Combining differentiable models of other non-repeatable processes with GP models can extend the applicability of BO techniques to a wider range of optimization problems.
Using GP models to represent smaller units of accelerator processes, as is done in the case here, increases their accuracy with smaller data sets and improves their interpretability.

\section{Acquisition Function Definition} \label{sec:acq_def}
The definition of an acquisition function $\alpha(\mathbf{x})$ guides the Bayesian optimization process by defining the potential value of future measurements given a predictive surrogate model.
During BO, input parameters that maximize the acquisition function will be chosen for evaluation during the next iteration.

Almost all acquisition functions aim to perform global optimization by balancing two optimization strategies, often referred to as ``exploration" and ``exploitation".
Exploration refers to placing high value on choosing points in parameter space that will add information to the GP surrogate model, often in regions of parameter space where the model has high uncertainty.
Exploitation on the other hand, places a high value on points in parameter space that the surrogate model predicts to be optimal.
By balancing the weighting between these two strategies in the acquisition function (either implicitly or explicitly) during optimization, BO can increase the chances of efficiently finding global solutions to the optimization problem, instead of being stuck in local extrema.

As opposed to other standard optimization algorithm definitions, acquisition functions in BO are often defined with the assumption that objective functions are to be maximized.
In order to use these acquisition functions for objective function minimization, transformations are applied to the model predictions before they get passed to the acquisition function.
This approach is preferable to modeling negated objective values with the GP, as it makes model interpretation more challenging.

In this section, we first describe basic acquisition functions used for general purpose optimization.
Then we describe complex acquisition functions and modifications used to solve accelerator physics problems in online control and simulation.

\subsection{Basic Acquisition Functions}
\begin{figure}
	\includegraphics[width=\linewidth]{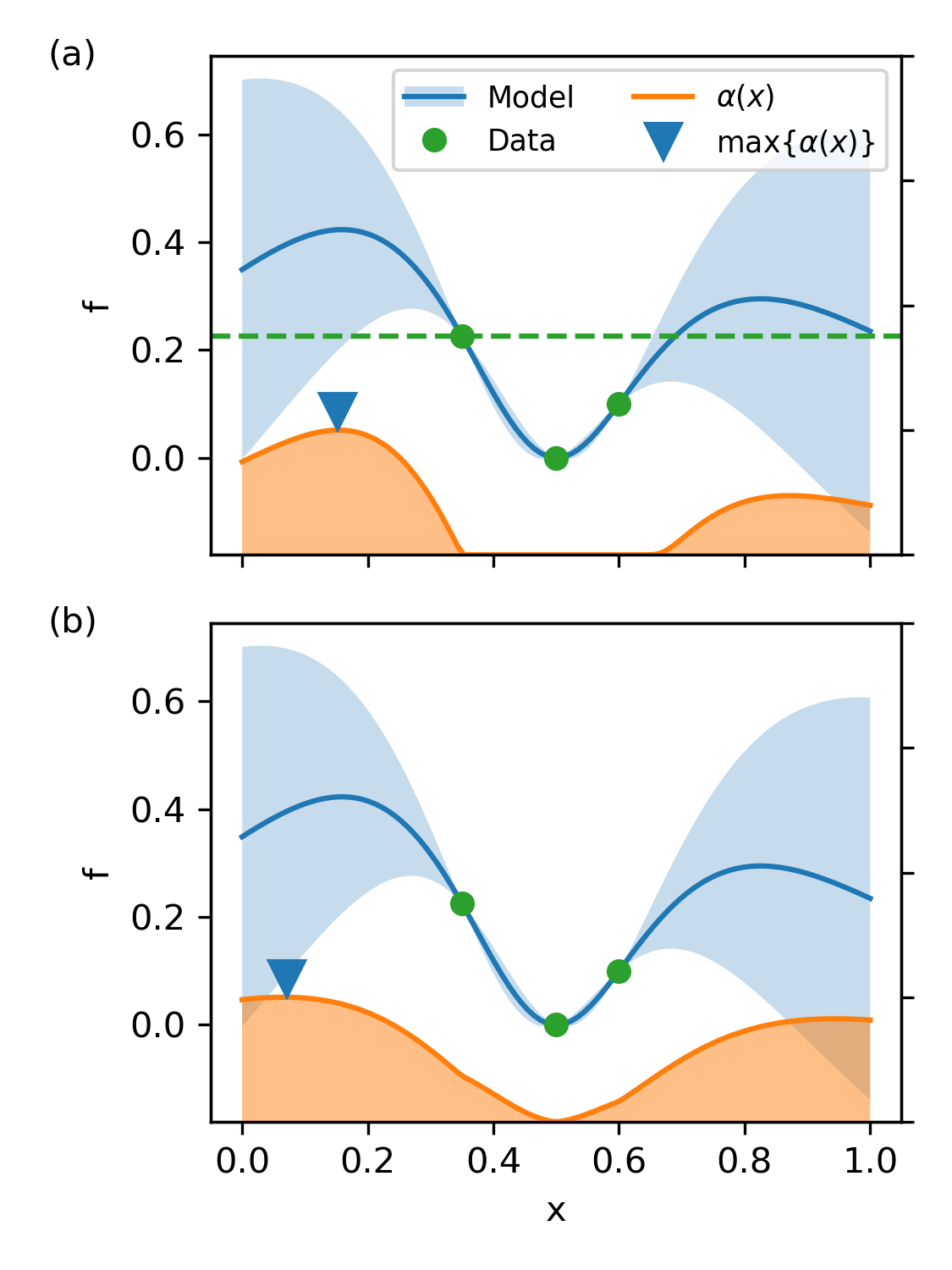}
	\caption{Examples of the EI and UCB acquisition functions for objective function maximization given the same GP model and training data. (a) EI acquisition function, where the dashed horizontal line denotes the best previously observed value $f(\textbf{x}^*)$. (b) UCB acquisition function. \label{fig:ei_ucb}}
\end{figure}

The two most commonly used acquisition functions for performing optimization are Expected Improvement (EI) and Upper Confidence Bound (UCB)~\cite{brochu_tutorial_2010}.
These simple acquisition functions, illustrated in Fig.~\ref{fig:ei_ucb}, are often the starting point for optimizing general problems and generally provide similar convergence speeds.

As its' name suggests, EI uses the GP model to calculate an expectation value of the improvement $I(\textbf{x}) = \mathrm{max}\{f(\textbf{x}) - f(\textbf{x}^*), 0\}$ over the optimal previously observed value of the objective function $f(\textbf{x}^*)$.
For a GP model with a Gaussian likelihood, the expected improvement can be calculated analytically:
\begin{align*}   
	EI(\textbf{x}) &= \mathbb{E}[I(\mathbf{x})] \\
	&= \sigma(\mathbf{x})(z\Phi(z) + \phi(z)) \label{eqn:ei} \\
	z &=\frac{\mu(\textbf{x})-f(\textbf{x}^*)}{\sigma(\textbf{x})}
\end{align*}
where $\Phi(\cdot)$ and $\phi(\cdot)$ denote the Cumulative Density Function (CDF) and Probability Density Function (PDF) of a Normal distribution.
As shown in Fig.~\ref{fig:ei_ucb}, EI emphasizes choosing observations that are predicted to be optimal, have large variance, or a combination of both, thus balancing exploration and exploitation.

UCB explicitly specifies a trade-off between exploitation and exploration by using a linear combination of the predicted mean and variance from the GP model with a weighting factor $\beta$:
\begin{equation}
	UCB(\mathbf{x}) \equiv \mu(\mathbf{x}) + \beta \sigma(\mathbf{x})
	\label{eq:UCB}
\end{equation}
For most optimization problems, a default value of $\beta = 2$ works well to balance exploitation and exploration.
Defining UCB with a larger $\beta$ value favors exploration, while smaller values of $\beta$ prioritize exploitation.
If the objective function is expected to be convex or unimodal, smaller values of $\beta$ may speed up convergence by prioritizing exploitation (often referred to as ``greedy optimization").

EI and UCB often provide similar levels of performance in terms of convergence speed for most optimization problems.
However, EI can sometimes become difficult to numerically optimize if large regions of the input space have zero probability of improving over the best observed point.
In this case gradient based optimization of the acquisition function (see Section \ref{sec:acq_opt}) can struggle to escape regions with zero gradients, often referred to as the ``vanishing gradient problem".
However, it has been suggested that taking the log of the EI acquisition function before optimizing can address this issue \cite{ament2023unexpected}.
On the other hand, UCB can create problems when using it in combination with advanced acquisition function modifications since it is not a strictly positive function.

\subsection{Advanced Acquisition Functions}
Here we describe definitions and modifications that tailor the behavior of BO in order to solve problems in accelerator physics.
In some cases, these acquisition functions are not analytically tractable, and are thus evaluated by using Monte Carlo sampling.
Calculations of the acquisition function in these cases is done by drawing function samples from the GP model and averaging over their individual contributions. 
A detailed discussion of this formalism can be found in \cite{wilson_reparameterization_2017}.

\subsubsection{Unknown function characterization}
\label{subsec:bayesian_exploration}
\begin{figure}
	\includegraphics[width=\linewidth]{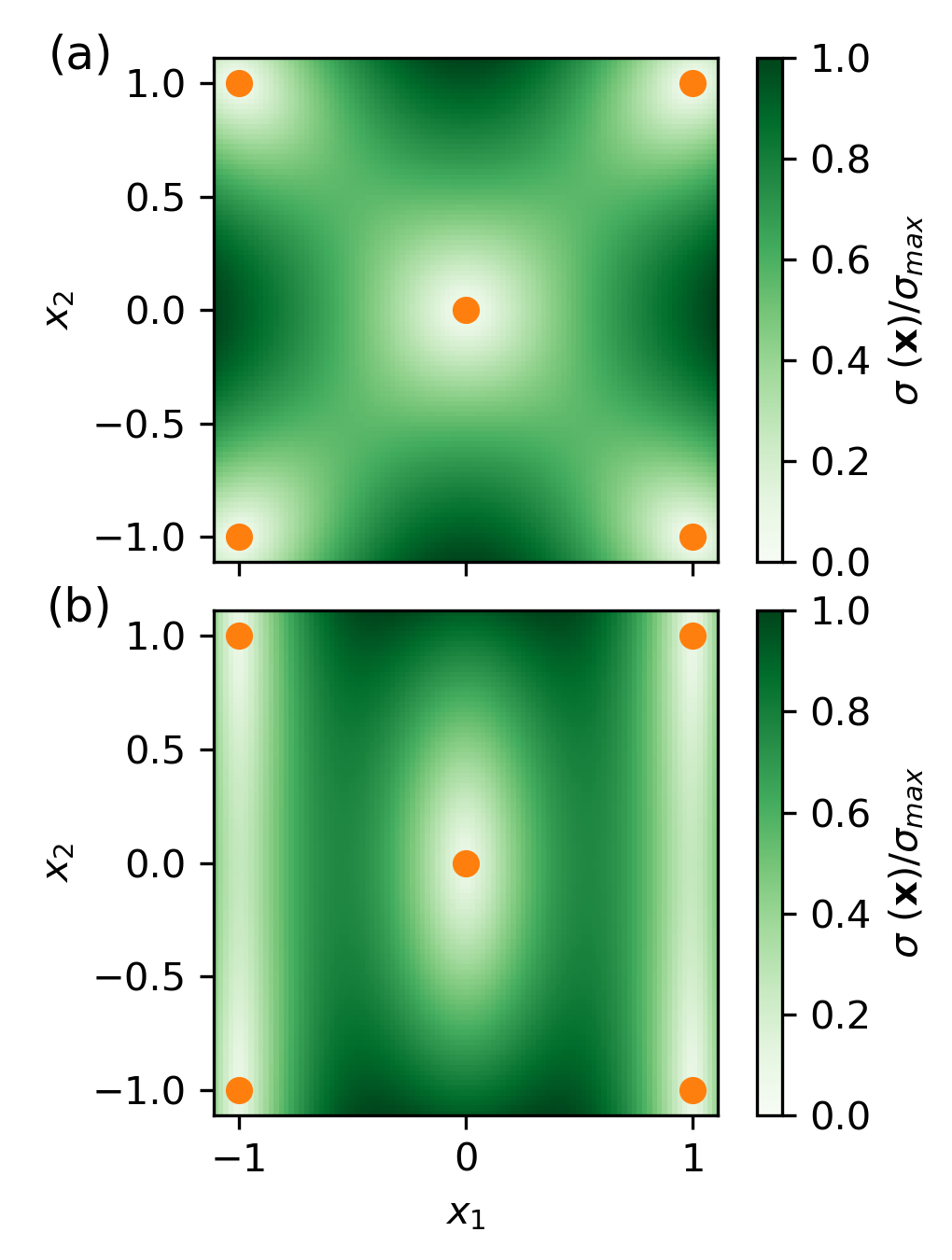}
	\caption{Example of sampling behavior of Bayesian exploration (BE). 
		(a) The BE acquisition function is maximized at locations in parameter space where the model uncertainty is highest, usually at locations farthest away from previous measurements.
		(b) In cases where the function is less sensitive to one parameter ($x_2$ in this example) the model uncertainty is smaller along that axis, resulting in less frequent sampling along that dimension.
		\label{fig:bayesian_exploration}}
\end{figure}

In some cases, instead of finding an solution to an optimization problem we aim to characterize an unknown function to learn its structure and evaluate sensitivities to individual parameters.
So-called ``active learning" acquisition functions can be defined too choose points that optimally characterize an unknown function instead of finding the extrema.
A simple example of this is known as uncertainty sampling or ``Bayesian Exploration" (BE) \cite{roussel_turn-key_2021}.
In this case, the acquisition function is defined as
\begin{equation}
	\alpha_{BE}(x) = \sigma(x).
	\label{eq:bayes_exp}
\end{equation}
When using this acquisition function, BO will sample locations where the model uncertainty is maximized, usually at points in parameter space that are farthest from previous evaluation locations, as shown in Fig.~\ref{fig:bayesian_exploration}(a).
Combining this acquisition function with a GP model that uses \textit{automatic relevance determination} (see \ref{par:kernels}) makes this technique especially powerful for performing high dimensional characterization of previously unknown functions.
In this case, the sampling pattern will change as the relative sensitivities of the target function with respect to each optimization parameter are learned, as shown in Fig.~\ref{fig:bayesian_exploration}(b).
Bayesian exploration and similar active-learning techniques have been used to perform a wide variety of characterization studies in both accelerator experiments and simulations \cite{roussel_turn-key_2021}, as well as in automating experiments in X-ray and neutron scattering experiments \cite{noack_kriging-based_2019,noack2021gaussian}, and performing material discovery \cite{noack_autonomous_2020}.

\subsubsection{Incorporating unknown constraints}
\label{subsec:constraints}

\begin{figure*}
	\includegraphics[width=\linewidth]{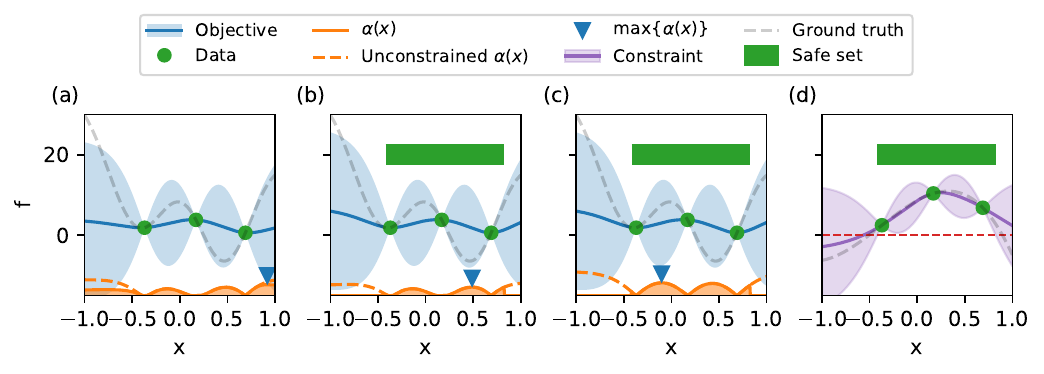}
	\caption{Comparison between different constrained Bayesian optimization algorithms. 
		(a) Weighting the acquisition function by the probability of satisfying the constraining function \cite{Gardner_2014}. (b) Acquisition function optimization within a safe set using MoSaOpt in exploitation mode \cite{Luebsen_2023} and (c) SafeOpt \cite{Sui_2015}. (d) The constraint function, where valid regions satisfy $c(x) > 0$.}\label{fig:SafeBO_Comp}
\end{figure*}

Constraints play a crucial role in accelerator operation for guaranteeing safe operation without damaging expensive equipment or increasing optimization time due to interlock violations. 
If the constraints are known, they can be incorporated by constraining optimization to a feasible subdomain of parameter space \cite{Boyd_2004,Berkenkamp_2021}. 
However, if the constraint functions $c_i(\mathbf{x})$ are unknown (as is often the case in accelerator physics), they need to be actively learned alongside the objective function $f(\mathbf{x})$ during optimization.

Several approaches have been developed in order to tackle the task of constrained optimization. 
These approaches aim to limit the number of times that constraints are violated during optimization, with varying degrees of ``safety".
This refers to the likelihood that constraints are violated during optimization (``safer" algorithms are less likely to violate constraints). 
A review regarding safe optimization techniques, inside and outside the context of BO, is given in \cite{Kim_2020}.
Two approaches of interest that have been used in accelerators are as follows.

The first approach is to modify the acquisition function by biasing it against selecting points that violate the set of constraints.
This is done by weighting the acquisition function by the probability that the constraints are violated, calculated by integrating over GP models of the constraining functions \cite{Gardner_2014}.
While this method is straightforward to implement and interpret, it does come with some disadvantages.
First, it requires that the unconstrained acquisition function is strictly positive (which can be achieved via a transform or offset) such that the constrained acquisition function has a minimal value of zero when constraints are not satisfied.
Second, in tightly constrained spaces there are large regions of parameter space where the constrained acquisition function is nearly zero, thus resulting in large areas where the derivative of the acquisition function is also nearly zero, making it difficult to optimize with gradient descent methods.
This issue can also potentially be addressed by taking the log of the acquisition function prior to optimization, as is suggested in \cite{ament2023unexpected}.
Finally, while biasing the acquisition function in this way does minimize the probability that parameters which violate the constraint are chosen for future measurements, it is possible using this technique that constraints are violated during optimization.
This technique has been applied towards both online and offline accelerator optimization problems with a variety of unconstrained acquisition functions, including Bayesian exploration \cite{roussel_turn-key_2021, roussel2023demonstration} and multi-objective BO \cite{roussel_multiobjective_2021}, 

The second approach was originally presented under the name \textit{SafeOpt} in \cite{Sui_2015,Berkenkamp_2021}. 
Instead of modifying the acquisition function, predictions from the GP models of the constraints are used to define an arbitrarily shaped but compact safe set, within which the constraints are predicted to be satisfied with a desired confidence level under the assumption of Lipschitz continuity and knowledge of an upper bound of the Lipschitz constant \cite{sohrab2003basic}. 
Slightly less conservative but less intuitive conditions for the safety guarantees are given in the respective papers \cite{Sui_2015,Berkenkamp_2021}.  
The acquisition function is then optimized inside this safe set, guaranteeing safety at a chosen confidence level. 
While in the original work, only one-dimensional constraint functions were considered, multiple constraints can also be incorporated \cite{Berkenkamp_2021}. 

The safety guarantees, achieved by the constrained optimization problem for evaluating the acquisition function come at a cost however, as defining and optimizing over the irregular valid sub-domain of parameter space is difficult, especially in high dimensional spaces. 
Different paths have been taken in order to increase the efficiency. 
The high dimensionality issue was addressed by \textit{LineBO}, where the global BO problem is decomposed into a sequence of one-dimensional sub-problems \cite{Kirschner_2019}. 
Stage-based procedures, i.e., \textit{StageOpt} and \textit{MoSaOpt}, where the expansion of the safe set (exploration), and exploitation phases are staged \cite{Sui_2018, Luebsen_2023}. 
This allows adjustment of hyperparameters in exploitation while still guaranteeing safety. 
Efficiency can be further improved by using goal-oriented safe exploration (GOOSE) where expansion only takes place if necessary \cite{Turchetta_2019}. 

Given the safety guarantees, this branch of approaches originated in safety-critical fields such as robotics, but has also been successfully applied in the control of accelerators starting with \cite{Kirschner_2019}. 
Here, SafeOpt was applied for the beam intensity optimization at SwissFEL for up to 40 optimization variables. 
A lower threshold on pulse energy was considered for safety and the high dimensionality was addressed by using LineBO. 
Defining the valid sub-domain in a 1D space greatly simplified the problem and provided useful visual feedback to operators during optimization, without significantly degrading optimization performance.

Further adaptions to this application are made in \cite{Kirschner_2022}, where in addition application results to the High Intensity Proton Accelerator (HIPA) are presented targeting to minimize the overall beam losses around the machine using 16 optimization variables and ensuring safety via 224 constraints coming from different interlocks. 
The stage-wise procedure MoSaOpt was applied in simulation to the optical synchronization system as well as a laboratory setup at the European XFEL in \cite{Luebsen_2023} in order to minimize the timing jitter by tuning up to 10 controller variables as optimization variables and ensuring an upper threshold on the timing jitter in order for the lasers to not lose the lock.

A comparison of the algorithms used for performing constrained BO on a simple test problem is shown in Fig.~\ref{fig:SafeBO_Comp}.
Figure~\ref{fig:SafeBO_Comp}(a) shows that weighting the acquisition reduces the chances of violating the constraint, although there are no guarantees the constraint violations will not occur.
On the other hand, methods that restrict the optimization of the acquisition function to within a valid sub-domain of the parameter space, such as MoSaOpt (Fig.~\ref{fig:SafeBO_Comp}(b)) and SafeOpt (Fig.~\ref{fig:SafeBO_Comp}(c)), do not allow points that are predicted to violate the constraint to be sampled, ensuring safety.

It is important to note that both of these approaches to constrained optimization rely on accurate models of the constraining functions to effectively reduce the number of violations during optimization.
As a result, most constraint violations happen during the initial stages of optimization, where few observations of the constraining functions are available to create an accurate GP model.
In order to prevent this, it is critical to start with a valid point in parameter space and conservatively explore the local region in the initial first steps or include prior information about the constraining functions into the GP model of the constraints.

Finally, it is reasonable to expect that concepts from the two methods currently used for constraining BO in accelerator physics can be combined into a single algorithm that contains the benefits provided by both methods.
Additionally, characterization of the trade-offs between safety tolerance and optimization speed should also be investigated.

\subsubsection{Multi-objective optimization}
\label{subsec:mobo}
In accelerator physics, it is often the goal of optimization to simultaneously minimize or maximize more than a single objective, referred to as multi-objective optimization.
These objectives can compete with one another, requiring trade-offs between objectives to reach an optimal solution.
For example, it is difficult to simultaneously maximize the lifetime and dynamic aperture of electron storage rings \cite{SONG2020164273}, or minimize the bunch size and beam emittance in a photoinjector due to space charge \cite{bazarov2005multivariate, edelen2020machine}.
One strategy to solve this problem is to combine the objectives into a single objective by weighting the contribution of each objective to a single term, a process known as \textit{scalarization}.
However, the goal of multi-objective optimization is to determine what is known as the \textit{Pareto front} (PF).
A PF represents a set of non-dominated solutions, where no other solution can improve one objective without degrading at least one other objective. 
These solutions are considered \textit{Pareto-optimal} because they form the best compromise among the multiple conflicting objectives.

One of the most popular methods for solving multi-objective optimization problems is the use of evolutionary algorithms \cite{maier2019introductory}, which use evolutionary heuristics to generate a large population of candidate points in parameter space from the previous generation to search for the PF.
While these algorithms are easy to implement and use, they are incredibly inefficient, requiring the use of massively parallelized evaluation of many candidate points to converge to a solution set.
As a result, multi-objective optimization is computationally expensive in the case of simulated optimization of beam dynamics and nearly impossible to use during beamline operations.

\begin{figure}
	\includegraphics[width=\linewidth]{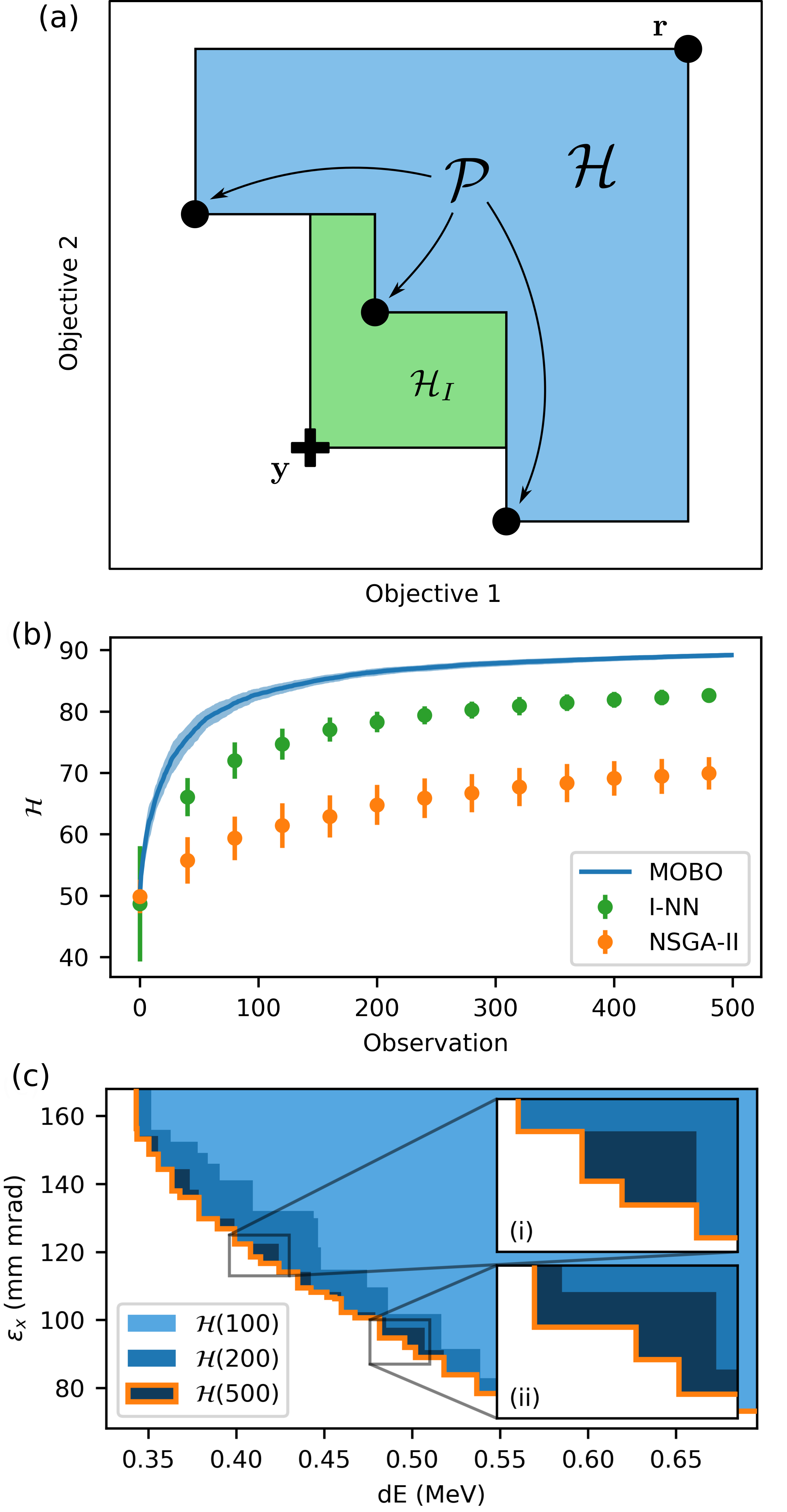}
	\caption{Summary of multi-objective BO (MOBO) using expected hypervolume improvement (EHVI). (a) Given Pareto front $\mathcal{P}$ and corresponding hypervolume $\mathcal{H}$, the increase in hypervolume $\mathcal{H}_I$ due to a new measurement $\mathbf{y}$ is given by the shaded green area. (b) Comparison between multi-objective optimization algorithms for optimizing the AWA injector problem. NSGA-II is a standard evolutionary algorithm \cite{deb_fast_2002}, I-NN is surrogate model assisted NSGA-II \cite{edelen2020machine}. (c) Projected hypervolume after a set number of MOBO iterations with insets showing hypervolume improvement due to fill in points (i) and measurement of newly dominant points (ii). Reproduced from \cite{roussel_multiobjective_2021}. \label{fig:mobo_summary}}
\end{figure}

Special acquisition functions in BO have been developed to quickly identify the PF solution in multi-objective optimization problems.
These acquisition functions rely on a metric known as the PF \textit{hypervolume} (denoted $\mathcal{H}$), shown in Fig.~\ref{fig:mobo_summary}(a).
The hypervolume is a widely used quality indicator in multi-objective optimization and is particularly useful for problems with more than two objectives. 
It measures the size of the dominated space, i.e., the portion of the objective space that is not covered by the PF. The larger the hypervolume, the better the set of solutions is considered because it indicates a better coverage of the objective space and a higher degree of Pareto optimality.
To calculate the hypervolume, a reference point is specified in the objective space, typically set to be a point with worst values for all objectives. 
Then, for each non-dominated solution in the PF, the hypervolume is computed as the volume of the space dominated by the reference point and the current solution. 
The total hypervolume of the entire PF is the sum of these individual hypervolumes.
Once additional observations of the objective values no longer increase the hypervolume then the current PF is said to have been identified.

The Expected Hypervolume Improvement (EHVI) \cite{daulton2020differentiable} acquisition function uses the notion of an increase in PF hypervolume to select points in parameter space.
Starting with a PF containing previous measurements of the objectives, EHVI predicts the average expected increase in hypervolume (as shown in Fig.~\ref{fig:mobo_summary}) as a function of optimization parameters using GP models for each objective.
As a result, BO using EHVI will select points that are more likely to maximally increase the hypervolume of the PF than other algorithms, whereas genetic algorithms select points only based on their optimality.
When applied to identifying the PF of the AWA photoinjector containing 7 objectives (beam sizes, beam emittances, and energy spread), EHVI was able to converge to a maximum hypervolume several orders of magnitude faster than evolutionary algorithms, as shown in Fig.~\ref{fig:mobo_summary}(b).

EHVI is able to increase the PF hypervolume through two means, shown in Fig.~\ref{fig:mobo_summary}(c).
One method ``fills-in" the multi-dimensional surface of the PF, leading to hypervolume increase that improves the detail described by the Pareto set.
The second method increases the hypervolume by selecting observations that will likely dominate current non-dominated points in the PF.

A major advantage of EHVI over genetic algorithms is that it can be used in serial optimization contexts where objectives cannot be evaluated in parallel, for example, determining the PF during online accelerator operations, as was done at the SLAC MeV-UED beamline \cite{ji_multi_2022}.
However, a downside of EHVI acquisition function is the computational expense associated with calculating the hypervolume improvement.
The cost of partitioning the PF hypervolume into hyper-rectangles scales exponentially with the number of objective functions.
As a result, this can be a significant roadblock towards using EHVI in practice when a large number of objectives are to be optimized.
This motivates the use of alternate acquisition function optimization algorithms in certain instances when a large number of objectives are present (see Sec.~\ref{sec:acq_opt} for details).


A final consideration when using EHVI is the specification of the reference point.
The reference point specifies the worst case value for each objective, thus any objective observations that are worse than the corresponding reference point values will not contribute to the PF hypervolume.
As a result, the PF explored by EHVI will be limited to within the boundary specified by the reference point, thus ignoring objective function values beyond the reference point.
However, specifying a reference point that is too far from perceived optimal values of the objective functions will reduce the detail of the PF as different points will have vanishingly small contributions to the total hypervolume.

The multi-objective Bayesian Optimization scheme has been demonstrated experimentally at the SLAC-MeV Ultrafast Electron Diffraction (UED) facility \cite{ji_multi_2022, ji_multi_2024}. 
For MeV-UED, different scientific experiments often pose different requirements on multiple electron beam properties, such as electron pulse length, spot size at sample and momentum space resolution (q-resolution). However, it is difficult to simultaneously minimize these beam properties due to space charge forces. The goal of online optimization is to determine the PF which represents the system performance limit and gives the trade-offs between key beam properties. 
In practice, the evaluation time cost is high ($\sim60$s per data point) so that evolutionary algorithms are nearly impossible to use. The MOBO scheme is much more data efficient and can converge to the PF at least one order of magnitude faster than evolutionary algorithms. As a result, MOBO is the most suitable solution for beam optimizations at MeV-UED.

During optimization, gun phase and solenoid strengths were varied to explore the response of electron pulse length, spot size at sample and q-resolution and obtain PF giving trade-offs between them.
MOBO was able to obtain PF within 150 measurements ($\sim3$ hour). The PF offers an unprecedented overview of the machine’s performance limitations and can greatly assist human scientists in rapid decision-making. The achieved performance was comparable with that obtained by experienced human operators and requires a significantly fewer measurements compared with traditional exploration methods such as a Grid Search (GS).
During the experiment, the extra computation time associated with GP fitting and EHVI acquisition function optimization is small (below 5 s per iteration) relative to the reduction in beam property evaluation time associated with faster convergence of HV.
Currently, the major limitation is the time taking electron beam diagnostics, by implementing highly efficient single-shot, non-destructive and automated electron beam diagnostics, $> 10^3$ data points could be obtained within a shorter time. This enhancement could improve the accuracy of GP and fully exploit the advantages of the MOBO algorithm.

\subsubsection{Multi-point optimization and virtual objectives}
\label{subsec:virtual_objectives}
In some optimization tasks, each acquisition requires a secondary scan in a separate domain to calculate the objective function. 
In engineering, this type of measurement process is referred to as a \emph{multi-point query} (see e.g. \citep{liem2015}). 
Consider, for example, the task of aligning particle beams through the magnetic center of quadrupole focusing magnets.
If the beam is misaligned with respect to the magnetic center of quadrupoles in the beamline, scanning the quadrupole strength results in a centroid kick causing further misalignments downstream.
This can be corrected through the use of steering magnets which provide an angular kick to the beam such that it intercepts the center of the quadrupole, resulting in no kick as the quadrupole strength is varied.
However, determining the optimal steering strength requires either beam position monitors at the quadrupole location, or constant scanning of the quadrupole strength while varying the steering parameter to estimate the beam misalignment.
This is relatively simple for a single quadrupole but becomes increasingly complex when using multiple steering elements to align through multiple quadrupoles.

\begin{figure*}
	\includegraphics[width=\linewidth]{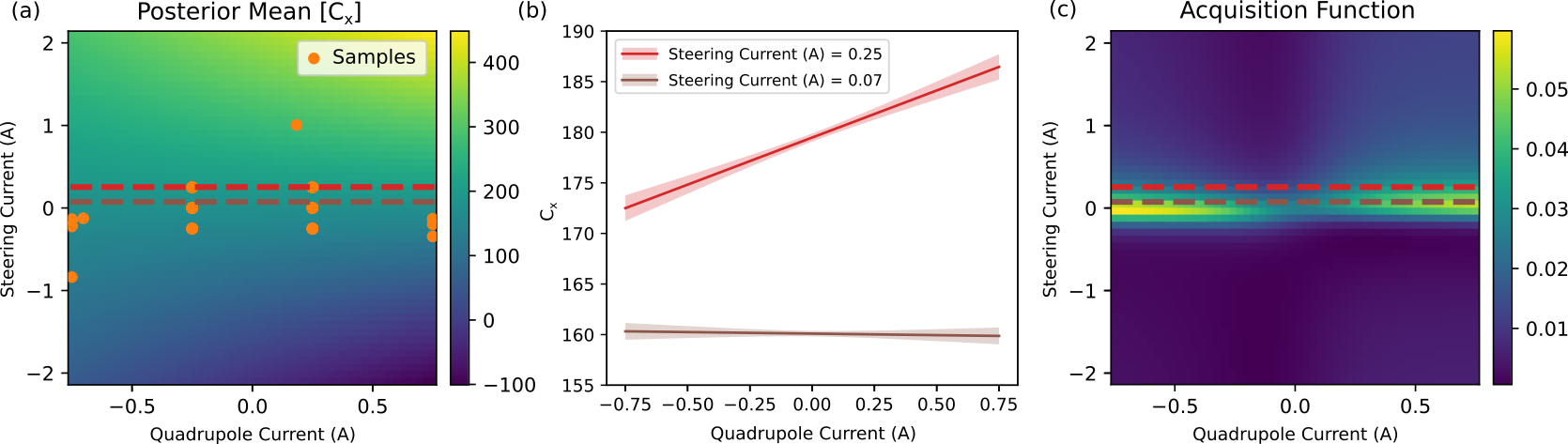}
	\caption{Visualization of the BAX process for beam steering through quadrupole magnets. (a) Experimental measurements are used to build a GP model of the horizontal beam centroid position at a downstream screen $C_x$ as a function of the quadrupole strength and steering parameter. Note that the GP model is built with a 1st order polynomial kernel, constraining predictions to planar surfaces. Dashed lines denote cross sections of the GP model shown in (b). (c) The BAX acquisition function which predicts the information gained about the ideal steering current by making future measurements.}
    \label{fig:bax_alignment}
\end{figure*}

To address this problem using BO techniques, an acquisition function known as Bayesian Algorithm Execution (BAX) \cite{neiswanger2021bayesian} has been developed which uses a so-called ``virtual" objective to make control decisions.
In the quadrupole alignment problem, the virtual objective is to minimize the slope of the beam centroid with respect to the quadrupole strength, which is proportional to the beam misalignment.
Instead of directly measuring this slope every time the steering parameter is varied, BAX builds a model of the beam centroid as a function of both the quadrupole strength and the steering parameter, as shown in Fig. \ref{fig:bax_alignment}(a).
This model of the beam centroid is then used to predict the magnitude of centroid deflection as the quadrupole strength is varied (slope) as a function of the steering parameter, shown in Fig. \ref{fig:bax_alignment}(b).
The BAX acquisition function uses these predictions to evaluate which future measurements will provide the most information about the steering current that leads to a minimization of the centroid deflection.
This aspect is seen in Fig.~\ref{fig:bax_alignment}(c), where the maximum of the acquisition function is at the edges of the quadrupole parameter domain (which provides the most information about the slope) and close to the optimal steering parameter.
In the limit of many measurements, BAX will continue to make measurements close to the optimal steering parameter in order to improve model confidence in that region of parameter space.

This method of using virtual objectives can be extended to more complex situations.
For example, performing alignment through multiple quadrupoles in both horizontal and vertical directions can be done by simply adding or multiplying multiple virtual objectives together into a single objective.
BAX also supports more complex virtual objectives such as transverse beam emittance \cite{miskovich2024multipoint}.
In this case, the virtual objective involves fitting polynomials to the beam size squared as a function of quadrupole strength using predictions from the GP model.
At FACET-II, BAX was able to match the best emittance found by hand-tuning, while at LCLS, the solution found by BAX produced about 25\% lower emittance than hand-tuning. 
In simulation studies, BAX minimizes the emittance using 20 times fewer beam size measurements than traditional BO. 
The dramatic improvement results from both increased sampling efficiency (by selecting single beam-size measurements at each acquisition) and from modeling the beam-size function rather than the noisier emittance values.

\subsubsection{Proximal Biasing}
\label{sec:proximal}
Unlike optimization problems in other fields, online particle accelerator optimization sometimes requires incremental traversal of parameter space to maintain accelerator stability.
Accelerator facilities often have many interconnected subsystems that are independently controlled through feedback systems to maintain accelerator parameters, such as water temperature, RF phase, and beam steering.
As a result, making rapid changes in accelerator parameters can negatively affect these feedback loops, causing instabilities in accelerator operation that can ultimately shut down the accelerator.
One strategy for mitigating this issue is to place a strict upper bound on the travel distance from the current location in parameter space.
Unfortunately, this in turn limits the exploration of parameter space needed to successfully find global extrema in BO.
While it is possible (and sometimes necessary in sensitive systems) to place this hard limit on the maximum travel distance during each optimization step, it is sometimes more useful to bias the acquisition function towards making smaller steps in parameter space.
This can be done through a technique known as ``proximal biasing" \cite{roussel_2021_proximal}.
Proximal biasing modifies a base acquisition function by adding a multiplicative term
\begin{equation}
	\Tilde{\alpha}(\mathbf{x}) = \alpha(\mathbf{x})\exp\Big(-\frac{(\mathbf{x} - \mathbf{x}_0)^2}{2l^2}\Big)
\end{equation}
where $\mathbf{x}_0$ was the last location in parameter space to be observed and $l$ is an algorithm parameter that controls how strongly biased the acquisition function is towards making small steps in parameter space.
This formalism places a restriction on the base acquisition function, requiring that $\alpha(\mathbf{x}) \geq 0$, however is satisfied for most acquisition functions (with the notable exception of UCB).
\begin{figure}
	\includegraphics[width=\linewidth]{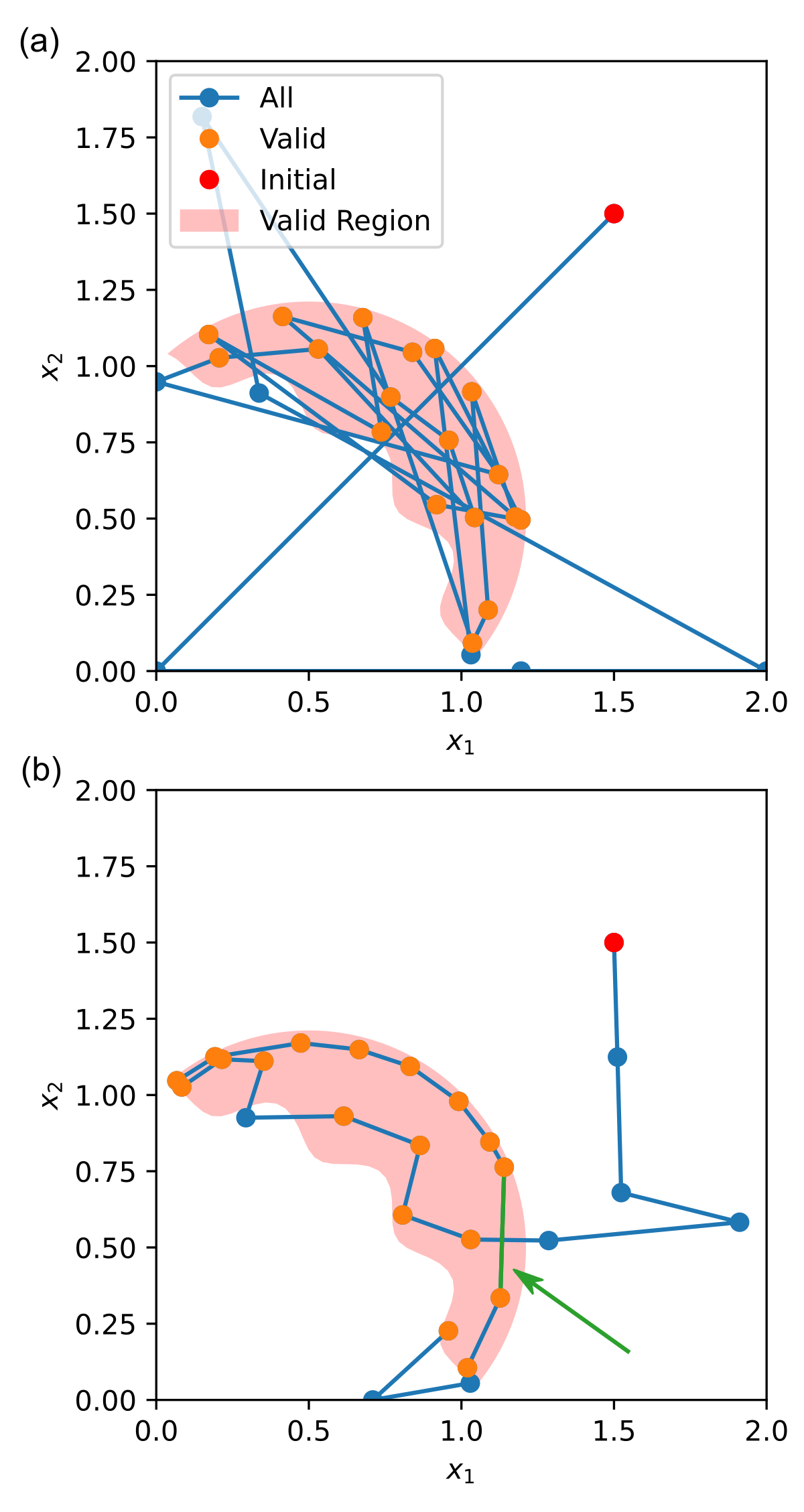}
	\caption{Demonstration of proximal biasing effects during Bayesian Exploration (BE) of the constrained TNK test problem. (a) Normal BE. (b) BE using proximal biasing with $l=0.1$. The green arrow highlights a step where a larger jump in parameter space was allowed by proximal biasing. \label{fig:prox_bias}}
\end{figure}

A visualization of how proximal biasing effects BO is shown in Fig.~\ref{fig:prox_bias}.
In this case, the goal is to characterize the first objective of the TNK test function \cite{deb_fast_2002} so the base acquisition function is Eq.~\ref{eq:bayes_exp}.
Figure~\ref{fig:prox_bias}(a) demonstrates that without proximal biasing the exploration process makes large steps in parameter space in order to aggressively explore the objective function within the valid region.
On the other hand, adding proximal biasing to the acquisition function significantly reduces the average step size, resulting in a smoother exploration of the parameter space, as shown in Fig.~\ref{fig:prox_bias}(b).
In addition, proximal biasing does allow for larger steps in parameter space when necessary, as evidenced by the step highlighted by the green arrow in Fig.~\ref{fig:prox_bias}.
If instead of proximal biasing, a hard limit on travel distance was set for this algorithm, it's likely that this larger travel distance would not have happened, resulting in a lack of exploration of the southernmost region of the valid domain.

\subsubsection{Multi-fidelity optimization}
\label{subsec:multi_fidelity}

In the case where data can be queried at different fidelities (quantified by a parameter $s$ ; see section \ref{subsec:multi_fidelity_modeling}), the BO algorithm needs to choose both the input parameter $\boldsymbol{x}$ and the fidelity $s$ for each evaluation of the objective function. 
In addition to balancing exploration and exploitation, the algorithm must also balance the cost of an evaluation at a given fidelity with the corresponding information gain at the target (highest) fidelity. 
For instance, in the case where $s$ represents the resolution of a numerical simulation, there is a trade-off between low-resolution simulations that provide low-fidelity information at a reduced computational cost, and high-resolution simulations that provide high-fidelity information at an increased computational cost. 
If low-fidelity objective function values are strongly correlated with high-fidelity objective function values, BO can leverage low-fidelity approximations of the objective function to reduce the cost of optimization.

One simple way to handle this trade-off is to opt to use repeated fixed-size batches of low-fidelity and high-fidelity evaluations \cite{letham_bayesian_2019}. 
For example, multi-fidelity Bayesian optimization was run using a multi-task GP model with repeated batches of 96 low-cost, low-fidelity simulations and 3 high-cost, high-fidelity simulations, in order to optimize the performance of a laser-plasma accelerator \cite{ferran_pousa_bayesian_2023}. 
In this case, the acquisition function was a modified version of EI \cite{letham_bayesian_2019}, whereby only the highest-fidelity evaluations are considered when determining the optimal previously observed point $f(\boldsymbol{x}^*)$. 
In this particular example, the multi-objective Bayesian optimization was observed to require 7$\times$ less computational resources to find an optimal accelerator configuration, compared to single-fidelity Bayesian optimization based only on high-cost, high-fidelity simulations \cite{ferran_pousa_bayesian_2023}.

However, in many cases, instead of using fixed-size batches of low-fidelity and high-fidelity evaluations, the fidelity $s$ is dynamically decided by the algorithm for each evaluation. 
Typically, one would want the algorithm to mostly use low-fidelity evaluations early on in the optimization (to get a cheap, coarse picture of the overall objective landscape) and to progressively use more high-fidelity evaluations as it narrows down on the optimal point. 
It is also desirable that the algorithm rapidly stops using low-fidelity evaluations, if the underlying multi-fidelity Gaussian Process model determines that low-fidelity evaluations are not representative of high-fidelity data (see section \ref{subsec:multi_fidelity_modeling}). 
This behavior can be obtained by using acquisition functions that incorporate both the cost and the information gain of an evaluation at a given fidelity; examples of such acquisition functions include multi-fidelity versions of Upper Confidence Bound \cite{kandasamy_multi-fidelity_2017} and Knowledge Gradient \cite{wu_continuous_2017}.
An alternative to these modified acquisition function is to instead cast the multi-fidelity optimization as a multi-objective optimization problem \cite{irshad_multi-objective_2023,irshad_leveraging_2023}.
In this scenario, a user-defined function assessing the reliability of a fidelity, denoted as $s$, is included as one of potentially multiple objectives. 
The Expected Hypervolume Improvement (EHVI) acquisition function, explained in section \ref{subsec:mobo}, is used to solve this multi-objective problem, with an added penalty for the evaluation cost \cite{irshad_leveraging_2023}.
This multi-objective, multi-fidelity algorithm resulted in lower optimization costs when used in simulation-based design optimization of laser-plasma accelerators \cite{irshad_multi-objective_2023}.

\section{Acquisition Function Optimization} \label{sec:acq_opt}
Conducting BO involves addressing a nested numerical optimization challenge to determine the point in parameter space that maximizes the acquisition function. 
The computational demands of numerically optimizing the acquisition function make it the most resource-intensive step in the BO process. 
This process necessitates repetitive evaluations and/or sampling from the GP surrogate model posterior, incurring computational expenses—albeit generally less than those associated with evaluating the objective function directly.
Adding to the complexity, acquisition functions are often non-convex and may exhibit numerous local extrema \cite{wilson2018maximizing}. 
As a result, the selection of the numerical optimization algorithm employed to optimize the acquisition function becomes pivotal in achieving optimal performance in BO.

In scenarios where several points, or multiple objectives and constraints can be measured concurrently, BO can also be used to propose multiple measurement candidates. 
This is accomplished by identifying multiple parameter sets that collectively maximize the acquisition function. 

In this section, we highlight a variety of approaches to optimize acquisition functions, which affect the execution speed, improve performance of BO algorithms, and tailor BO to specific use-cases.

\subsection{Basic Algorithms}
The simplest approaches to optimizing acquisition functions are brute-force methods, such as random sampling or sampling on a mesh grid of points.
These algorithms are usually poor choices for maximizing the acquisition function, due to their performance scaling to even modest numbers of free parameters.
However, in low-dimensional parameter spaces (1-2 dimensions) the number of acquisition function evaluations necessary to maximize the acquisition function can be similar to other iterative methods due to their complex nature (non-convexity).
Given that the acquisition function can be evaluated in parallel through the use of batched computations, using random or grid based sampling strategies can sometimes be faster than iterative optimization algorithms.

Iterative, black-box optimization algorithms, such as Nelder-Mead simplex and RCDS can also be used to maximize the acquisition function.
However, in most cases, maximizing the acquisition function is often best done using gradient-based optimization algorithms.
The most straightforward example of this is gradient descent algorithms such as Adam \cite{kingma2014adam}.
Higher order gradient algorithms, such as limited-memory-BFGS (L-BFGS) which uses an implicit estimation of the inverse Hessian, are also often commonly used to further speed up convergence.
In both cases, accurate calculations of the gradients can significantly reduce the number of iterations needs to reach convergence.
Acquisition function calculations that are differentiable can be used to quickly calculate accurate gradients to speed up optimization.
This is usually done by implementing the GP model and acquisition functions in a machine learning library that supports differentiability, such as PyTorch \cite{paszke_pytorch_2019}.
Unfortunately, these algorithms are themselves local optimization algorithms.
To improve chances of finding the global maximum of the acquisition function, parallel optimization from multiple random starting points is often used to explore diverse regions of parameter space.

\subsection{Trust region optimization}
\label{subsec:turbo}
One disadvantage of BO is that common acquisition functions tend to over-prioritize exploration over exploitation in high dimensional parameter spaces.
This is due to the relatively large posterior uncertainties of GP models that result from the exponential growth of parameter space volume with dimensionality (models in high dimensional space need more data to update prior function distributions).
As a result, BO tends to pick points at the extremes of the domain in high dimensional parameter spaces even if optimal points are found in a local region, see Fig.~\ref{fig:prox_bias}(a) for an example of this behavior.
In addition, GP models used in BO aim to create a global description of the objective function, which may not be appropriate for functions that have varying local characteristics in different regions of parameter space.

Trust region BO (TurBO) \cite{eriksson2019scalable} aims to address both of these issues by restricting optimization of the acquisition function to within a so-called ``trust region" around previous measurements where the model is expected to be the most accurate.
The trust region is a local region centered at the best previously observed measurement so far during optimization, with side lengths equal to a base length $L$ multiplied by the relative length scale of the GP model along each axis in parameter space.
As optimization progresses, the location and size of the trust region is continuously updated to be centered at the best measured point in parameter space and scaled to match length scales of the GP model.
Additionally, the base length of the trust region is increased or decreased based on the number of consecutive successes (improvements in the solution) or failures (no improvement) respectively.
As a result, the trust region shrinks in cases where the model does not correctly identify the location of optimal solutions or expands the trust region when the model is making accurate predictions that result in continuous improvements in the objective function value.
By limiting exploration of the parameter space within a local region, TurBO transforms BO from a global optimization algorithm into a local one, resulting in substantially faster convergence to local extremum in high-dimensional parameter spaces than conventional BO.

\begin{figure}
	\includegraphics[width=\linewidth]{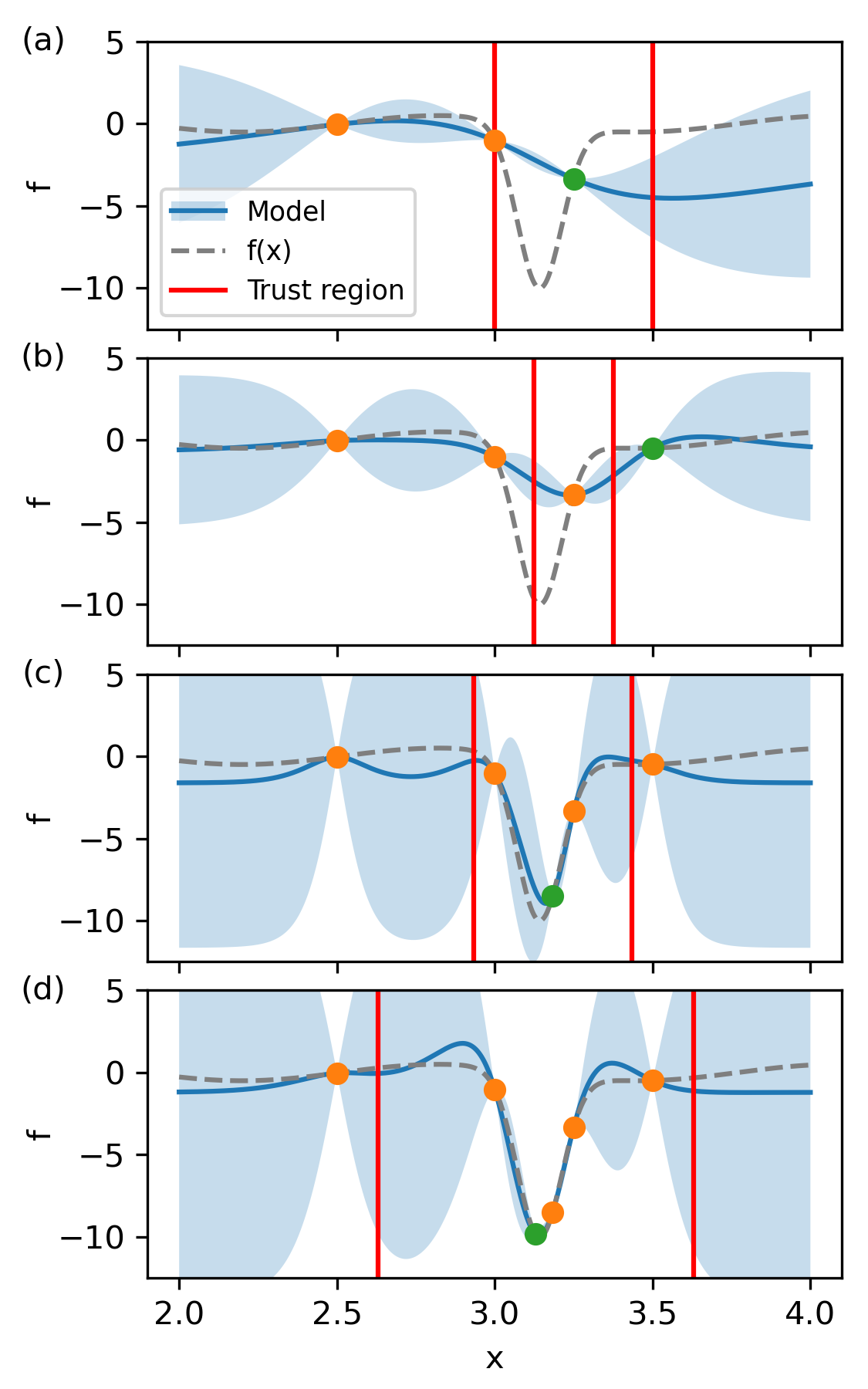}
	\caption{One dimensional visualization of trust region BO (TurBO) applied to a minimization problem with the UCB acquisition function. (a-d) Sequential evolution of the GP model and sampling pattern. Orange circles denote objective function measurements and green circles denote the most recent sequential measurement at each step. 
		\label{fig:turbo}}
\end{figure}

A 1-dimensional example of TurBO applied to a test minimization problem is shown in Fig.~\ref{fig:turbo}.
Despite large model uncertainties at the edge of the domain, which would normally cause BO to sample points on the boundary, TurBO chooses observations that are in the local trust region around the best observed solution.
In cases where the new observations do not improve over the best solution, the trust region contracts around the optimal point to increase model accuracy.
If new observations do improve over the previous optimal point, the trust region is re-centered at the location of those observations and expanded to find potential new solutions.
Throughout the course of optimization, TurBO will develop a locally accurate BO model near observed optimal solutions, instead of trying to accurately describe the global function behaviour.
While in this example TurBO shrinks and expands the trust region after every step, a threshold for successes and failures is usually set such that multiple failures or successes in a row are necessary to change the overall trust region size.

\begin{figure}
	\includegraphics[width=\linewidth]{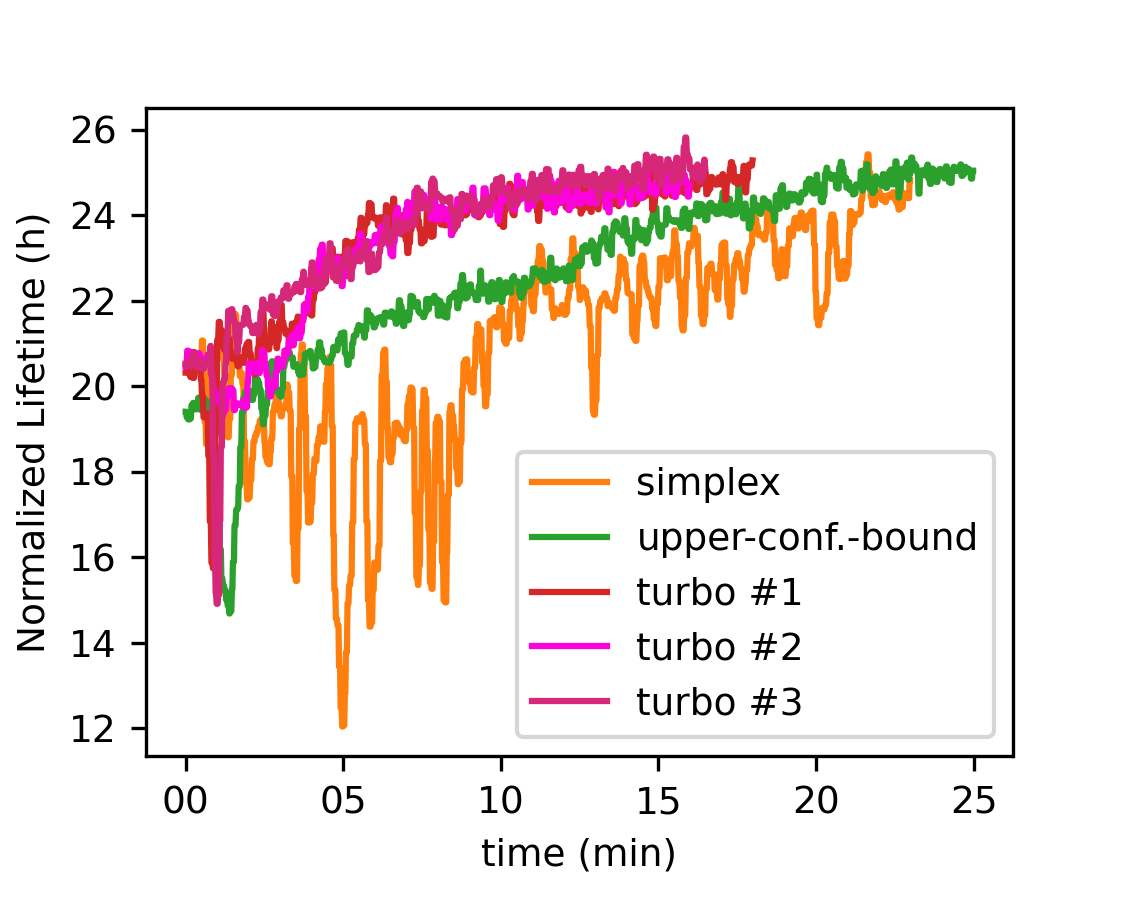}
	\caption{Trust region BO (TurBO), simplex, and UCB applied to the minimization of total losses (maximization of lifetime) at the ESRF-EBS storage ring. 
		\label{fig:turboEBS}}
\end{figure}

TurBO was used on the ESRF-EBS storage ring \cite{ebs} for the optimization of lifetime and compared to the existing optimization procedure. 
The 192 sextupoles and 64 octupoles available for the optimization of lifetime have been sorted and selected into 24 tuning parameters. 
To have fast and reproducible values for the optimization the sum of all signals from the 128 beam loss detectors was used as objective of the minimization rather than the lifetime value itself. 
Figure \ref{fig:turboEBS} shows the resulting lifetime during the optimization process performed with: TurBO, simplex and UCB. 
The same parameters and procedure for optimization are used in all cases, only the optimization algorithm is changed. 
More details on the measurement can be found in \cite{liuzzo:icalepcs-MO3AO01}. 
The TurBO optimization was repeated three times and led in all cases to similar lifetime values within the same optimization time and with comparable final sextupole and octupole settings. 
Also starting from degraded storage ring conditions, TurBO could quickly recover the optimal set point for the magnets. 

TurBO can also be slightly modified to improve exploration of tightly constrained problems where a majority of the input space violates one or more constraining functions.
In this case, the goal is to reduce the number of constraint violations during optimization through the use of a conservative trust region.
Instead of centering the trust region at the best observed solution, this approach centers the trust region at the average value of valid observations.
Then the trust region side lengths are varied depending on the frequency of constraint violations observed during optimization or exploration.
This prevents sampling at the extremes of the parameter space, which often results in measurements that violate the constraints.

\subsection{Parallelized optimization}
While in most cases BO is used in the context of serial optimization (one evaluation of the objectives/constraints is done at a time), it is possible for BO to propose a set of promising points that can be evaluated in parallel.
We describe three distinct strategies here that are relevant to generating small ($n<10$), medium ($10 < n < 100$), or large ($n >100$) sets of candidate points to evaluate in parallel.

\subsubsection{q-Sampling}
This strategy aims to generate a set of candidate points in parameter space that jointly optimize given acquisition functions \cite{ginsbourger2010kriging}.
Many common acquisition functions (EI, UCB) can be expressed as the expectation of some real-valued function outputs at some designed input space \cite{wilson_reparameterization_2017}. 
Evaluating the acquisition function in the context of parallel selection of candidate points requires evaluating integrals over the posterior distributions. 
However, this makes evaluating parallelized versions of acquisition functions analytically intractable.

An alternative is to use Monte-Carlo (MC) sampling to approximate the integrals. 
An MC approximation of acquisition function $\alpha$ at input space $\mathbf{x}$ using $N$ MC samples given the data observed so far is
\begin{equation}
	\alpha(\mathbf{x}) \approx \frac{1}{N} \sum_{i=1}^{N} a(\xi_i) \label{eq:mc}
\end{equation}
where $a(\cdot)$ is a real valued function and the samples $\xi_i$ are drawn from the posterior predictive distribution $p(y|\mathbf{x}, \mathcal{D})$.

To see this in action, we can examine the definition of parallelized Expected Improvement (qEI)~\cite{rezende_stochastic_2014, wilson_reparameterization_2017,wilson_maximizing_2018, balandat_advances_2020} which generates $q$ candidates that jointly optimize the EI acquisition function:
\begin{align}
	qEI(\mathbf{x}) & \approx \frac{1}{N} \sum_{i=1}^{N} \max_{j=1,...,q} \left\{ \max(\xi_{ij}-f^*, 0) \right\} \label{eq:qEI} \\
	&\xi_i \sim p(f|\mathcal{D}) \nonumber
\end{align}
where $f^*$ is the best observed objective value so far.

To maintain the inexpensive computation of gradients for MC-based acquisition functions using automatic differentiation, a technique known as the ```reparameterization trick" \cite{wilson_reparameterization_2017} is used.
Instead of sampling directly from the posterior of the GP model, samples are drawn from a unit Normal distribution, then scaled and shifted such that the distribution matches GP predictions.
This preserves differentiability by sidelining the stochastic generation of random samples.

\subsubsection{Local Penalization Techniques}
Local penalization is proposed as an alternative method for performing batched BO~\cite{gonzalez_batch_2016}. Instead of maximizing the joint distribution as in the q-sampling approach, it selects the samples in the batch sequentially and thus scales better with the input dimensions and batch sizes. The $i$-th sample is selected by maximizing the product of the acquisition $\alpha$ and the penalization $\phi = \prod_{1}^{i-1} \phi_j$, where $\phi_j \in (0,1]$ denotes the local penalization function around a previously selected point $x_j$ in the batch. 
It effectively excludes the region around previously chosen points and goes to 1 elsewhere. 
The behavior of the penalization is governed by the Lipschitz constant of the objective function, which could be inferred from the GP model.

The local penalization method has been used in the simulation study at the linear accelerator FLUTE for radiation optimization~\cite{xu_optimization_2022}. 
It enabled an efficient selection of parameters to run parallelized simulations in a high-performance computing cluster, resulting in better performance compared to using the genetic algorithm.

\subsubsection{Large scale parallelization}
In cases where objective functions can be evaluated in a massively parallelized fashion ($> 100$ simultaneous evaluations), i.e. in simulation on high performance computing clusters, optimizing the acquisition function using the strategies outlined above may exceed the computational cost of evaluating the objective itself.
As a result, it makes sense to use alternative methods for acquisition function optimization.
Evolutionary or genetic algorithms are extensively employed towards solving optimization problems using large-scale parallelization.
These algorithms use simple heuristics to generate candidate points, which is much cheaper than repeatedly numerically optimizing an acquisition function.
Thus, it is advantageous to generate a large number of candidate points using a genetic algorithm and then determine a subset of those candidate points using a model-based acquisition function to be evaluated in BO. 
Combining genetic algorithms with BO takes advantage of both of their strengths, generating large sample sizes in a relatively short amount of time while still incorporating model information and acquisition function definitions into the selection of candidates for evaluation.


The Multi-Objective Multi-Generation Gaussian Process Optimizer (MG-GPO) represents one such algorithm that takes advantage of this combination \cite{huang2021multi,zhang_online_2020}. 
This algorithm attempts to solve multi-objective optimization problems by first generating a number of candidate points using evolutionary heuristics (mutation~\cite{liagkouras_elitist_2013}, crossover~\cite{deb_simulated_1995}, and flocking operations).
A subset of candidate points are then selected to be evaluated on the real objective by using a GP surrogate model (based on previous measurements or simulation results) to predict which candidate points are likely to dominate over previous measurements.
By leveraging information in the learned GP surrogate model, the candidate points generated by MG-GPO are more likely to improve the Pareto optimal set when compared to model-free evolutionary algorithms (such as NSGA-II).

A slight modification can be made to MG-GPO to improve its performance by choosing a subset of candidates based on expected hypervolume improvement (as is done in conventional multi-objective BO) instead of predicted Pareto-optimality.
This has the added benefit of selecting candidates that not only will improve the Pareto front, but will maximize improvement according to the predicted increase in hypervolume once observed.


The MG-GPO method has been applied to design optimization of storage ring lattices~\cite{SONG2020164273}. 
It has also been applied experimentally to the SPEAR3 storage ring and the APS accelerator complex to demonstrate its online optimization capability with several important problems, including storage ring vertical emittance minimization with skew quadrupoles~\cite{zhang_online_2020}, nonlinear beam dynamics optimization with sextupoles~\cite{zhang_online_2020,Emery2022}, and linac front-end transmission tuning with steering and optics parameters~\cite{ShangIPAC2021}.
In each case it was shown that the algorithm can effectively improve the performance of the machine when compared to other algorithms.

\section{Discussion} \label{sec:discussion}
In this section we discuss several aspects of BO that are relevant to its use in accelerator physics.
We first describe the relationship between BO algorithms and other algorithms currently used in accelerator physics for optimization and control.
We then discuss how to interpret and monitor BO performance during optimization and general best practices for improving optimization performance.
Additionally, we highlight software packages, both inside and outside the accelerator physics field, that are used to implement BO algorithms.
Furthermore, we provide estimations of run time and computational memory usage for BO algorithms.
Finally, we describe future research avenues in BO methods for accelerator physics.

\subsection{BO in relation to other optimization algorithms}
\label{subsec:bo_relation_to_other_algorithms}

Here, we describe how classical BO relates to various other types of optimization and control algorithms. 
We also highlight the conceptual differences and similarities between online optimization and continuous control. 
Finally, we discuss the impact of different function approximations and ML model choices within those paradigms.

Note that we cannot make definitive, general statements about algorithm performance. 
The performance of a particular algorithm on a given accelerator problem is dependent on numerous factors, including, but not limited to, the specific algorithmic hyperparameters chosen, as well as the problem dimensionality, nonlinearity, convexity, multi-modality, and noise.

\subsubsection{Episodic optimization} 
Typically when describing ``optimization," we mean an episodic process of adjusting settings to reach an optimal combination, that then ideally remains fixed for some period of time. 
Aside from BO, various other optimization algorithms have been developed and are actively used in the accelerator physics domain. 
Generally, these algorithms can be split into gradient-based and gradient-free (black box) algorithms, and, additionally, algorithms which learn some underlying representation of the system and those that do not.

Gradient-based algorithms use direct information about the gradient of the cost function, or approximations of it (for example via finite difference methods), to determine setting changes during optimization. 
Gradient approximations on non-differentiable systems (whether in simulation or on an experiment) can be time-consuming to obtain, particularly as the number of variables increases. 
In some instances in accelerators, gradient information has been approximated from machine jitter, allowing small, minimally-invasive setting changes to slowly compensate for drift or move toward an optimum \cite{private_communication_Florian}. 
Gradient-based algorithms can also easily become stuck in local minima, although techniques do exist to work around this (e.g. providing warm starts from a system model or previously-known global solution, restarting the algorithm several times at different random starting points). 

Gradient-based algorithms, such as stochastic gradient descent and variants (e.g. Adam, RMSProp  \cite{kingma2014adam,goodfellow_deep_2016,Ruder2016AnOO}), can scale well to higher dimensions particularly in cases where the evaluation of the objective function is fast and gradients are directly available. Consequently, they are used frequently in ML for training neural networks. In that context, updates to model parameters using small batches of data help to avoid local minima by adding noise to the gradient.

Gradient-based methods can also be used in conjunction with differentiable models, e.g. through differentiable physics simulations~\cite{dorigo_toward_2023, ratner_recovering_2021,roussel_differentiable_2022,kuklev_differentiable_2023}, codes such as \textit{Bmad-X}~\cite{roussel_phase_2023} or \textit{Cheetah}~\cite{stein_accelerating_2022, kaiser_cheetah_2024}, or surrogate models based on function approximators such as neural networks~\cite{roussel_differentiable_2022, Edelen_using_2017}. 

Nelder-Mead Simplex (NM) \cite{nelder_1965_simplex} is a gradient-free heuristic method that has been used extensively in accelerators for tuning ~\cite{emery_2003_use,zhang_badger_2022,agapov_online_2017,shang_parallel_2005,huang_2018_robust}. It does not learn a model or use curve fitting, but adjusts a ``simplex'' in search space at each iteration. NM requires very little preparation prior to use and is typically computationally inexpensive. 
For examples of studies that have run NM and BO on the same problem, see
~\cite{duris_bayesian_2020, hanuka_physics_2021, miskovich_online_2022, kaiser_learning_2023, xu_xfel_2023}. 
Theoretically speaking, NM is best suited to convex and noise-free objective functions \cite{chong2010introduction}, but it is difficult to assess how this translates to real-world experience in accelerators, where NM has performed well in practice even on quite noisy objectives.

Robust conjugate direction search (RCDS)~\cite{huang_an_2013} has been used for numerous accelerator tuning problems, particularly in rings for nonlinear dynamics optimization. 
In RCDS, local curve fitting at each iteration is used to aid estimation of the curvature of the objective function and the corresponding optimal direction in which to move settings. 
The addition of curve fitting adds robustness in the face of measurement noises and occasional machine failures. 
A successor variant RCDS-S~\cite{zhang_optimization_2022} takes safety constraints and machine drifts into consideration. 

A similar approach is taken in the BOBYQA algorithm, which constructs a second-order local model of function values near a candidate set of optimal parameters \cite{powell2009bobyqa}.
This algorithm has been used to perform optimization in simulation \cite{neveu2017photoinjector, appel2019beam}.
These approaches are similar to BO in the way that they create local models of the objective function to inform parameter selection for episodic optimization.

From the domain of feedback and control, Extremum Seeking (ES) has been applied to many accelerator problems~\cite{scheinker_online_2020,scheinker_model_2013,scheinker_extremum_2022}. 
ES adjusts settings with specific amplitudes and frequencies to approximate the gradient of the cost function and gradient descent, meaning that it works well as a local optimizer.
Furthermore, ES parameter selection is much less expensive than BO methods, allowing it to be used to provide faster feedback than ABO approaches discussed in Sec.~\ref{subsec:time_dependence}.
However, ES can become stuck in local minima if not provided a sufficiently good starting point (e.g. provided by a system model \cite{scheinker_demonstration_2018}), and it does require careful adjustment of the main hyperparameters (the dither amplitude and frequency).  

Finally, Deep Reinforcement Learning (RL) has also found application in the accelerator domain ~\cite{kaiser_learningbased_2022,kain_sampleefficient_2020,meier_optimizing_2022,boltz_feedback_2019,kafkes_developing_2021}. 
While RL is traditionally used to train dynamic feedback controllers, it can also be used to train domain-specific optimization algorithms. 
In the case of RL, this may be referred to as Reinforcement Learning-trained Optimisation (RLO)~\cite{ke_learning_2016}. 
Deep RL is computationally cheap and sample efficient at application time, but requires significant upfront engineering effort to train. 
A case study comparing RL and BO on an accelerator tuning problem was conducted in~\cite{kaiser_learning_2023}.

\subsubsection{Relation to Continuous Control and Time-Dependent Control}

By ``continuous control", we refer to processes that are adjusting settings continuously as the accelerator is running (e.g. orbit feedback, corrections to LLRF phases and amplitudes to maintain the beam energy, etc).  
A further distinction can be made between algorithms that take into account the sequential nature or time-evolution of a problem and those that do not. 
In some classical control techniques such as model predictive control (MPC) \cite{Camacho_introduction_2007} and in reinforcement learning (RL) \cite{Sutton_reinforcement_1998_2018}, the sequentially-dependent nature of a system is formalized as a Markov Decision Process \cite{howard_dynamic_1960}, in which an observed system ``state'' is sufficient to predict the following system evolution. 
MPC and RL include direct consideration of the dynamic evolution of the system over a future time horizon when making decisions in the present. 
To accomplish this, these algorithms have access to or learn the dynamics of the system, and/or approximate solutions to the dynamic optimal control problem.

In contrast, classical BO assumes a stationary (i.e. non-drifting) system where the sequence of control actions is not taken into account in decision making. 
For example, when magnets are not affected by hysteresis, the problem of tuning magnets can be treated as non-sequential. 
When hysteresis effects are present, the sequence of magnet current settings affects the resultant magnetic field; as a result, the problem becomes sequential and this state information should be taken into account in decision making. 
Additionally, because BO is learning a stationary  model of the objective function, its performance can degrade when being run on a non-stationary (i.e. drifting) system; this is why adjustments such as the adaptive BO approaches described in earlier sections are needed in order to run BO continuously as a feedback.

\subsubsection{Relation to Feed-forward Corrections and Warm Starts} ``Warm starts'' or feed-forward corrections from learned models can be used both in continuous control and optimization in accelerators.  
For example, learned models can be used to provide fast setting changes when different setups are desired (e.g. see \cite{Edelen_using_2017,scheinker_demonstration_2018}), followed by fine-tuning with optimization algorithms such as BO. 
Indeed, the system model that provides the warm start can even be the GP model obtained from previous BO runs. 
Continuously-running feed-forward corrections using ML models have also been used in accelerators; for example, this type of approach has been used for source size stabilization in light sources by compensating for optics deviations induced by different insertion devices \cite{Leeman_demonstration_2019}.

\subsection{Model Choices} 
Bayesian optimization takes advantage of Gaussian Process models, which can learn functions that are suitable for interpolation from very few samples and provide fairly robust uncertainty estimates, to perform efficient optimization of expensive objective functions.
However, GP models do not scale as well as other model types, such as neural networks, to large data sets often required to solve high dimensional optimization problems.
As a result, they are more computationally expensive and typically slower to execute when solving high dimensional optimization problems. 
For high-dimensional optimization and faster execution, BO can use other types of ML models so long as an uncertainty estimate is also available, including, but not limited to, Bayesian neural networks, quantile regression with neural networks or neural network ensembles \cite{goodfellow_deep_2016,wilson2016stochastic,wilson2016deep,Springenberg_bayesian_2016}. 
Using different types of surrogate models inside BO can also facilitate inclusion of high-dimensional contextual information (such as initial beam images), which can improve convergence speed.

\subsection{Interpreting BO Performance}
\begin{figure}[h!]
	\includegraphics[width=\linewidth]{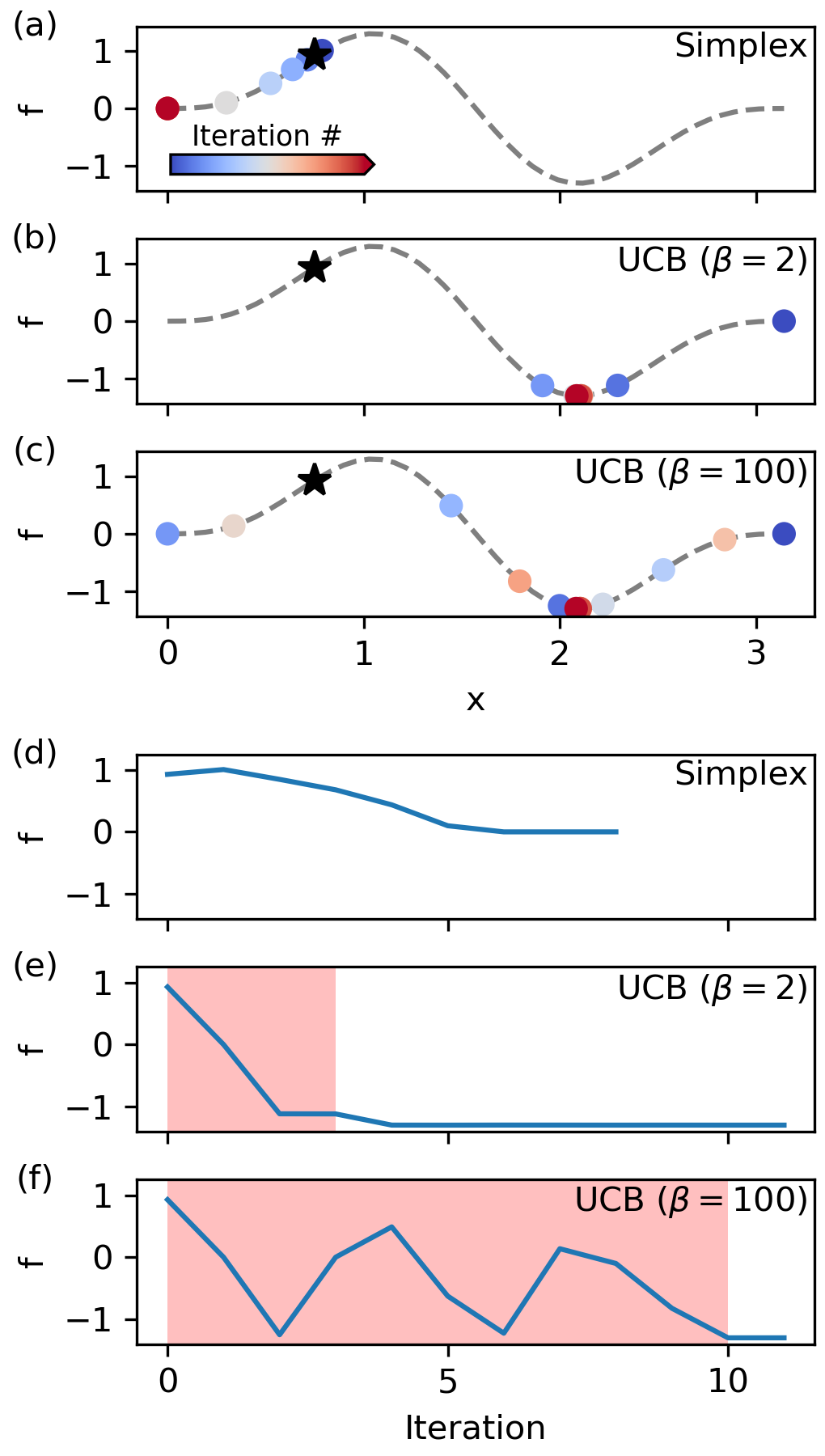}
	\caption{Comparison of optimization performance between a local optimization algorithm (Nelder-Mead simplex), BO using the UCB acquisition function ($\beta=2$), and BO using the UCB acquisition strongly weighted towards exploration ($\beta$=100). All algorithms are initialized with a single observation at $x=0.75$ and aim to minimize the objective function. (a-c) Observations of the objective function in parameter space for each algorithm. The dashed line denotes the true objective function. (d-f) Objective function values as a function of algorithm iteration. Red shading in (e-f) denotes iterations that are used to explore the objective function, ie. points where $\beta\sigma(x^*) > |\mu(x^*)|$ and $x_*$ is the maximum location of the acquisition function. Note that simplex terminates after reaching a convergence criteria.
		\label{fig:simplex_vs_bo}}
\end{figure}

Unlike other optimization algorithms commonly used in accelerator physics, basic BO algorithms are designed to solve global optimization problems.
This can sometimes lead to behaviors (shown in Fig. \ref{fig:simplex_vs_bo}) that are unfamiliar to users expecting to see strong convergence to optimal values during optimization.
Local optimization algorithms, such as Nelder-Mead simplex, often monotonically improve the objective function value, with small excursions around a local optimum to explore the objective function, as shown in Fig.~\ref{fig:simplex_vs_bo}(a,d).
As we see in Fig.~\ref{fig:simplex_vs_bo}(a) this can sometimes lead to converging to a local optimum instead of the global one.

In contrast, BO algorithms often explore the domain to build a global model of the objective function in parameter space before sampling in a local region around the predicted optimal point.
The number of iterations needed to perform this exploration can depend on the relative weighting of exploration vs. exploitation in the acquisition function, the dimensionality of the parameter space, and characteristics of the objective function. 
For example, when the UCB acquisition function is used with roughly even weighting between exploration and exploitation ($\beta=2$), BO briefly explores parameter space before exploiting regions the GP model predicts are likely to be optimal, as shown in Fig.~\ref{fig:simplex_vs_bo}(b,e).
Increasing weighting towards exploration, Fig.~\ref{fig:simplex_vs_bo}(c,f) increases the number of iterations used to explore the objective function before it samples in the globally optimal region of parameter space.
If the amplitude of the objective function far exceeds the predicted uncertainty by the GP model, exploration of the parameter space can cease relatively quickly compared to when optimizing more smoothly varying functions.
Conversely, if the optimum of the objective function is comparable to the predicted uncertainty, strong convergence to the optimal value will only occur once all other areas of parameter space have been explored. 
Increasing the dimensionality of input space further increases the number of iterations used to explore the objective function to build a global GP model.

As a result of the trade-off between exploration and exploitation, new users of BO algorithms who are used to seeing nearly monotonic improvements in the objective value might infer that BO optimization is performing poorly.
However, it is important to keep in mind that this is a direct result of continuously searching for global extremum and does not signify an issue with the optimization algorithm.
If the objective function is expected to be strongly convex, ie. having a single global extremum, BO can be strongly biased towards exploitation through a variety of methods, most notably TuRBO (see Sec.~\ref{subsec:turbo}).
In this case, strong convergence to a fixed location in parameter space is expected.

\subsection{Practical strategies for best performance}
Here we discuss some best practices to improve the performance of BO methods in the field.
\paragraph{Normalizing training data}
As is standard in most machine learning algorithms, it is critical that input data passed to the GP model is transformed prior to training in order to maintain stability of hyperparameter optimization.
It is standard practice to normalize the parameter space to the unit domain $[0,1]$ and to standardize objective function values such that they have a mean of zero and a unit standard deviation.
There are two major benefits to this.
First, transforming training data in this way conditions the derivatives of the marginal log likelihood with respect to hyperparameters to be of unit magnitudes, increasing the stability of gradient descent optimization of the hyperparameters.

Second, data that has been normalized and standardized is more consistent with prior notions incorporated into GP models.
The prior of a GP model is often stated as a distribution of functions with a zero mean and unit standard deviation.
Having data that agrees with this initial prior assumption also improves the robustness of maximizing the marginal log likelihood as well as ensuring that covariance matrices are well-conditioned. 
Finally, it is often advantageous to place reasonable priors on hyperparameters such as the kernel length scale and likelihood noise to regularize hyperparameter training.
Applying these priors to arbitrary modeling problems requires that incoming data is normalized and standardized.

\begin{figure}[h!]          
	\includegraphics[width=\linewidth]{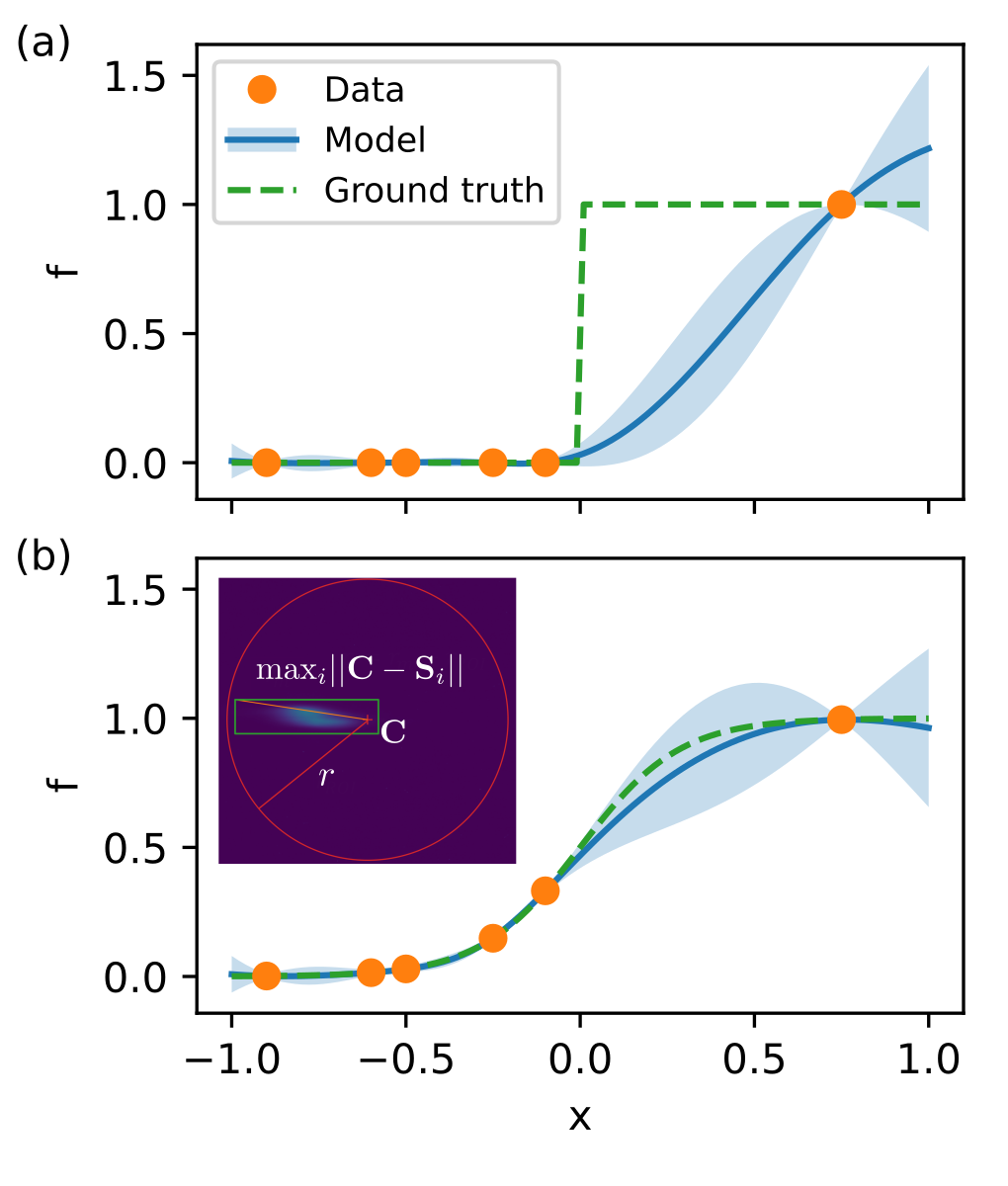}
	\caption{Comparison between GP modeling of hard and soft constraining functions. (a) GP modeling of a heaviside constraining function does not accurately predict constraint values due to a single sharp feature that cannot be learned without dense sampling on either side of the constraint boundary. (b) Smooth constraining functions with a single characteristic length scale are more accurately modeled with GP modeling. Inset: Visualization of bounding box constraint function $f(x) = \max_i \{||\mathbf{C}-\mathbf{S}_i(x)||\}$ used to keep beam distributions inside an ROI, where $r$ is the radius of a circular ROI, $\mathbf{C}$ is the center coordinates of the ROI, and $\mathbf{S}_i$ are corner coordinates of a bounding box around the beam.}
	\label{fig:roi_constraint}
\end{figure}

\paragraph{Defining smoothly varying objectives and constraints} The accuracy of GP predictions relies on learned correlations between function values at different points in parameter space.
This implies that when defining objective functions and constraining functions for BO, it is crucial to ensure that these correlations exist.
An example of where this becomes relevant in accelerator physics is maintaining a beam distribution inside a region of interest (ROI) on a diagnostic screen.
One way to define this constraining function is to return a value of one if the beam is fully within the ROI and a zero otherwise.
However, this is not ideal since it is difficult for the GP model to predict where the boundary between valid and invalid measurements is given a limited set of data values (since function values in space are poorly correlated), as demonstrated in Fig.~\ref{fig:roi_constraint}(a).
On the other hand, if the constraining function measures how close the beam is to violating the constraint, as shown in Fig.~\ref{fig:roi_constraint}(b) and discussed in \cite{roussel2023demonstration}, the GP model can accurately predict extrapolated constraining function values with fewer measurements, which reduces the number of constraint violations during optimization. 

\paragraph{Interpolated measurements}
In some instances during accelerator operations, it is faster to make multiple measurements of the beam distribution than making control decisions using BO algorithms.
Additionally, modifying the accelerator control parameters is more time-consuming than making measurements as well, especially if large changes in parameter values takes a longer amount of time.
This is often the case when tuning magnet parameters, since power supplies take a non-zero amount of time to change the applied current, or in mechanical actuators, where stepper motors take time to traverse the operational range.
Bayesian optimization can reduce the costs of making large changes in input space when performing optimization (see Sec. \ref{sec:proximal}), however this requires re-optimizing the acquisition function at every step, which can become costly.

\begin{figure}[h!]          
	\includegraphics[width=\linewidth]{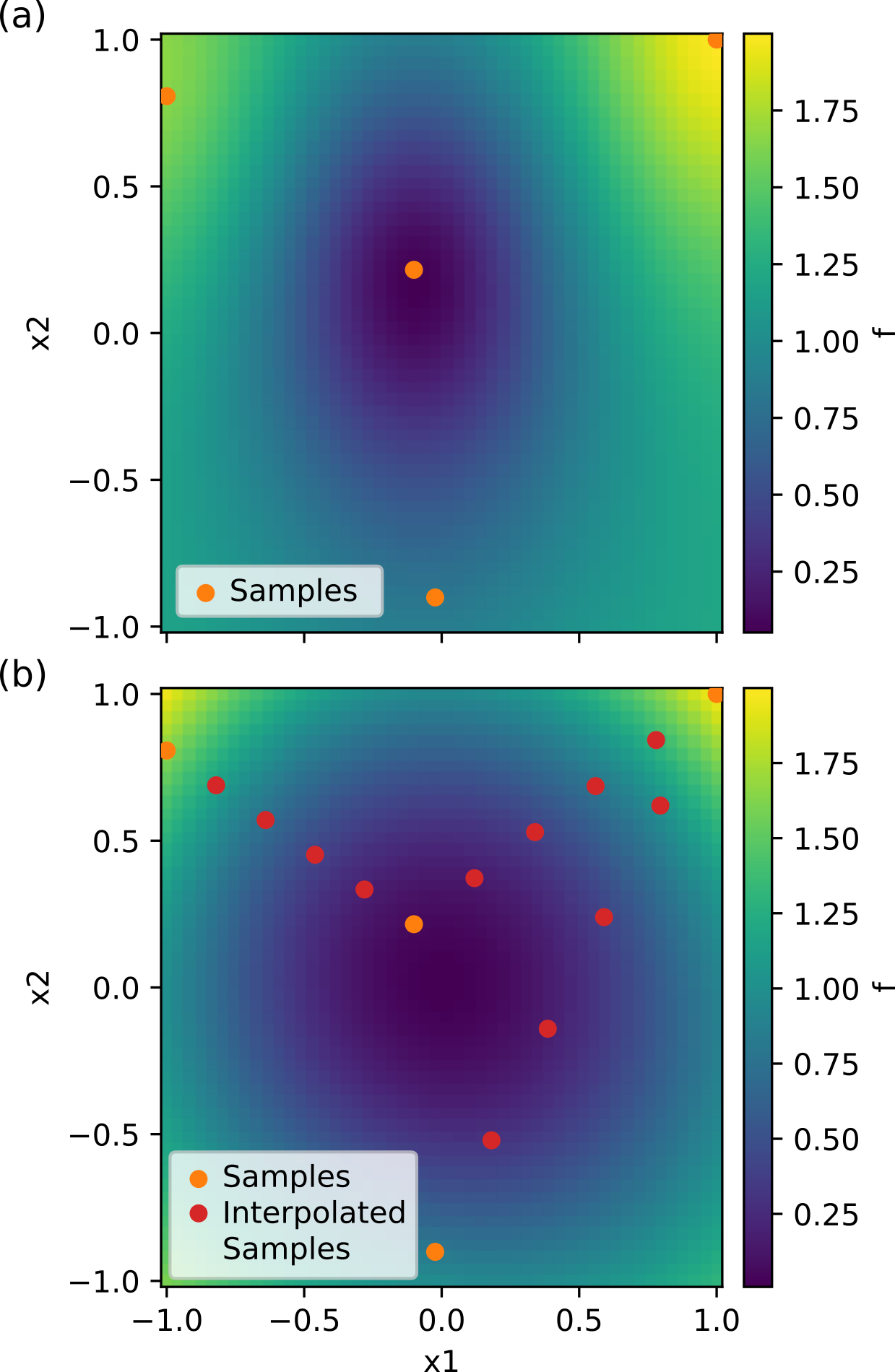}
	\caption{Comparison between GP modeling of the 2-dimensional sphere function $f(x_1,x_2)=x_1^2 + x_2^2$ with and without interpolated measurements. (a) Shows the posterior mean of the GP model with 4 measurements taken sequentially. (b) Shows the same 4 measurements taken sequentially but with interpolated points in-between each measurement. Incorporating interpolated points in the data set leads to higher modeling accuracy, leading to accurate identification of the sphere function minimum at the origin.}
	\label{fig:interpolate}
\end{figure}

An alternative approach to minimizing measurement costs on the total optimization cost it to pre-compute a set of measurements during each optimization step.
Instead of immediately jumping to the next point proposed by BO during optimization, we can generate a set of future observation points that interpolate between the current set-point and the future set-point, as shown in Fig. \ref{fig:interpolate}.
This allows multiple measurements to be taken quickly throughout the input space without the need to wait long periods of time for accelerator parameters to change or to re-optimize the acquisition function for each measurement.
As a result, the GP model BO uses increases in accuracy faster due to the additional data, enabling BO to make better decisions with fewer optimization iterations.
This does however increase the computation time associated with training the GP model and optimizing the acquisition function, requiring careful consideration of the trade-offs associated with using this technique.

\paragraph{Leveraging batch computations}
To address modern challenges in high performance computation, significant effort by those in the machine learning community has focused on developing hardware and software that enables fast matrix manipulations.
For example, GPUs are specifically designed to perform difficult matrix computations extremely quickly due to massive hardware parallelizations.
Bayesian optimization computations are well suited to take full advantage of these developments as most computations involved in making GP predictions or computing acquisition function values involve matrix manipulations.
Extending the evaluation of GP models or acquisition functions in parallel using \textit{batched computations} (which adds new dimensions to matrices used in evaluations) plays a critical role in leveraging modern computing hardware and software to improve performance.

A core application of batched computation is acquisition function optimization.
Optimizing acquisition functions is often a challenging problem, since they usually are not convex and can contain many local extrema.
Multi-restart optimization can be used in this case to improve the search for a global maximum by restarting optimization at a number of different initial starting points.
Batched computation allows this process to happen in parallel, significantly reducing the computation time needed to maximize the acquisition function while leveraging the advantages provided by fast matrix computational techniques.
As a result, high performance software libraries that implement BO take advantage of this technique (see Sec.~\ref{subsec:implementations}).

\subsection{Implementations of BO}
\label{subsec:implementations}
There are several open-source software packages that implement GP modeling and BO, mostly using the Python programming language for API and C/C++ for computations.
It is strongly recommended that practitioners of BO do not ``reinvent-the-wheel" when trying to implement BO algorithms to solve their specific optimization problems.
Current implementations of BO (as described below) have capabilities that cover a wide range of accelerator physics problems and applications.
If further modifications are needed to tackle a specific problem it is strongly recommended that these modifications are built from existing software packages with the intent to contribute back to the existing package for others in the community to use.
This will accelerate the state-of-the art for all parties, and prevent a fractured landscape of competing implementations that hinder algorithmic development and application to optimization problems.

\paragraph{Scikit-learn}
The Scikit-learn general machine learning package \cite{scikit-learn} provides a simple implementation of basic GP modeling and BO while also providing good documentation, making it a good resource for gaining experience using basic BO algorithms.

\paragraph{BoTorch, GPyTorch, and Ax}
This set of open source packages are developed and maintained by Cornell University, Columbia University, University of Pennsylvania, New York University, and Meta \cite{balandat_botorch_2020} and provide implementations of state-of-the-art GP modeling and BO.
They are built upon the PyTorch \cite{paszke_pytorch_2019} machine learning language that implements automatic differentiation and GPU computing, both of which significantly improve the speed and performance of BO.
BoTorch relies on a lower level package, GPyTorch \cite{gardner2018gpytorch}, to implement GP models, allowing significant customization of all aspects of GP modeling, including custom kernels, priors and likelihoods.
BoTorch also takes advantage of batched Monte Carlo sampling to maximize performance when computing and optimizing acquisition functions. With recent improvement in PyTorch like the JIT compiler, BoTorch stack is very competitive in performance benchmarks and is highly amenable to GPU acceleration.
BoTorch is complemented by Ax, which provides an accessible user interface to BoTorch.

\paragraph{Xopt} 
The Xopt Python package \cite{roussel2023xopt} is a high-level optimization package developed at SLAC that connects advanced optimization algorithms to arbitrary optimization tasks (in both simulations and experiments), with a focus on solving problems in accelerator physics.
The object-oriented structure of Xopt allows for significant flexibility in defining and executing optimization processes, including specification of optimization runs through simple text files (YAML,JSON), asynchronous evaluation of objective functions, model introspection during optimization, and human-in-the-loop optimization. 
Xopt implements a number of algorithms for easy off-the-shelf use, including most of the BO algorithms and techniques discussed in this review.
These algorithms can be easily tailored towards solving specific optimization problems through the use of sub-classing.
Xopt has been developed by the SLAC machine learning (ML) group specifically to address optimization problems in accelerator science, with the ability for extension and customization for other scientific fields.
It has been used to perform online accelerator control at a number of facilities including LCLS, LCLS-II, FACET-II (SLAC), AWA, ATLAS (Argonne), FLASH, FLASHForward, European XFEL, Petra-III (DESY), RHIC, NSLS-II (BNL), ESRF, and LBNL. 
It has also been used to perform optimization in simulation at Cornell University, University of Chicago, and on high performance computing (HPC) clusters such as NERSC.

\paragraph{Badger}
The Badger package \cite{zhang_badger_2021,zhang_badger_2022}, also developed by the SLAC ML group as a successor to DESY's \textit{Ocelot Optimizer}~\cite{agapov_online_2017}, provides an easy to use graphical user interface for accelerator control rooms to interface with algorithms implemented by Xopt.
It provides an extendable interface for communicating with a variety of accelerator control systems and can be customized with extensions to provide online analysis of optimization performance and algorithm introspection.

\paragraph{Optimas}
The Optimas package \cite{ferran_pousa_bayesian_2023}, focuses on optimization workflows using numerical simulations at varying computational scales, from laptops to high-performance computing platforms. Optimas relies on the library libEnsemble \cite{hudson_library_2022} to orchestrate multiple simulations running concurrently as part of the optimization, and to allocate appropriate multi-CPU and multi-GPU resources to each of these simulations (as well as GPU resources, if needed, for the Bayesian optimizer).
Optimas provides multiple algorithms for parallel parameter exploration and optimization such as single- and multi-objective Bayesian optimization, including multi-fidelity and multi-task options. It is also highly interoperable with the Ax library.
Optimas has been used on large-scale clusters such as Perlmutter (NERSC) and JUWELS (JSC), and is developed by a collaboration between DESY, Lawrence Berkeley National Laboratory, and Argonne National Laboratory.

\paragraph{APSopt}
The APSopt package \cite{kuklev_apsopt_2023} is being developed by the APS accelerator physics and operations group to integrate internally and externally developed BO, RL, and classical methods into a robustly tested tool for both API-based use by physics experts and GUI-only use by operators. It aims to provide a coherent optimization environment through a number of advanced features for data management, distributed client-server operation, automatic initialization with machine-tuned algorithms, parameter hints, and human-in-the-loop interactive model review and refinement. It has been experimentally tested in the APS injector, APS storage ring, NSLS-II storage ring, Fermilab IOTA/FAST complex, and is being used extensively for the APS-Upgrade commissioning.

\paragraph{GeOFF}
The Generic Optimisation Framework (GeOFF) \cite{geoff} is being developed by the data science teams at CERN and GSI. It allows to easily integrate RL, BO and numerical optimisation in the control room or for offline optimisation on e.g. simulation. The optimisation problem definition 
can handle arbitrarily complex controls or simulation processes as long as a Python interface is available. GeOFF also comes with a plug-and-play graphical user interface to test and run the optimisation or continuous control problems. It is routinely used at CERN for the entire accelerator complex and for various problems at GSI.  

\subsection{Computational requirements and scaling}
\label{subsec:computational_benchmarks}
Because of the complexity involved in efficiently implementing the low-level mathematical routines, above BO packages rely on linear algebra or machine learning frameworks, predominantly PyTorch. 
While the theoretical complexity and storage scaling of BO methods is well understood and has been discussed previously, in practice the performance of the PyTorch/GPyTorch/BoTorch libraries can deviate significantly from ideal behavior.

\begin{figure}[h]
	\includegraphics[width=\linewidth]{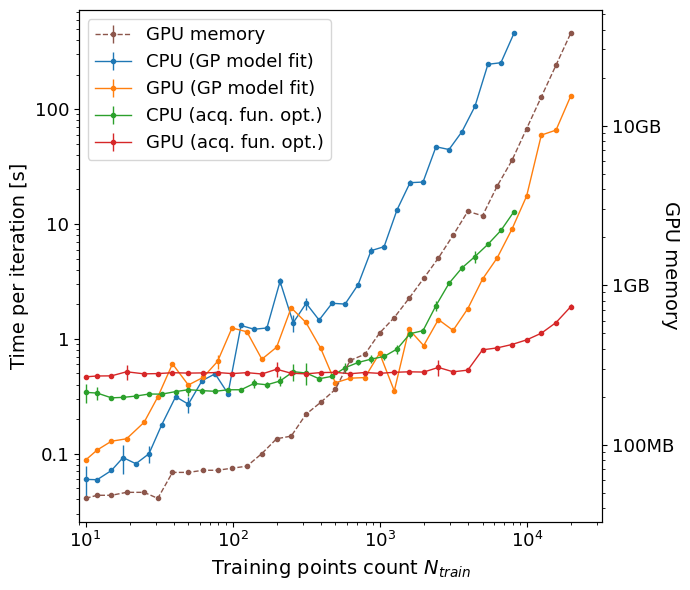}
	\caption{Performance scaling with dataset size for BoTorch/GPyTorch (0.9.4/1.11) libraries on a single-objective optimization run. Synthetic 5-variable quadratic objective was used with Monte Carlo version of UCB acquisition function and 100 Adam optimizer iterations. GPU memory usage only applicable to GPU runs.
		\label{fig:bo_hw_performance}}
\end{figure}

Due to the high costs of the specialized GPU hardware it is critical to understand what tasks are computationally feasible in practice. 
In Fig. \ref{fig:bo_hw_performance} we provide benchmark results for a `mid-range' 2023 ML hardware setup consisting of 16 cores (32 threads) from AMD EPYC 7742 CPU and a single A100-40GB GPU. We do not consider multi-GPU configurations, but they are supported by PyTorch - this yields only a slight increase in the feasible problem sizes.

We benchmark three typical components of the BO process - GP model fitting, GP model evaluation, and acquisition function optimization (which involves model evaluation and auto-grad operations as part of the optimizer loop). The overall scaling is consistent with expectations, with model fitting and evaluation showing $\mathcal{O}(n^3)$ growth as a function of number of collected points $n$. However, the progression is not smooth, with repeatable deviations at particular sizes due to different bottlenecks and code paths that are encountered depending on internal PyTorch configuration. Note also that there is a constant time floor of $100-1000$ ms per BO loop due to initialization, data copies, and Python overhead - in practice this limits BO applications to making sub-1Hz decisions (although data acquisition can take place at a higher rate).

The ultimate limit on number of model points is determined by available memory, and is encountered at $\sim 25$k points on a 40GB GPU (at which point CPU is too slow even if there is sufficient RAM). Approximate GP methods can extend this limit, but are not particularly popular in BO applications. Our practical recommendation is to limit problem sizes to 10k points with a GPU and 3k with a CPU-only machine, and apply BO only in cases when objective evaluation time is sufficiently long to amortize computational costs for your particular choice of model, acquisition function, and hyperparameters (see Sec.~\ref{sec:opt_algorithm_selection}). This ensures that BO use is worthwhile in terms of overall wall-clock convergence speed.

\subsection{Future directions for BO research in accelerator science}
While BO algorithms have been shown to be able to solve a wide variety of accelerator physics problems in an efficient manner, there are still ample opportunities for future improvements towards using BO in accelerator science.

First and foremost is continuing research in the integration of physics information into GP models.
As has been highlighted in several sections of this review, improving the accuracy of GP modeling improves decision-making during optimization, leading to faster convergence to optimal solutions and reductions in the number of constraint violations.
Incorporating information into GP models before performing optimization is especially critical in making good decisions during the first few iterations.
Furthermore, if uncertainties exist in the sources of information used, these uncertainties should be incorporated into the GP model as well.

In both online accelerator operations and in simulated optimization, improving the orchestration of objective function evaluation, GP model generation, and acquisition function maximization is another source of major potential improvements.
The development of a centralized control framework that dispatches these tasks contained in BO on parallel resources could lead to major reductions in the overall cost of performing optimization.
A potential example of this would be an online accelerator control program that would send current and/or future potential machine states to be evaluated on high performance computing clusters outside the control room.
Results collected from these physics simulations could be used to inform online control in real-time, similar to what is done in \cite{gulliford2013demonstration}.

\section{Conclusion} \label{sec:conclusion}
In conclusion, BO algorithms are an effective, extendable way of solving a wide variety of optimization challenges in accelerator physics.
BO algorithms are particularly valuable when dealing with optimization challenges that involve significant resource expenses, such as beam time, personnel, or computational resources.
These algorithms use statistical surrogate models based on gathered data to inform optimization, reducing the number of objective function evaluations versus other black box optimization schemes.
As a result, the BO framework provides a straightforward and robust way to incorporate prior knowledge (either from past measurements or physics information) or approximate measurements/computation into the modeling process to further improve optimization convergence speed.
By modifying standard acquisition functions, BO algorithms can be customized to solve a wide variety of single, multi-objective, and characterization problems in accelerator physics.
Using BO algorithms can reduce the overall cost of performing optimization when compared to conventional black box optimization algorithms, allowing accelerator scientists to address more complex optimization challenges.



\begin{acknowledgments}

This work was supported by the U.S. Department of Energy, Office of Science, Office of Basic Energy Sciences, under Contract No. DE-AC02-76SF00515.
This manuscript acknowledges the support of Fermi Research Alliance, LLC under Contract No. DE-AC02-07CH11359 with the U.S. Department of Energy, Office of Science, Office of High Energy Physics.
This work was supported by the U.S. Department of Energy, Office of Science, Office of Basic Energy Sciences, under Contract No. DE-AC02-06CH11357, through a “Data, Artificial Intelligence, and Machine Learning at DOE Scientific User Facilities” Grant from the DOE’s Office of Nuclear Physics.
This work has in part been funded by the IVF project InternLabs-0011 (HIR3X) and the Initiative and Networking Fund by the Helmholtz Association (Autonomous Accelerator, ZT-I-PF-5-6). The authors acknowledge support from DESY (Hamburg, Germany) and KIT (Karlsruhe, Germany), members of the Helmholtz Association HGF. 
Work supported by Brookhaven Science Associates, LLC under Contract No. DE-SC0012704 with the U.S. Department of Energy.
M.J.V.S. acknowledges support from the Royal Society URF-R1221874.
W.L. acknowledges support from the U.S. National Science Foundation under Award PHY-1549132.

Conceptualization, R.R., A.S.G., and A.L.E.; Data curation, R.R.; Visualization, R.R., T.B., M.J.V.S., J.O.L., N.K., A.L.E., S.M.L, R.L.; Writing---original draft, R.R., D.K., Y.G., W.L., T.B., C.X., A.S.G, J.M., J.K., A.E., J.O.L., N.K., V.K., W.N., Z.Z., R.L., and A.L.E.; Writing---review and editing, R.R., D.K., W.N., N.M.I, Y.G., W.L., T.B., M.J.V.S., A.E., J.K., C.X., A.S.G., B.M., N.K., V.K., X.H., D.R., J.S.J, and A.L.E. All authors have read and agreed to the published version of the manuscript.
\end{acknowledgments}

%


\end{document}